\documentclass[twocolumn,floatfix]{aastex631}

\hypersetup{linkcolor=blue,citecolor=blue,filecolor=cyan,urlcolor=magenta}

\usepackage{subfigure} 
\usepackage{placeins} 
\usepackage{amsmath}
\usepackage{soul}

\received{November 15, 2023}
\revised{February 24, 2024}
\accepted{March 11, 2024}

\begin{document}

\title{Identifying coronal sources of L1 solar wind disturbances using the Fisk heliospheric magnetic field and potential field extrapolations during three solar minima}

\correspondingauthor{Ruhann Steyn}
\email{Ruhann.Steyn@wnu.ac.za}

\author[0000-0003-2099-8093]{P. J. Steyn}
\affiliation{Centre for Space Research, North-West University, Potchefstroom, South Africa, 2522}

\author[0009-0008-6440-0173]{D. Johnson}
\affiliation{Department of Mathematics, Physics and Electrical Engineering, Northumbria University, Newcastle upon Tyne, NE1 8ST, United Kingdom}
\affiliation{School of Mathematics and Statistics, University of St Andrews, St Andrews, Fife, KY16 9SS, United Kingdom}

\author[0000-0002-5915-697X]{G. J. J. Botha}
\affiliation{Department of Mathematics, Physics and Electrical Engineering, Northumbria University, Newcastle upon Tyne, NE1 8ST, United Kingdom}

\author[0000-0001-8954-4183]{S. R\'egnier}
\affiliation{Department of Mathematics, Physics and Electrical Engineering, Northumbria University, Newcastle upon Tyne, NE1 8ST, United Kingdom}

\begin{abstract}
The solar minima between solar cycles $22$-$23$, $23$-$24$ and $24$-$25$ are the best observed minima on record. In situ solar wind and interplanetary magnetic field measurements by the WIND and ACE spacecraft at L$1$ with one-hour cadence are explored using wavelet analyses for the most quiescent year during each minimum. Times of local peaks in periodicities are identified in the solar wind velocity, magnetic field components, and proton number densities. The measured radial velocities at these times are used to trace magnetic field lines to the photosphere using two models. The first is the Fisk heliospheric magnetic field that traces field lines from L$1$ to the photosphere. They connect exclusively to solar poles and in $88\%$ instances to locations of polar coronal holes. The second model uses the Parker spiral to trace from L$1$ to the solar source surface and potential field extrapolations from the source surface to the photosphere. These field lines terminate at equatorial and mid-latitude coordinates of which some are located close to coronal holes. This study connects for the first time coronal hole signatures in the ecliptic plane at L$1$ with polar coronal holes using the Fisk field. It shows how sources from both the solar equator and poles influence the solar wind at L$1$ and how the two models compliment each other to identify these sources.
\end{abstract}


\section{Introduction} 
\label{Sec:Introduction}
Epochs of solar minima present unique opportunities to study the properties of the solar wind. While direct in situ measurements of the solar wind were first made in the early $1960$s \citep{Neugebauer1962}, continuous, high-resolution data were not available until the mid $1990$s. As a result, the last three solar minima, between cycles $22$-$23$ ($1996$/$1997$), $23$-$24$ ($2008$/$2009$), and $24$-$25$ ($2019$/$2020$) are the best observed minima on record \citep{Jian2011, Carrasco2021, Riley2022}. The aim of this study is to investigate the observed solar wind disturbances at the L$1$ Lagrangian point upstream of Earth during the previous three solar minima and their connection to solar coronal holes (CHs). Previous studies, such as \cite{SpaceWeather2022}, report that low to mid-latitude CHs dominate upstream plasma measurements and has an impact on space weather. 

It is known that the $27$-day synodic solar equatorial rotation period is observed in the solar wind \citep{Murula1996}. Due to the differential rotation of the photosphere, which in turn influences the corona, periodicities between $25$ and $30$ days are observed during low solar activity \citep{SpaceWeather2022}. 

This paper uses the observed periodicities in the radial, tangential, and normal components of both the solar wind velocity and the interplanetary magnetic field (IMF) as a proxy to trace field lines between the photosphere and L$1$. Proton number densities are also investigated. The photospheric origin of these components are of particular interest. A Fisk heliospheric magnetic field (HMF) model is assumed between L$1$ and the solar wind source surface (SWSS) after which the field lines are mapped to the photosphere using heliographic and heliomagnetic coordinate transformations around the tilt angle $\alpha$ \citep{Fisk1996,Zurbuchen1997,Fisk1999}. For comparison, a standard Parker HMF \citep{Parker1958} configuration is assumed between L$1$ and the SWSS after which the field lines are mapped to the photosphere incorporating potential field source surface (PFSS) extrapolations. The assumption of a background Parker field in the ecliptic plane has been used in numerous studies \citep{Li_2016, Strauss_2017, Zhao2017}.
\cite{Posner2001} used the Fisk HMF to study magnetic reconnection processes at the boundaries of CHs by investigating solar wind composition measurements from the Ulysses spacecraft at large radial distances ($\sim 4$ AU near the streamer belt). The scope of this paper, however, is constrained to the ecliptic plane and L$1$.

Due to ease of use, many studies use PFSS from the solar surface to the SWSS and model the Parker HMF beyond the SWSS \citep{Owens2013, Balogh2013}. Alternative methods tracing magnetic field lines from the solar surface to the SWSS exist, such as non-linear force free fields and steady-state magnetohydrodynamics \citep{Parenti2021}, magnetofrictional relaxation \citep{Yeates2018}, force-free electrodynamics \citep{Contopoulos2013} and the solar-interplanetary space-time conservation element and solution element magnetohydrodynamic (SIP-CESE MHD) model of \cite{Feng2012}. Alternative methods to trace the HMF from the solar surface into the heliosphere are relaxation of magnetic field lines \citep{Gilbert2007}, steady-state magnetohydrodynamic models \citep{Garraffo2013} and the Fisk model \citep{Fisk1996,Zurbuchen1997}. \cite{Zhao2017} point out that in spite of its approximations, the PFSS and Parker combination is useful to identify solar wind source regions from solar wind measurements on the ecliptic plane. This study will compare its results with the standard non-modified Fisk HMF between L$1$ and the photosphere.

The solar wind can be characterised by proton speed and heavy ion charge states, such as O$^{7+}$/O$^{6+}$ and O$^{6+}$/O$^{5+}$ oxygen ratios \citep{Zhao2017}. The solar wind speed varies as it moves towards L$1$ \citep{Jain2024}. In addition, the solar wind originating from equatorial CHs contains fast as well as slow components \citep{Wang2009,Stakhiv2015} and the various plasma populations interact with one another \citep{Stansby2019}. In this study, as will be shown later, a constant solar wind speed is assumed. The constraints of assuming a constant speed are mitigated by investigating measurements over a time interval at L$1$, rather than using a velocity at one time instant. The obtained results are then corroborated by taking oxygen ratios over the same time interval.

The data retrieval and processing is explained in Section \ref{section:Data}, an introduction to the two models is described in Section \ref{section:Magnetic field models}, the wavelet analyses of the three solar minima are presented in Section \ref{section:SW and Bfield properties at L1}, the Carrington maps are shown in Section \ref{section: Carrington Maps}, and the concluding remarks are made in Section \ref{Sec: Discussion and Conclusion}.

\section{Data}
\label{section:Data}
\subsection{Solar wind and interplanetary magnetic field}
\label{subsection:SW and Bfield}
One-hour cadence solar wind velocity component measurements and proton number density measurements for the three observed solar minima are from the Wind/SWE data sets \citep{Ogilvie1995,WIND2021}. The IMF measurements are from the Wind/MFI instrument \citep{Lepping1995} for the $1996$ solar minimum and from the ACE Magnetic Field Experiment data sets \citep{Smith1998} for the remaining solar minima. Time periods of one year, based on the lowest monthly smoothed sunspot number \citep{SILSO}, are used for each solar minimum epoch investigated in this study. The solar wind velocity data are converted from the GSE (geocentric solar magnetospheric) system of coordinates to the RTN (radial, tangential, normal) system of coordinates as follows, 
\begin{equation}
    V_r = -V_x, \hspace{0.3cm} V_t = -V_y, \hspace{0.3cm} V_n = V_z,
    \label{eq:RTNcoordinates}
\end{equation}
where $V_r$, $V_t$, and $V_n$ are the radial, tangential and normal solar wind velocity components respectively. The same conversions are followed for the IMF data. Known interplanetary coronal mass ejections (ICMEs) are removed from the solar wind and magnetic field data sets according to the ICME database of \cite{Richardson2010}. We replace the ICME data with Gaussian white noise with a mean and standard deviation computed from the data before and after the ICME. This successfully removes their signal from the wavelet analyses.

\subsection{Carrington Maps}
\label{subsection: carrington Maps Data}

For the $1996$ solar minimum, a set of Carrington maps is available from the SOHO/EIT Synoptic Map Database \citep{Benevolenskaya2001,CarringtonMaps1996}, which provides maps for Carrington rotations (CRs) $1911$ to $2055$. These Carrington maps are restricted to $\pm83^{\circ}$ due to Earth's changing vantage point \citep{Hamada2018}. Carrington maps for the $2008$/$2009$ and $2019$/$2020$ solar minima are assembled from full-disk solar images obtained from the European Space Agency's SOHO Science Archive database \citep{CoronalHoles} using the $195$\AA \hspace{0.1cm}wavelength. The Carrington maps are assembled following the methods of \cite{Thompson1997} and \cite{Thompson2006} where slices of $13.3^{\circ}$ longitude (corresponding to one day’s CR length scale) are extracted centred on the central meridian. The tilt angle $\alpha$ for each CR is from \cite{Hoeksema1995_Tiltangles}\footnote{\url{http://wso.stanford.edu/Tilts.html}}. 

\subsection{Photospheric magnetic field}
\label{subsection:Photospheric BField}
This study uses the PFSS packages\footnote{\url{https://www.lmsal.com/~derosa/pfsspack}} included in SolarSoft \citep{SSW1998, DeRosa2003}. The magnetograms used in the PFSS model to determine the configuration of the photospheric magnetic field are from SOHO/MDI \citep{SOHO1995,SOHO_MDI} for the solar minima of $1996$/$1997$ and $2008$/$2009$. Magnetograms from the HMI instrument \citep{SDO_HMI} onboard SDO \citep{SDO} are used for the $2019$/$2020$ solar minimum. The magnetograms used by this PFSS model is not controlled by the user.  

\subsection{Coronal holes}
\label{subsection:Coronal Hole data}
The locations and polarity of CHs are confirmed by the daily National Oceanic and Atmospheric Administration (NOAA) solar synoptic analysis charts \citep{SynopticCharts}. In addition, the heavy ion charge state ratio (O$^{7+}$/O$^{6+}$) obtained from ACE observations is used to confirm the locations of coronal holes.

\section{Magnetic field models}
\label{section:Magnetic field models}
\subsection{Fisk heliospheric magnetic field model}
\label{subsection:Fisk HMF Model}

\begin{figure}[t]
\centering
\includegraphics[width=0.45\textwidth]{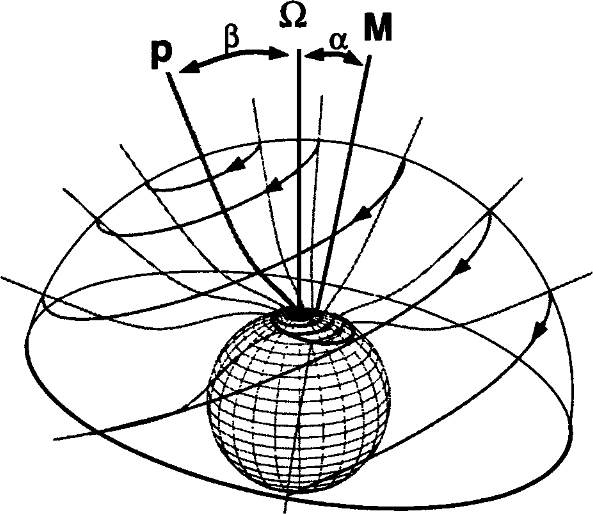}
\caption{An illustration of the expansion of magnetic field lines from the photosphere to the SWSS from rigidly rotating PCHs according to the Fisk HMF model. The direction of the footpoint trajectories are shown on the source surface. The rotation axis $\mathbf{\Omega}$, the magnetic axis $\mathbf{M}$, the $\mathbf{\hat{p}}$ axis, the tilt angle $\alpha$, and the Fisk-angle $\beta$ are shown. Figure adapted from \cite{Zurbuchen1997}.}
\label{fig:FiskBall}
\end{figure}

The tracing of Fisk magnetic field lines expanding from polar coronal holes (PCHs) are discussed in this section. \cite{Fisk1996} introduced a novel HMF model in an effort to explain the recurrent energetic particle events observed at high latitudes by the Ulysses spacecraft \citep{Simpson1995}. Figure \ref{fig:FiskBall} shows the geometry of the model proposed by \cite{Fisk1996}. The rotational axis $\mathbf{\Omega}$ is separated from the magnetic axis $\mathbf{M}$ by a tilt angle $\alpha$. The $\mathbf{\hat{p}}$-axis is defined by the magnetic field line originating from the solar pole where no differential rotation is assumed and is separated from the rotational axis by an angle $\beta$. Fisk-type field lines are assumed to originate from rigidly-rotating PCHs (the source of the fast solar wind) and to expand from the photosphere to the source surface symmetrically about the magnetic axis. The differential rotation of the field line footpoints on the photosphere and the super-radial expansion of field lines to the source surface cause the Fisk model to display large excursions in heliographic latitude. This unique characteristic makes this HMF model a good candidate to trace field lines from L$1$ back to their coronal, and subsequently, their photospheric origin. The expansion of the field lines form footpoint trajectories on the source surface symmetric about the $\mathbf{\hat{p}}$-axis. \cite{Zurbuchen1997} report the expression of the field to be 
\begin{equation}
    \mathbf{B} =B_0\left( \frac{r_o}{r}\right)^2\left[\mathbf{\hat{e}}_r - \frac{r\omega}{V} \mathbf{\hat{e}}_{\theta} - \frac{(\Omega-\omega)r\sin\theta}{V}\mathbf{\hat{e}}_{\phi}\right],  
    \label{eq:Fisk}
\end{equation}
where $r$ represents the radial distance away from the sun, $B_0$ is the field strength at $r_0$, $V$ is the solar wind speed, $\omega$ the differential rotation rate of the photosphere (typically assumed to be a constant fraction of the solar equatorial rotation rate $\Omega$, i.e., $\Omega$/$4$), and $\theta$ the heliographic co-latitude from which the field line expands. Note that for $\theta=90^{\circ}$ (the solar equator) and $\omega=0$ rads/sec (no differential rotation) equation (\ref{eq:Fisk}) reduces to the standard Parker HMF expression \citep{Parker1958}. Since each field line trace is unique, field lines can be traced from the source surface to the photosphere using this method. \cite{Steyn2020} report this tracing in three distinct processes. Firstly, a transformation about the tilt angle $\alpha$ from heliographic to heliomagnetic coordinates on the source surface is performed (equations (\ref{eq:FiskTrans1}) and (\ref{eq:FiskTrans2})). Next, the tracing from the source surface to the photosphere in both latitude and longitude is done based on the Divergence Theorem (equations (\ref{eq:FiskTrans3}) and (\ref{eq:FiskTrans4})), and finally a transformation about the tilt angle from heliomagnetic to heliographic coordinates are performed (equations (\ref{eq:FiskTrans5}) and (\ref{eq:FiskTrans6s})). This process is described by the following expressions
\begin{equation}
    \theta_{hm}^{ss} = \cos^{-1}(\cos\theta_{hg}^{ss}\cos\alpha+\sin\theta_{hg}^{ss}\sin\alpha\cos\phi_{hg}^{ss}),
    \label{eq:FiskTrans1}
\end{equation}
\small
\begin{equation}
    \phi_{hm}^{ss}=  \cos^{-1}\left(\frac{\sin\theta_{hg}^{ss}\cos\phi_{hg}^{ss}\cos\alpha-\cos\theta_{hg}^{ss}\sin\alpha}{\sin\theta_{hm}^{ss}}\right),
    \label{eq:FiskTrans2}
\end{equation}
\normalsize
\begin{equation}
    \theta^{ph}_{hm}=\sin^{-1}\left(\sqrt{\frac{(1-\cos\theta_{hm}^{ss})\sin^2\theta_{mm}^{ph}}{(1-\cos\theta_{mm}^{ss})}}\right),
    \label{eq:FiskTrans3}
\end{equation}
\begin{equation}
     \phi_{hm}^{ph} = \phi_{hm}^{ss},
    \label{eq:FiskTrans4}
\end{equation}
\begin{equation}
    \theta_{hg}^{ph}= \cos^{-1}(\cos\theta_{hm}^{ph}\cos\alpha-\sin\theta_{hm}^{ph}\cos\phi_{hm}^{ph}\sin\alpha),
    \label{eq:FiskTrans5}
\end{equation}
\small
\begin{equation}
\phi_{hg}^{ph}= \\
\cos^{-1}\left(\frac{\cos\theta_{hm}^{ph}\sin\alpha+\sin\theta_{hm}^{ph}\cos\phi_{hm}^{ph}\cos\alpha}{\sin\theta_{hg}^{ph}}\right),
    \label{eq:FiskTrans6s}
\end{equation}
\normalsize
where the subscripts $hg$ and $hm$ refers to heliographic and heliomagnetic coordinates respectively, and the superscripts $ss$ and $ph$ refers to the source surface and photosphere respectively, and where $\sin\theta_{hm}^{ss}\neq 0$, $\cos\theta_{mm}^{ss}\neq 1$, and $\sin\theta_{hg}^{ph}\neq 0$. The boundaries of the PCHs in heliomagnetic coordinates on the photosphere and source surface are represented by $\theta_{mm}^{ph}$ and $\theta_{mm}^{ss}$ and their values are chosen to be $24^{\circ}$ and $70^{\circ}$ respectively, in accordance to the example shown in \cite{Fisk1996}.

\begin{figure}[t]
\centering
\includegraphics[width=0.47\textwidth]{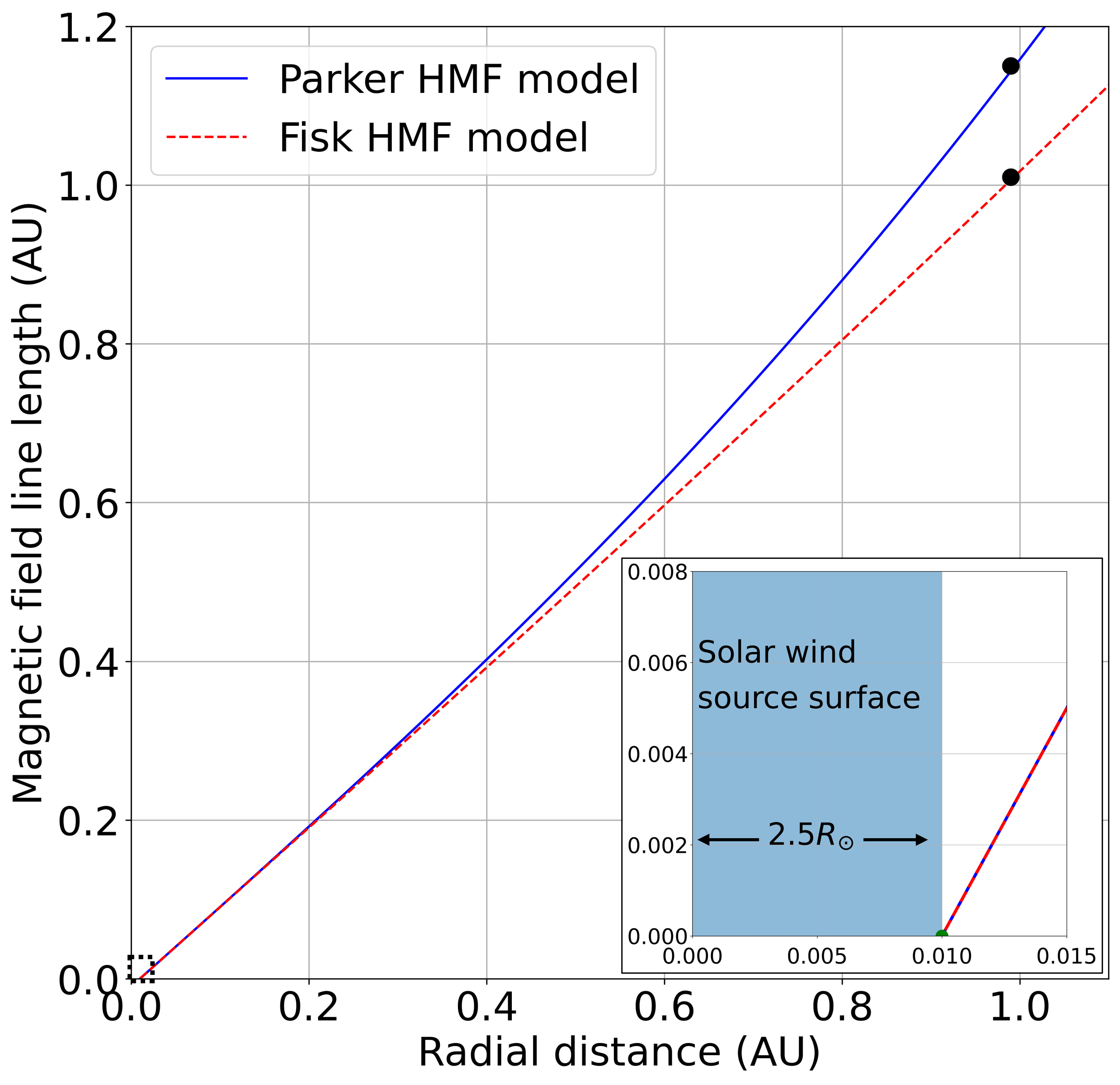}
\caption{Magnetic field line lengths of the Parker (solid blue) and Fisk (dashed red) HMF models between the SWSS and L$1$. The black dots refers to the L$1$ position at $r=0.99$ AU. The insert graph is of the square at the origin of the main graph showing the starting point of the magnetic field lines at the SWSS boundary located at $r=2.5R_{\odot}=0.01$ AU.}
\label{fig:ArcLengths}
\end{figure}

This study assumes that the measured radial solar wind velocity ($V_r$) remains constant between L$1$ and the SWSS during the time of tracing a field line. The length of a magnetic field line from L$1$ at $r=0.99$ AU to the SWSS at $r_0=2.5R_{\odot}=0.01$ AU is determined by $S=\int_{r_0}^{r}|d\mathbf{\ell}|$ where $|d\mathbf{\ell}|^2= dr^2 + r^2 d\theta^2 + r^2 \sin^2 \theta d\phi^2$ in spherical coordinates. $V_r$ measured at the time of the local maximum in the period at L$1$ is used together with $S$ to calculate the difference between Sun-time and L$1$-time. The small, but nonetheless non-zero, difference between the Parker ($1.14$ AU) and Fisk ($1.01$ AU) field line lengths due to different $d\mathbf{\ell}$ expressions for the two HMF configurations are shown in Figure \ref{fig:ArcLengths} for a constant solar wind velocity of $400$ km/s.

\subsection{Potential field source surface model}
\label{subsection:PFSS model}
\begin{figure}[t]
\centering
\subfigure[]{\includegraphics[width=0.49\textwidth]{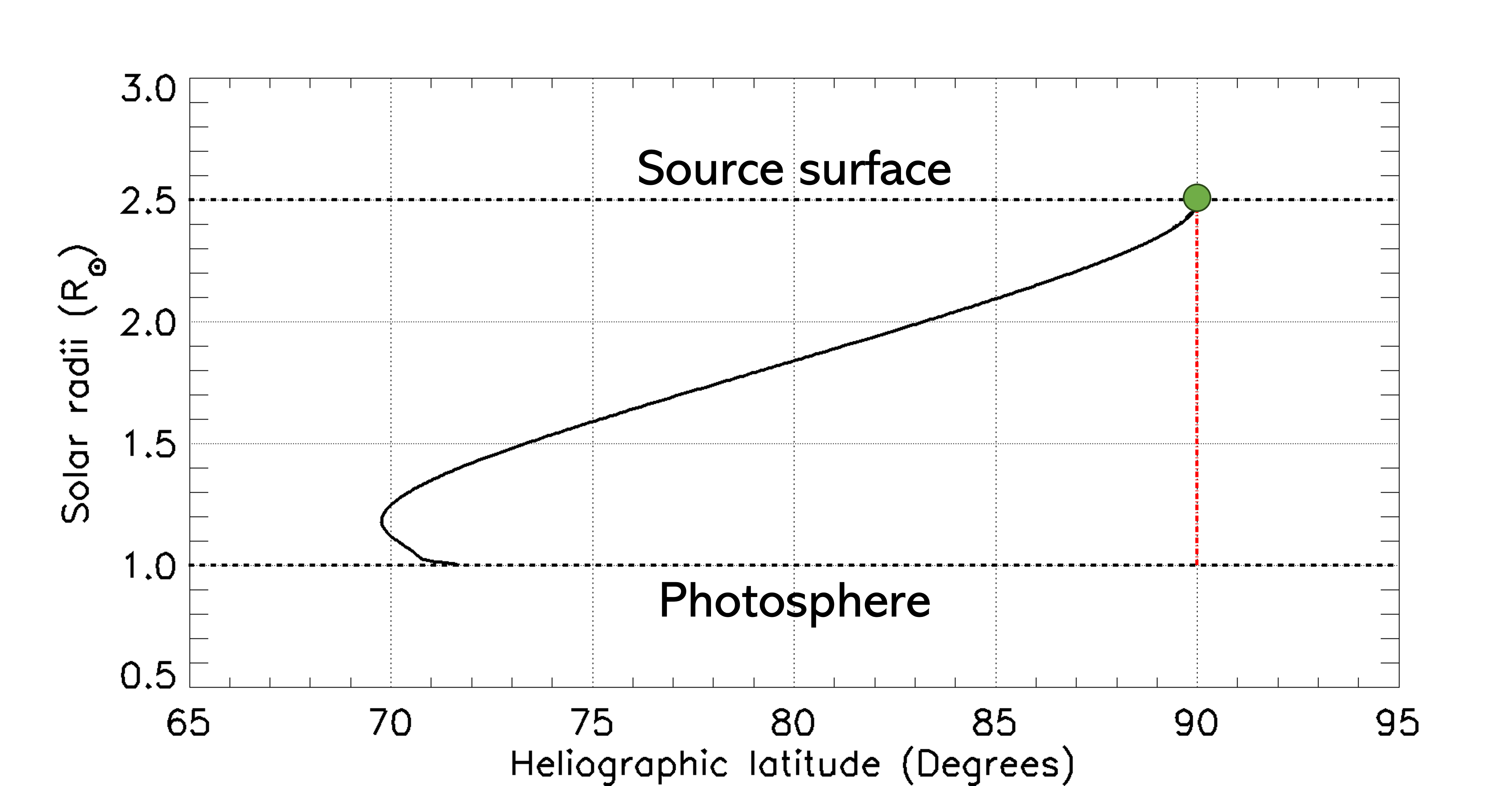}\label{fig:theta}}  
\subfigure[]{\includegraphics[width=0.49\textwidth]{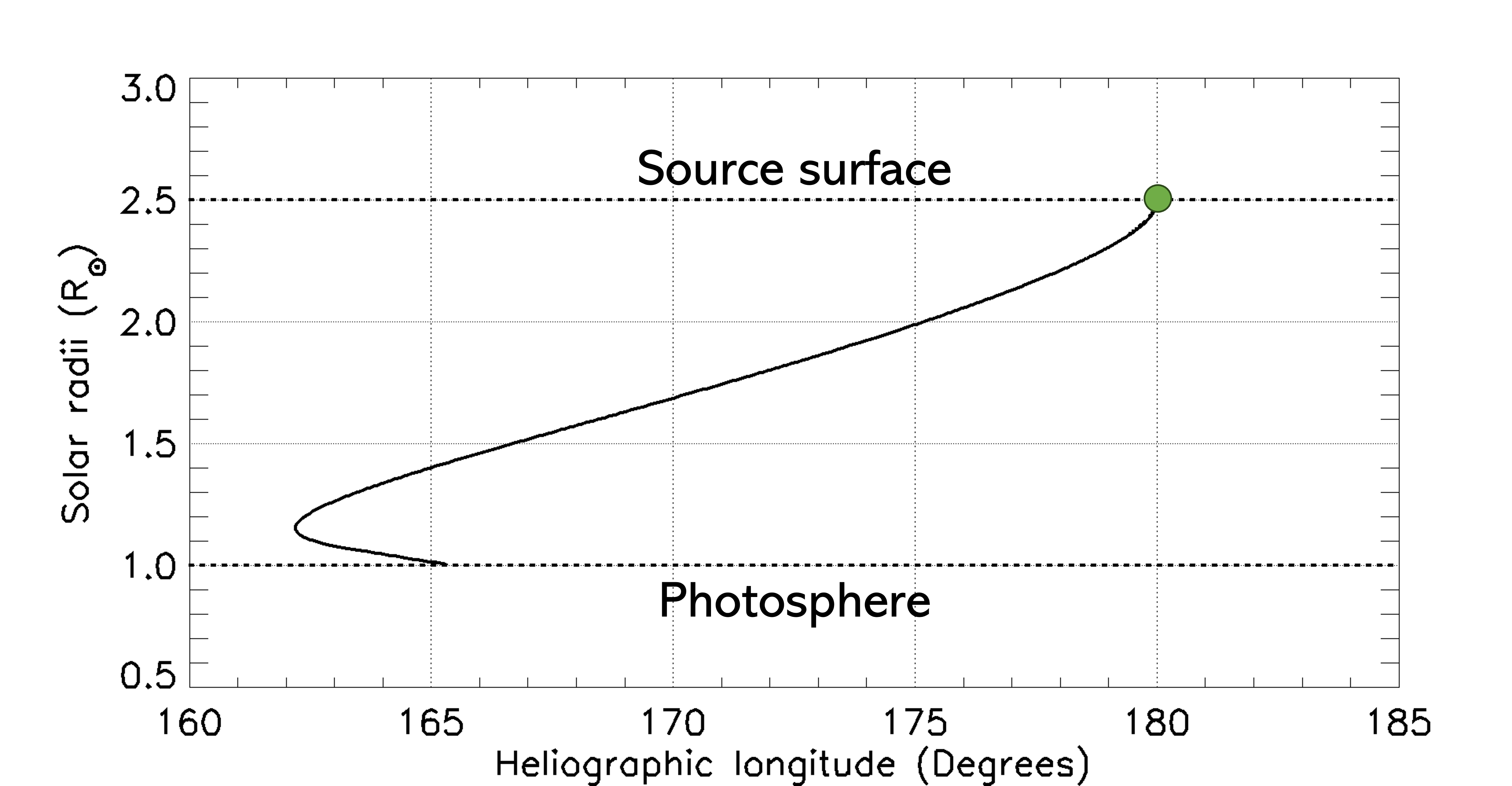}\label{fig:phi}} 
\caption{A magnetic field line trace from the source surface at $2.5R_{\odot}$ down to the photosphere at $1.0R_{\odot}$. The green dot represents the entry point at the SWSS. (a) shows the trace from $\theta_{hg}^{ss}=90^{\circ}$ to $\theta_{hg}^{ph}=71.6^{\circ}$. The equator is indicated by the vertical red dot-dashed line. (b) shows the same trace in heliographic longitude from $\phi_{hg}^{ss}=180^{\circ}$ to $\phi_{hg}^{ph}=165.6^{\circ}$.}
\end{figure}

The PFSS extrapolation method was developed and refined by \cite{Schatten1969}, \cite{Altschuler1969}, \cite{Hoeksema1984}, and \cite{Wang1992}. In this study, the upper boundary condition of the PFSS model is located at the SWSS, assumed to be spherical and a radial distance $2.5R_{\odot}$ from the photosphere. The lower boundary condition is the photospheric magnetic field generated from full-disk magnetograms. The magnetograms used in this study are from two different instruments as mentioned in Section \ref{subsection:Photospheric BField} which influences the comparability of the PFSS results between solar minima. Previous studies, such as \cite{SpaceWeather2022}, use magnetograms from a single instrument. Furthermore, \cite{Hofmeister2019} report a connection between long-lived (more than $40$ hours) photospheric magnetic elements and the magnetic flux of CHs. \cite{Wang1996} explain that the locations of CHs can be reproduced and better understood by applying extrapolation models to measurements of the photospheric magnetic field. These authors show that the solar wind speed and solar wind flux densities can be related to the magnetic field strength of CHs. Therefore, the PFSS model is advantageous to trace field lines from the SWSS to the photosphere since it considers the prevailing configuration of the photospheric magnetic field and, in principle, convey information about the state of the coronal magnetic field. Figures \ref{fig:theta} and \ref{fig:phi} illustrate an example of the tracing of a continuous magnetic field line from the SWSS to the photosphere in both heliographic latitude ($\theta_{hg}$) and longitude ($\phi_{hg}$) using the PFSS model. Small latitudinal and longitudinal perturbations, such as shifting $\theta_{hg}^{ss}$ and $\phi_{hg}^{ss}$ by $1^{\circ}$, have a minimal effect on the results of $\theta_{hg}^{ph}$ and $\phi_{hg}^{ph}$ which implies stable results for the investigated solar minima. A standard Parker field is assumed between L$1$ and the SWSS when tracing a field line for the PFSS model. The length of the Parker magnetic field line is shown in Figure \ref{fig:ArcLengths} and the difference between sun-time and L$1$-time is calculated in the same way as in Section \ref{subsection:Fisk HMF Model}, again assuming that the measured $V_r$ remains constant between L$1$ and the SWSS.  

\section{Solar wind and IMF properties at L1}
\label{section:SW and Bfield properties at L1}
\subsection{1996/1997 solar minimum }
\label{subsection: 1996/1997 Solar Min}
\begin{figure*}[t]
  \centering
  \subfigure[Radial solar wind velocity $V_r$]{\includegraphics[width=0.49\textwidth]{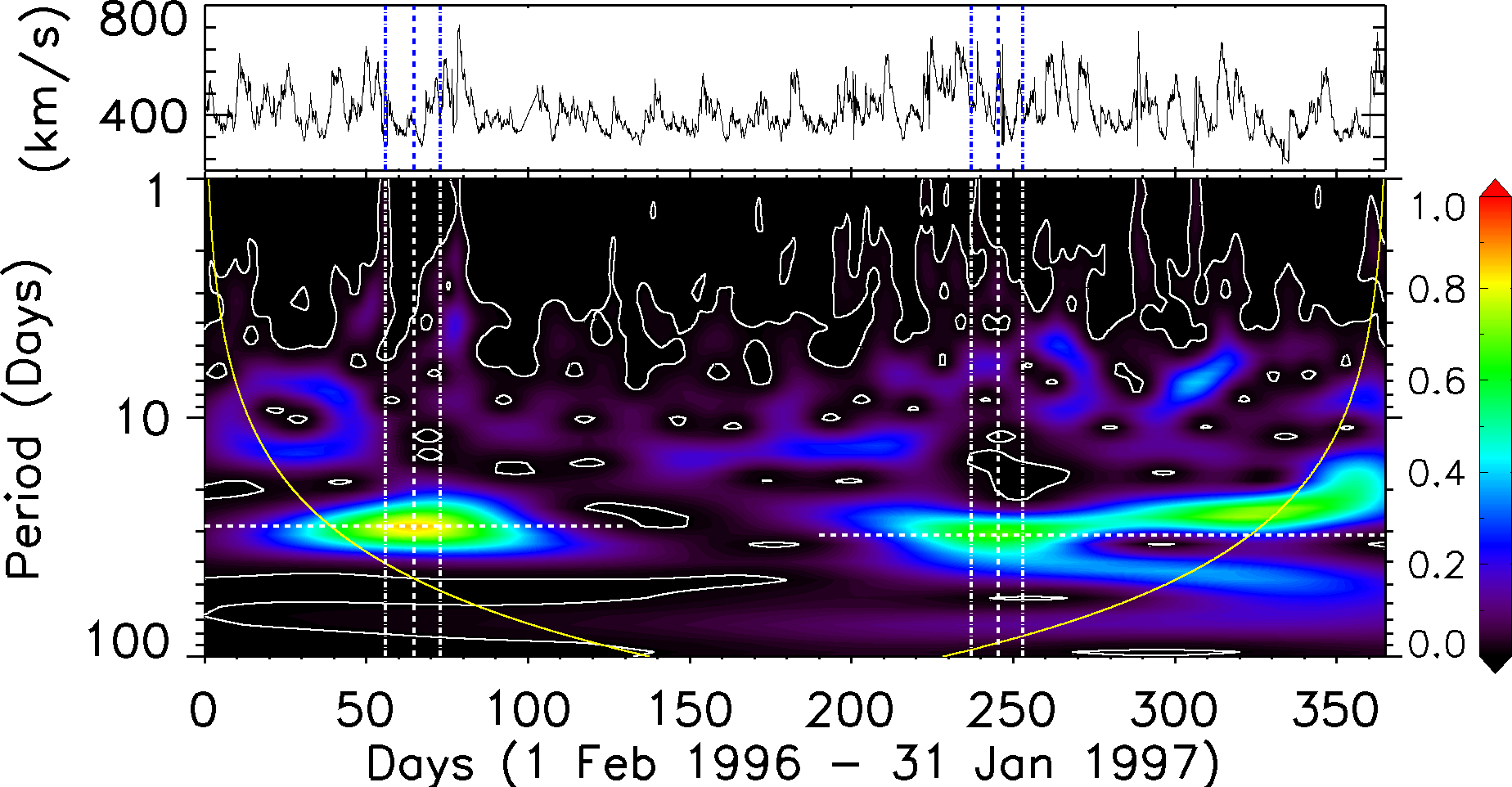}\label{fig:RSW1996}}  
  \subfigure[Proton number density $n_i$]{\includegraphics[width=0.49\textwidth]{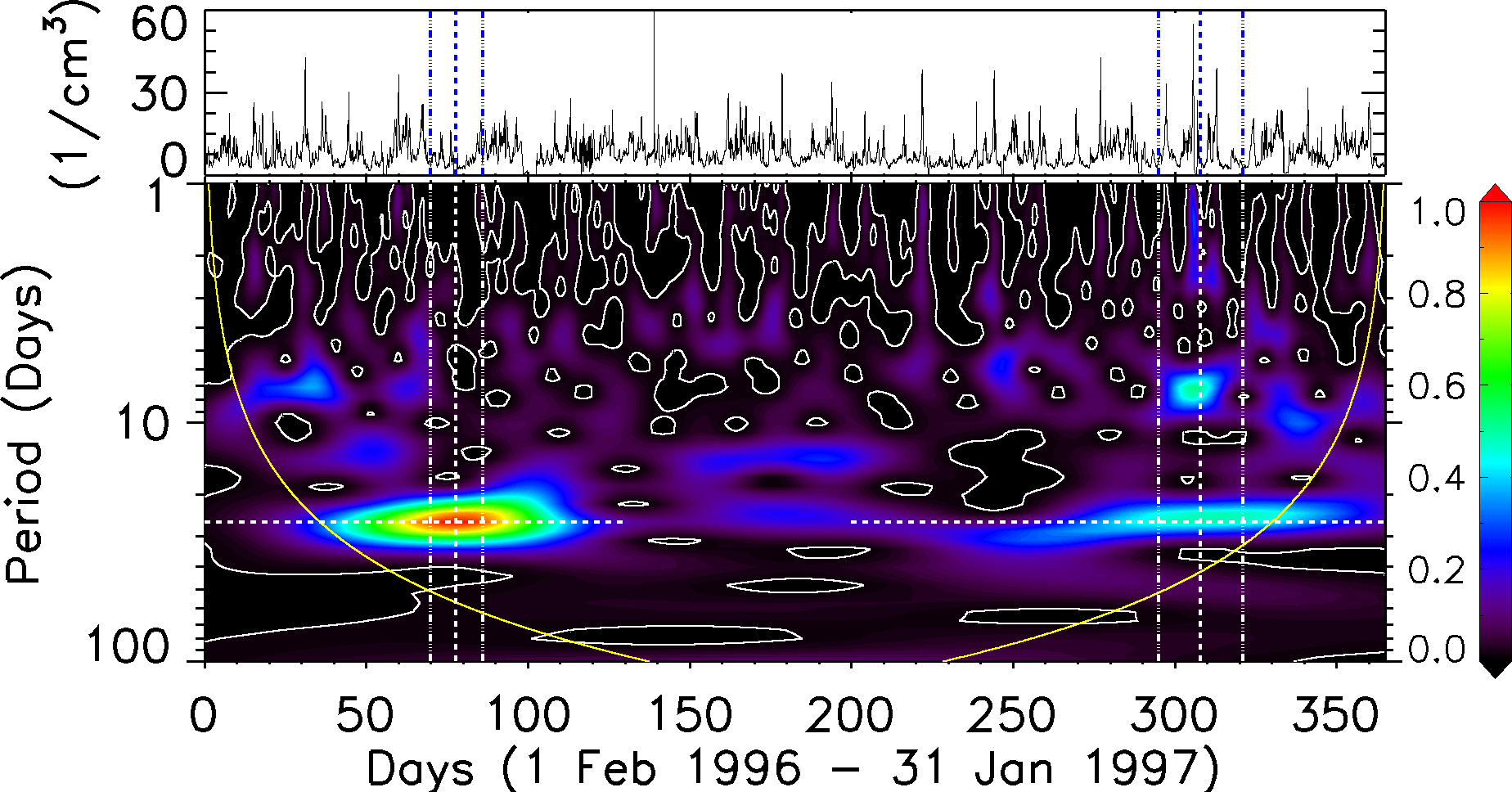}\label{fig:Proton1996}} 
  \subfigure[Normal magnetic field $B_n$]{\includegraphics[width=0.49\textwidth]{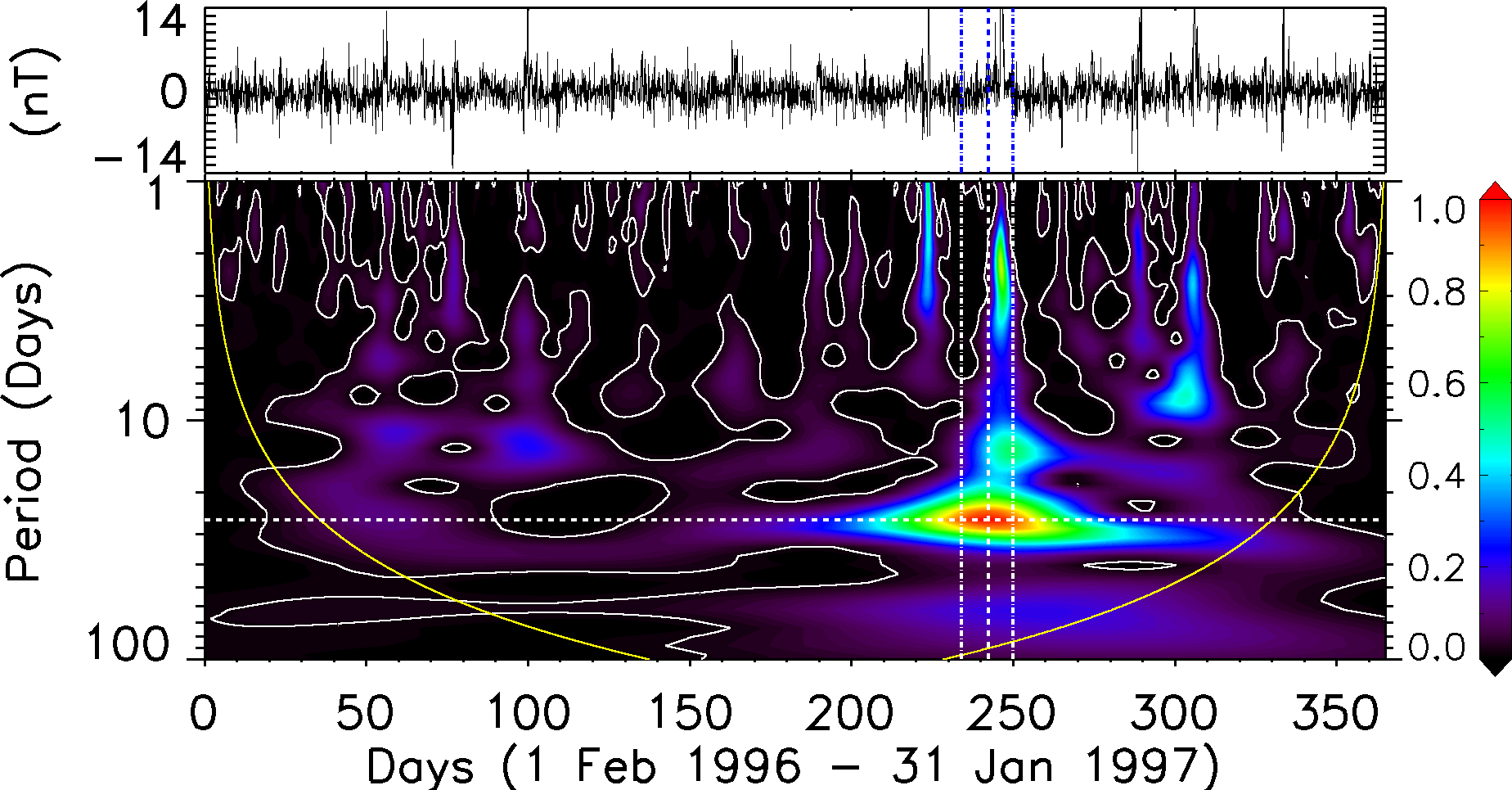}\label{fig:BN1996} }
  \subfigure[Tangential magnetic field $B_t$]{\includegraphics[width=0.49\textwidth]
  {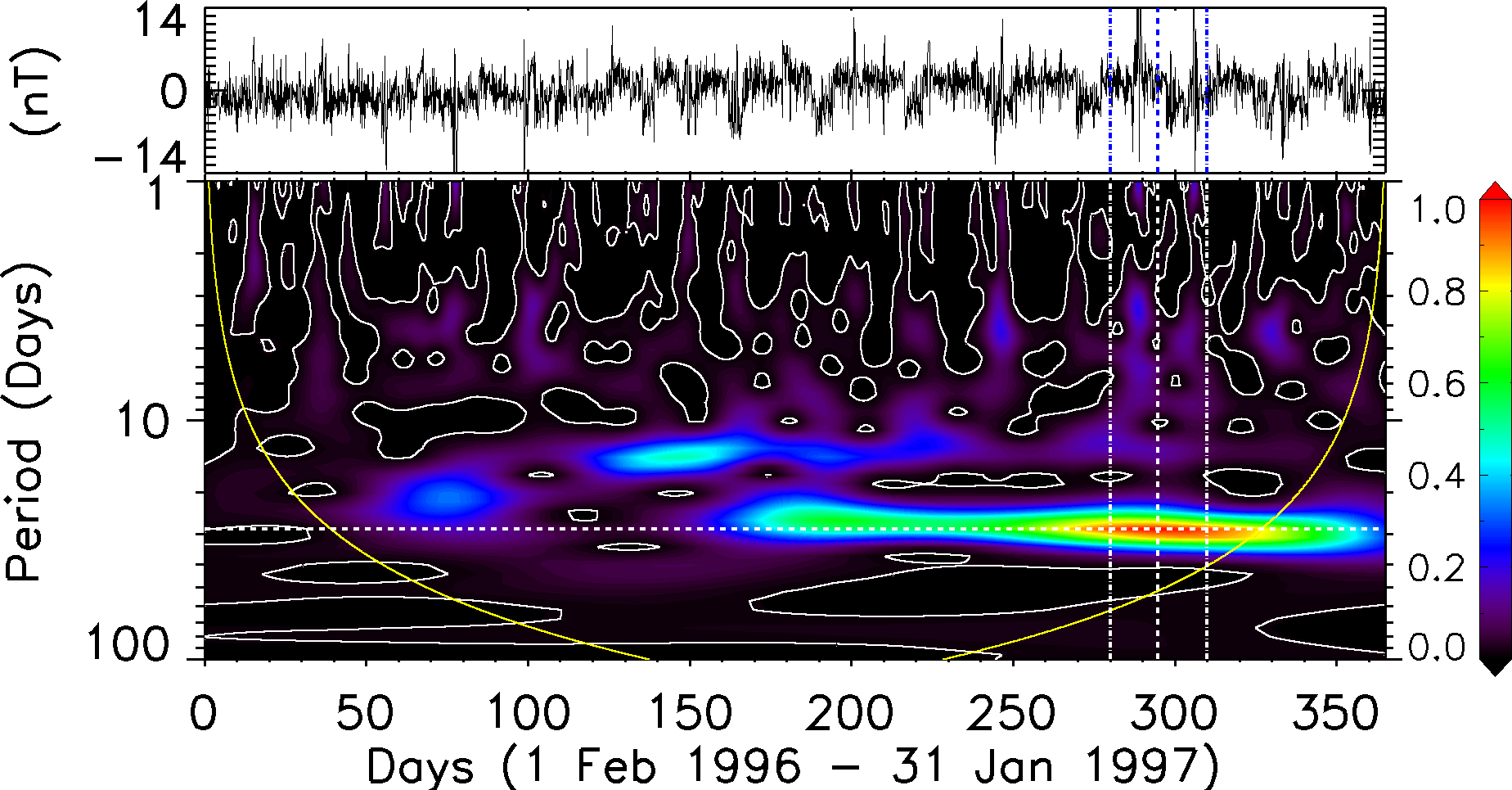}\label{fig:BT1996} }
  \caption{Wavelets indicating the signals and periodicities of the (a) radial solar wind velocity, the (b) proton number density, the (d) normal and (e) tangential magnetic field components between $1$ February $1996$ and $31$ January $1997$ during the $1996$/$1997$ solar minimum epoch. The panel above each wavelet shows the input data. The three vertical white dashed lines show the time of local maximum power (central line) of the dominant periodicity with the two side lines showing the times of a $5\%$ reduction from peak power. The same are shown in blue in the input data panel. The horizontal white dashed lines indicate the dominant periodicity. The normalised power of each periodicity is identified by the colour scale, while the cone of influence is indicated in yellow. The white contour shows the $99.5\%$ confidence level.} 
\label{fig:1996}
\end{figure*}

The following section focuses on the solar minimum period between $1$ February $1996$ to $31$ January $1997$ ($365$ days). Figure \ref{fig:1996} shows the wavelet results for the radial solar wind velocity $V_r$ (Figure \ref{fig:RSW1996}), the proton number density $n_i$ (Figure \ref{fig:Proton1996}), the normal $B_n$ (Figure \ref{fig:BN1996}), and tangential $B_t$ (Figure \ref{fig:BT1996}) magnetic field components. Table \ref{table:1996} summarises the results from the wavelet analysis, including the date at which the maximum period was observed at L$1$, the maximum period in days (ranging between $26.2$ and $31.2$ days), and the corresponding Carrington rotation (CR) during which the maximum occurred. The pairs of wavelets with maxima at approximately the same date include $V_r$ and $n_i$, $V_r$ and $B_n$, and $n_i$ and $B_t$ as seen in Figure \ref{fig:1996}. The first maxima observed in $V_r$ and $n_i$ on $5$ April $1996$ (day $64$) and $18$ April $1996$ (day $77$), respectively, are part of CRs $1907$ and $1908$. Although these Carrington maps are not available (see Section \ref{subsection: carrington Maps Data}), the maxima in $V_r$ and $n_i$ are included in Figure \ref{fig:1996} and Table \ref{table:1996} to illustrate the observed periodicity. The maximum observed in $B_n$ on $3$ October $1996$ (day $245$ of Figure \ref{fig:BN1996}) during CR$1914$ corresponds to the second maximum observed in $V_r$ on the same day (Figure \ref{fig:RSW1996}). The maximum observed in $n_i$ on $4$ December $1996$ (day $307$ of Figure \ref{fig:Proton1996}) is coupled together with the maximum observed on $21$ November $1996$ (day $294$ of Figure \ref{fig:BT1996}) in $B_t$. In each group of wavelets for all three solar minima studied in this paper, the dates of the maxima observed in several of the solar wind and magnetic field components are used to trace field lines between L$1$ and the SWSS. However, some components are not traced between L$1$ and the SWSS since they share a maximum at approximately the same time as another component and rather used as a confirmation of the signal observed. For example, $V_r$ and $B_n$ in Table \ref{table:1996} have their maxima on the same date ($3$ October $1996$) and only $V_r$ is traced to the SWSS. The double dash in the last column of Table \ref{table:1996} indicates that this field line is not traced. The convention is followed throughout the study. 

\begin{deluxetable}{ccccc}[t]
\tabletypesize{\footnotesize}
\tablecolumns{5}
\tablewidth{0pt}
\tablecaption{$1996$/$1997$ solar minimum with details of maxima in periodicity power at L$1$ as shown in Figure \ref{fig:1996}. \label{table:1996}}
\tablehead{
\colhead{} & \colhead{Date of max} &  \colhead{Period (Days)} &\colhead{CR} &\colhead{Figures} } 
\startdata
$V_r$ & $5$ Apr. $1996$    & $28.5$  & $1907$ & \ref{fig:RSW1996} ; $--$ \\ 
      & $3$ Oct. $1996$    & $31.2$  & $1914$ & \ref{fig:RSW1996} ; \ref{fig:CR1914} \\
$n_i$ & $18$ Apr. $1996$     & $26.2$  & $1908$  & \ref{fig:Proton1996} ; $--$\\
      & $4$ Dec. $1996$  & $26.2$  & $1916$-$1917$  & \ref{fig:Proton1996} ; \ref{fig:CR1916_1917}\\  
$B_n$  & $3$ Oct. $1996$    & $26.2$  & $1914$ & \ref{fig:BN1996} ; $--$\\
$B_t$       & $21$ Nov. $1996$     & $28.5$  & $1915$-$1916$ & \ref{fig:BT1996} ; \ref{fig:CR1915},\ref{fig:CR1916_1917}\\
\enddata
\end{deluxetable}

\subsection{2008/2009 solar minimum}
\label{subsection: 2008/2009 Solar Min}
\begin{figure*}[h]
  \centering
  \subfigure[Radial solar wind velocity $V_r$]{\includegraphics[width=0.49\textwidth]{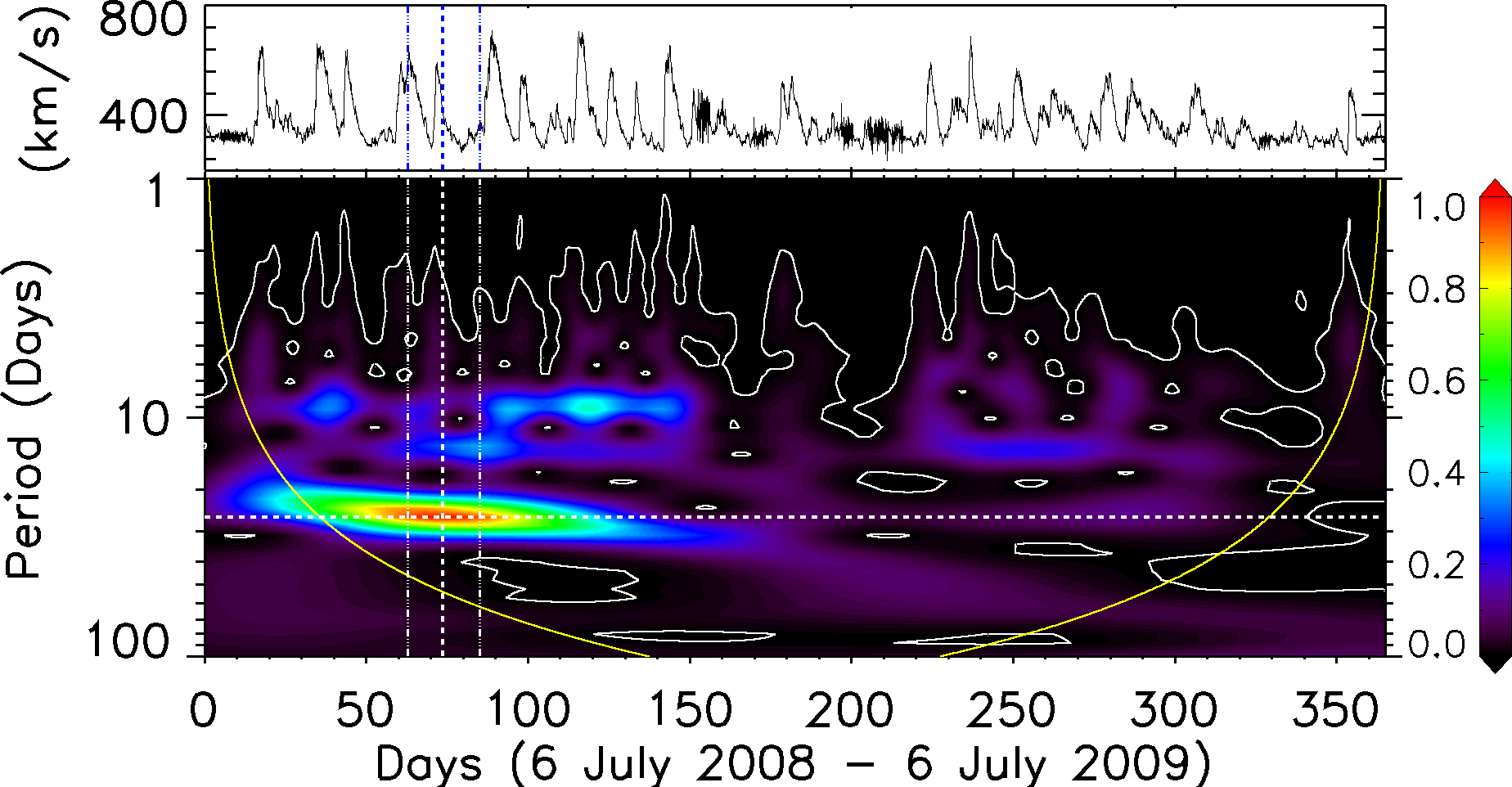}\label{fig:RSW2009}}  
  \subfigure[Tangential solar wind velocity $V_t$]{\includegraphics[width=0.49\textwidth]{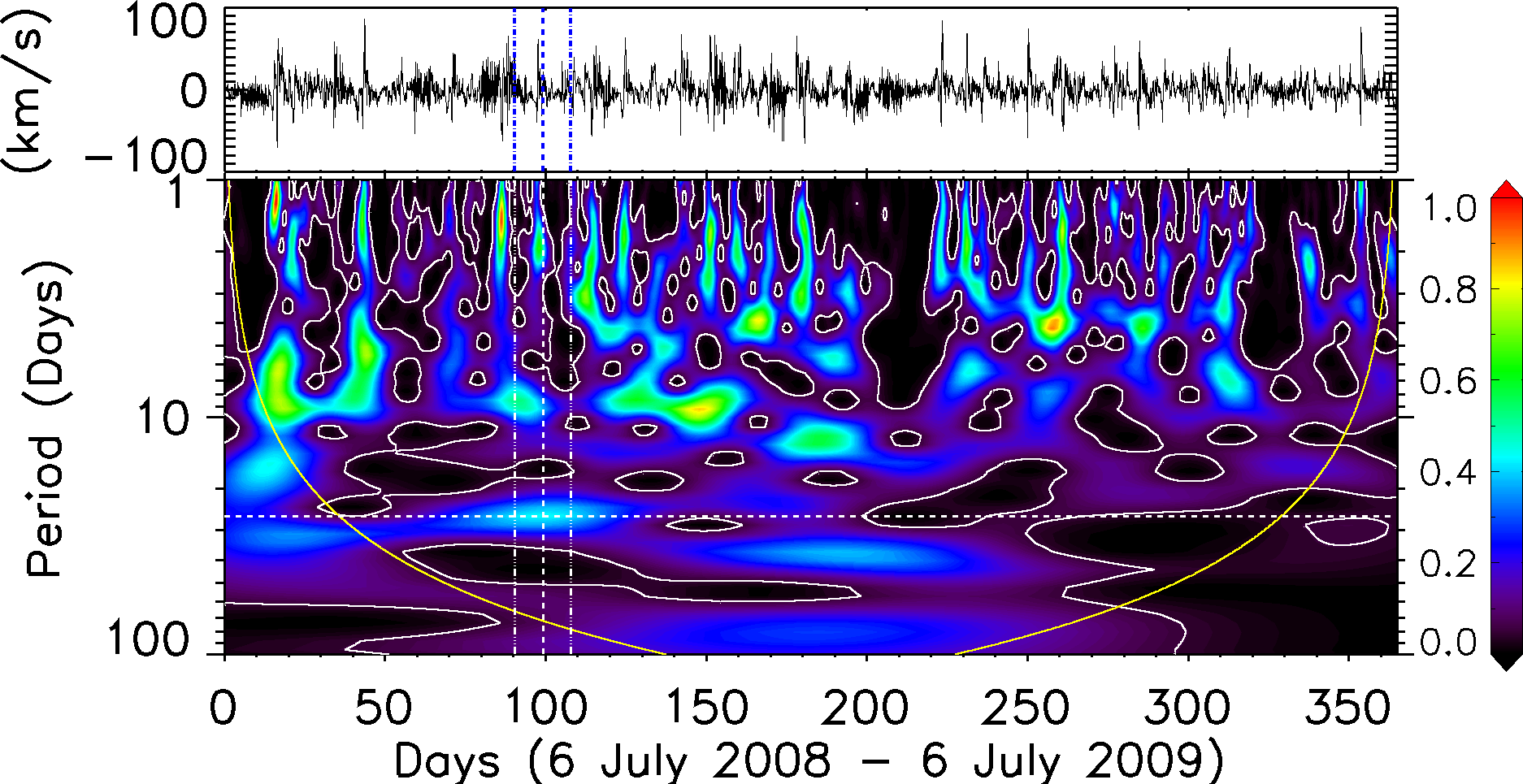}\label{fig:TSW2009}} 
  \subfigure[Normal solar wind velocity $V_n$]{\includegraphics[width=0.49\textwidth]{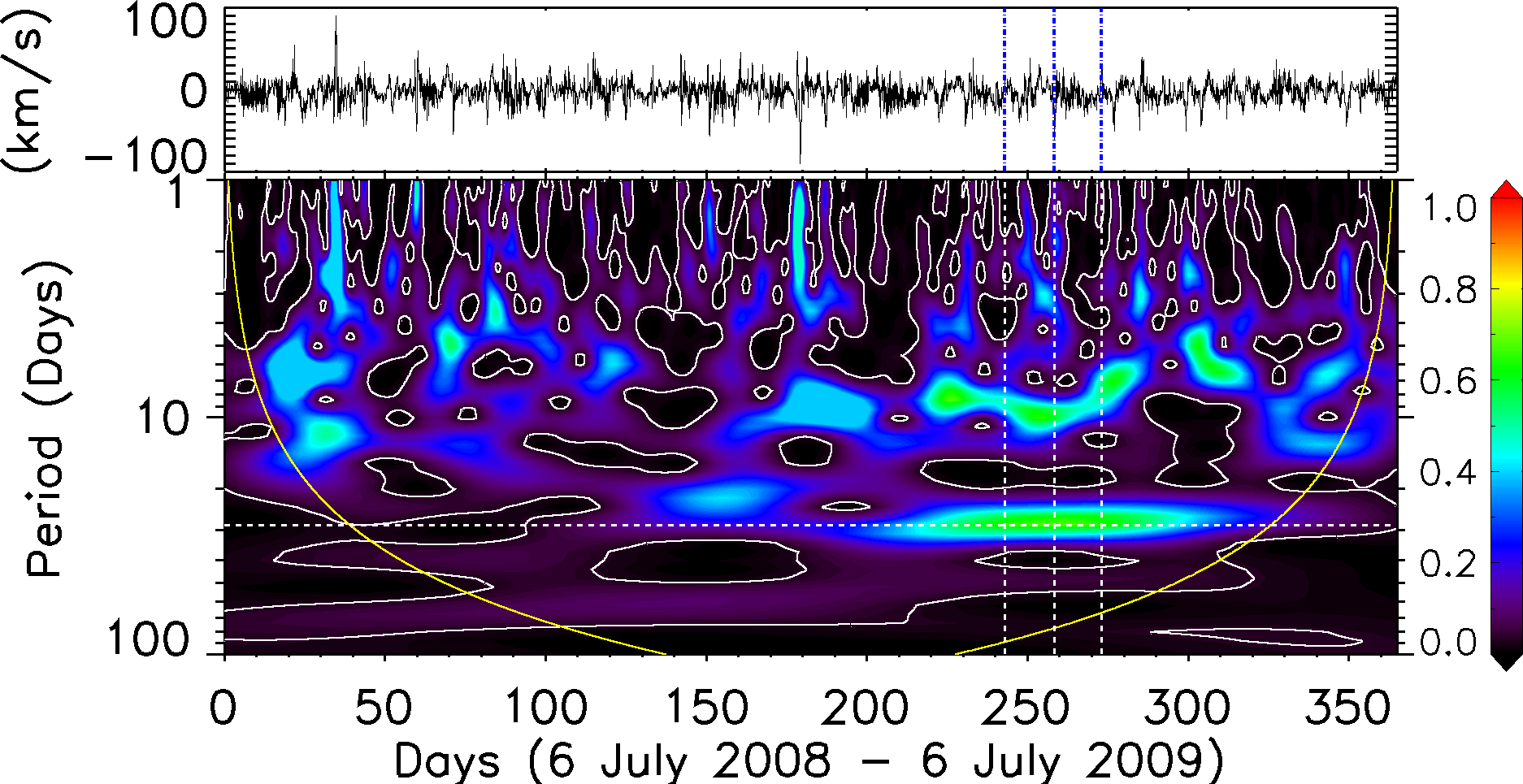}\label{fig:NSW2009} }
  \subfigure[Proton number density $n_i$]{\includegraphics[width=0.49\textwidth]{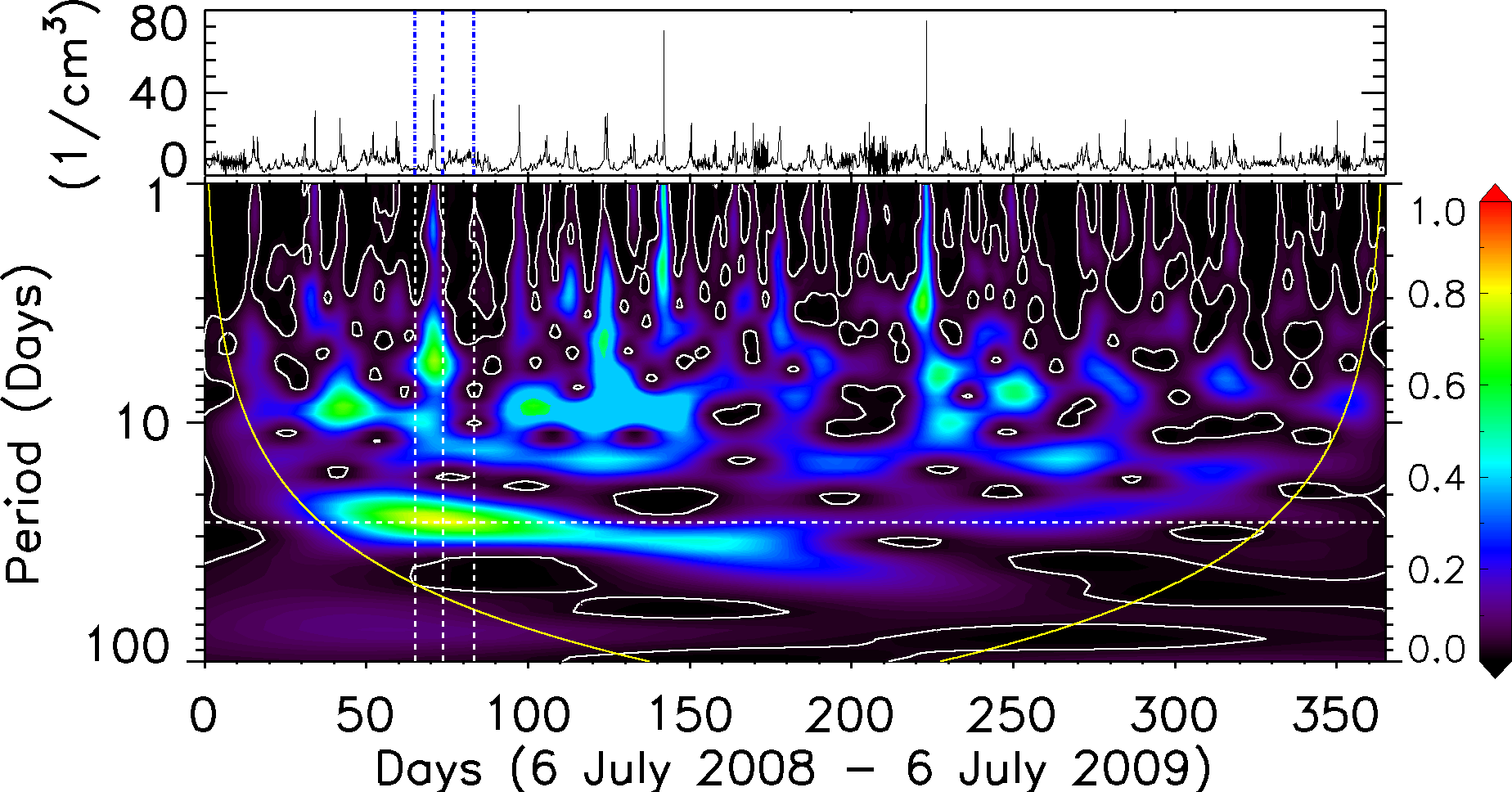}\label{fig:proton2009}}
  \subfigure[Radial magnetic field $B_r$]{\includegraphics[width=0.49\textwidth]{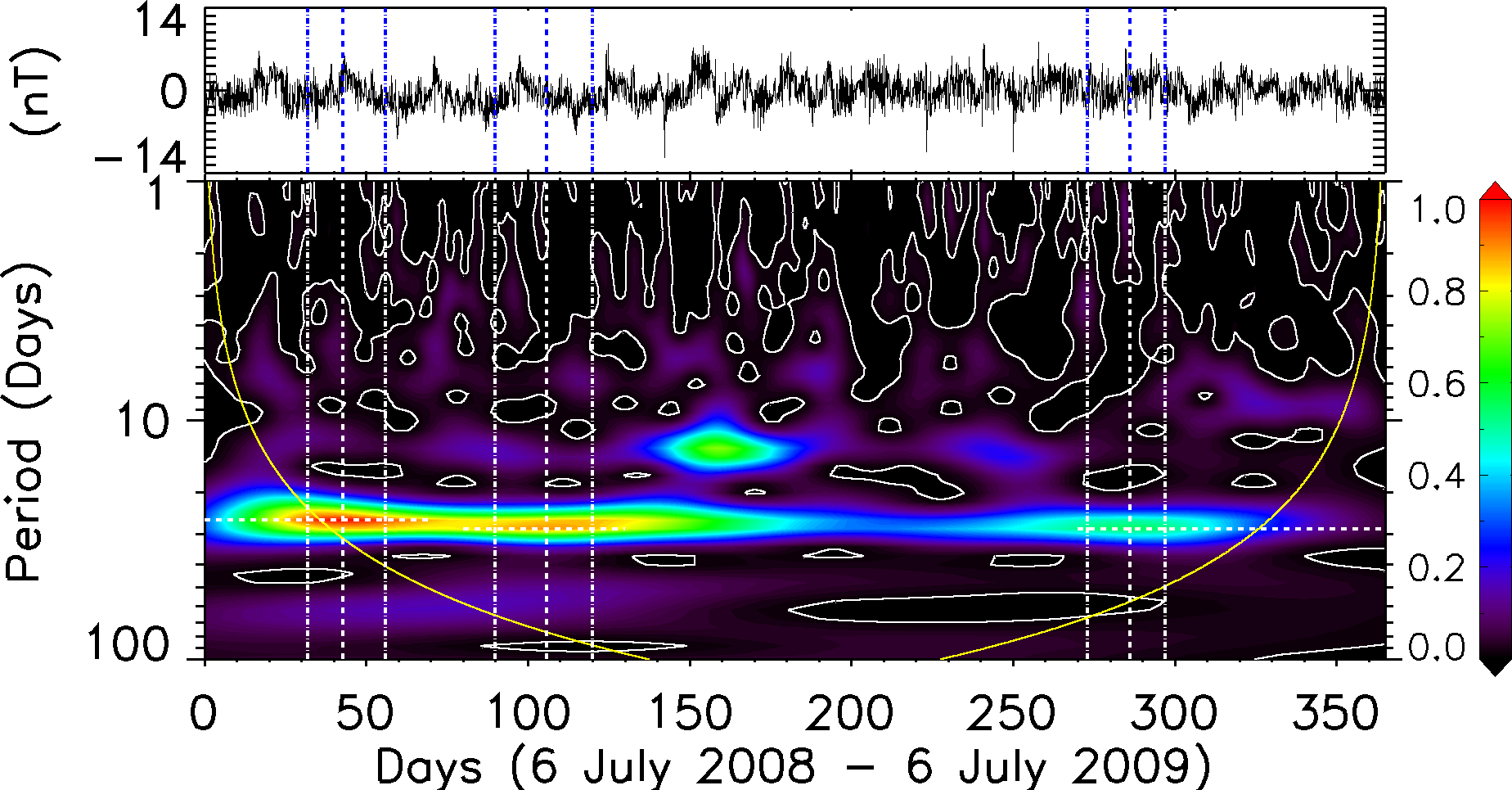}\label{fig:BR2009}} 
  \subfigure[Tangential magnetic field $B_t$]{\includegraphics[width=0.49\textwidth]{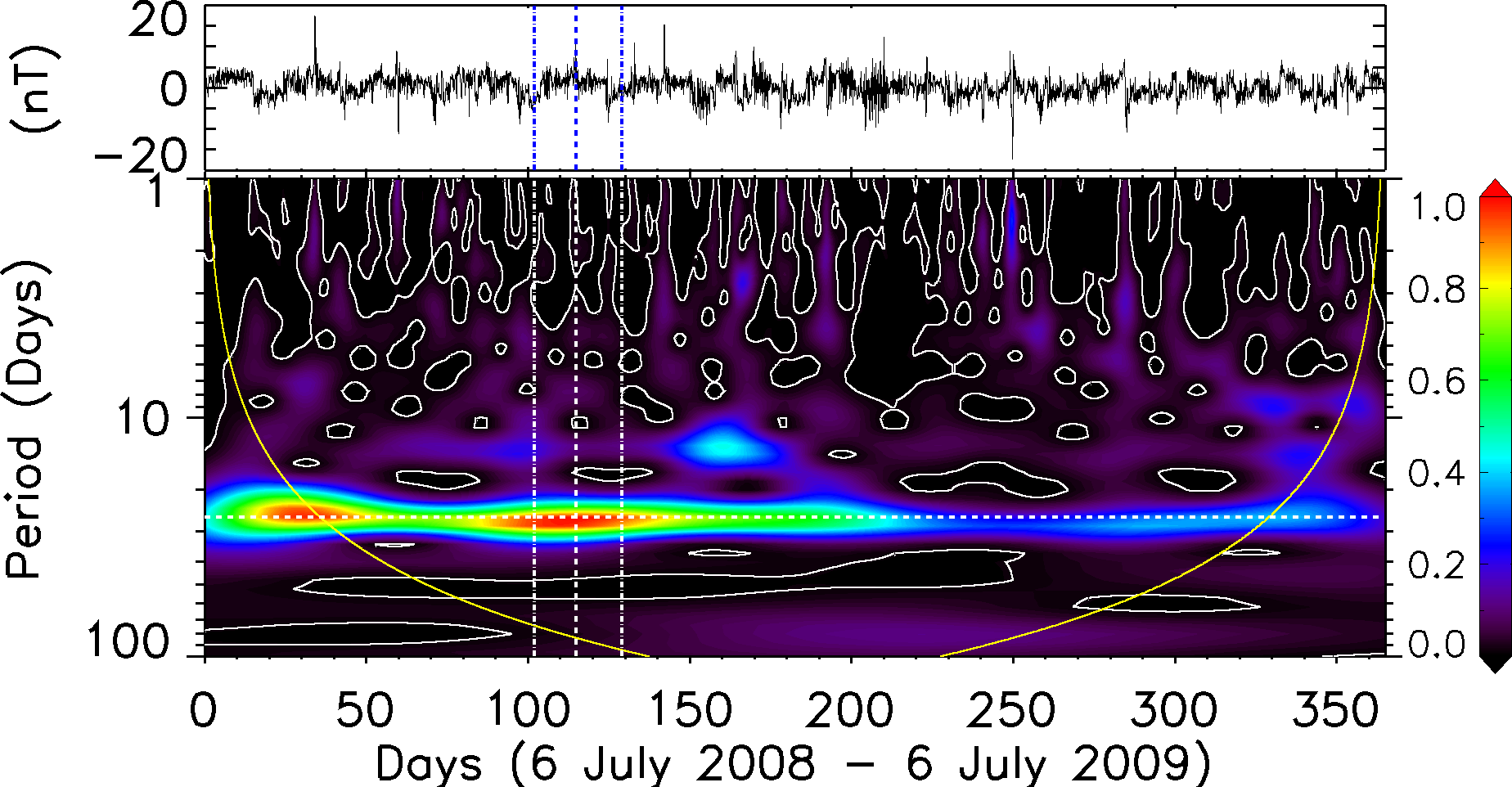}\label{fig:BT2009}}
  \caption{Wavelets indicating the signals and periodicities of the (a) radial, (b) tangential and (c) normal solar wind velocities, (d) the proton number density, the (e) radial and (f) tangential magnetic field components between $6$ July $2008$ and $6$ July $2009$ during the $2008$/$2009$ solar minimum epoch. Refer to Figure \ref{fig:1996} for a description of the dashed lines and contours.} 
\label{fig:2009}
\end{figure*}

The following section focuses on the solar minimum period between $6$ July $2008$ and $6$ July $2009$ ($365$ days) and describes the same process followed in Section \ref{subsection: 1996/1997 Solar Min}. Figure \ref{fig:2009} shows six wavelets including $V_r$ (Figure \ref{fig:RSW2009}), $V_t$ (Figure \ref{fig:TSW2009}), $V_n$ (Figure \ref{fig:NSW2009}), $n_i$ (Figure \ref{fig:proton2009}), $B_r$ (Figure \ref{fig:BR2009}), and $B_t$ (Figure \ref{fig:BT2009}). Table \ref{table:2008} summarises the details of this solar minimum. 

The wavelets from Figure \ref{fig:2009} that share periodicities at approximately the same time include $V_r$ and $n_i$ on $17$ September $2008$ (day $70$), the trio of measurements $V_t$ on $13$ October $2008$ (day $99$), $B_r$ on $21$ October $2008$ (day $107$) and $B_t$ on $28$ October $2008$ (day $114$), and finally $V_n$ on $21$ March $2008$ (day $258$) and $B_r$ on $18$ April $2008$ (day $286$). Note that the first maximum of $B_r$ (day $43$ on $17$ August $2008$ of Figure \ref{fig:BR2009}) is included in Table \ref{table:2008}, but is not used in further analyses since the $5\%$ decrease in intensity to the left of the maximum is inside the cone of influence (COI). There is also no confirmation of this signal since the maximum intensity of the periodicity identified at approximately the same time in $B_t$ (Figure \ref{fig:BT2009}) is also inside the COI. $V_n$ (Figure \ref{fig:NSW2009}) and the third maximum of $B_r$ (Figure \ref{fig:BR2009}) have a $28$-day gap between their respective local maxima shown in Table \ref{table:2008}. Due to the long duration of the observed periodicity in both $V_n$ and $B_r$, they might share a common physical process giving rise to the signal at L$1$.   

$V_r$ shows a period of $26.2$ days on $17$ September $2008$ (day $73$ of Figure \ref{fig:RSW2009}). $V_t$ shows a period of $26.2$ days approximately a month later on $13$ October $2008$ (day $99$ of Figure \ref{fig:TSW2009}). Additionally, Figure \ref{fig:TSW2009} also shows short term ($\sim 2.5$ days), short periodicity ($\sim 1.5$ days) vertical structures with high-power signals. The top panel confirms that these strong, short-term periodicities are caused by peaks in $V_t$. Although all known ICMEs were removed from the data sets, it is possible for different transient events to be observed in the data. These events were not removed for the sake of consistency and transparency. $V_n$ shows a period of $28.5$ days on $21$ March $2009$ (day $258$ of Figure \ref{fig:NSW2009}). $n_i$ follows the behaviour of $V_r$ closely with a period of $26.2$ days also occurring on $17$ September $2008$ (day $73$). Peaks in $n_i$ are accompanied by strong, short-term power signals. Next, Figures \ref{fig:BR2009} and \ref{fig:BT2009} show $B_r$ and $B_t$ results respectively. Both figures show long duration power signals at periods of $28.5$ days ($B_r$) and $26.2$ days ($B_t$). Three local maxima are observed in $B_r$ and are indicated on Figure \ref{fig:BR2009}. Table \ref{table:2008} shows the maxima of $B_r$ occurring on $17$ August $2008$ (day $42$), $21$ October $2008$ (day $107$), and $18$ April $2009$ (day $286$), and the maximum of $B_t$ occurring on $28$ October $2008$ (day $115$).Note that the periodicities are only identified in order to select the date and time the field lines are traced back from L$1$ to the SWSS and then mapped to the photosphere. It is well-known that the harmonic structure of the solar synoptic rotation has an imprint in the solar wind \citep{Prabhakaran2002,Singh2019,Tsichla2019}. In this study we focus on the fundamental frequency only and use its maximum power to determine the time of maximum disturbance at L1.

\begin{deluxetable}{ccccc}[t]
\tabletypesize{\footnotesize}
\tablecolumns{5}
\tablewidth{0pt}
\tablecaption{ $2008$/$2009$ solar minimum with details of maxima in periodicity power at L$1$ as shown in Figure \ref{fig:2009}. \label{table:2008}}
\tablehead{
\colhead{} & \colhead{Date of max} &  \colhead{Period (Days)} &\colhead{CR} &\colhead{Figures} } 
\startdata
$V_r$       & $17$ Sept. $2008$   & $26.2$  & $2074$ & \ref{fig:RSW2009} ; \ref{fig:CR2074} \\
$V_t$       & $13$ Oct. $2008$    & $26.2$  & $2075$ & \ref{fig:TSW2009} ; \ref{fig:CR2075_76}\\
$V_n$       & $21$ March $2009$   & $28.5$  & $2081$ & \ref{fig:NSW2009} ; \ref{fig:CR2080},\ref{fig:CR2081_82}\\
$n_i$  & $17$ Sept. $2008$   & $26.2$  & $2074$ & \ref{fig:proton2009} ; $--$\\
$B_r$   & $17$ Aug. $2008$    & $26.2$  & $2073$ & \ref{fig:BR2009} ; $--$\\
       & $21$ Oct. $2008$    & $28.5$  & $2075$ & \ref{fig:BR2009} ; $--$\\
        & $18$ Apr. $2009$    & $28.5$  & $2082$ & \ref{fig:BR2009} ; \ref{fig:CR2081_82}\\
$B_t$       & $28$ Oct. $2008$    & $26.2$  & $2075$ & \ref{fig:BT2009} ; \ref{fig:CR2075_76}\\
\enddata
\end{deluxetable}

\subsection{2019/2020 solar minimum}
\label{subsection: 2019/2020 Solar Min}

\begin{figure*}[t]
  \centering
  \subfigure[Radial solar wind velocity $V_r$]{\includegraphics[width=0.49\textwidth]{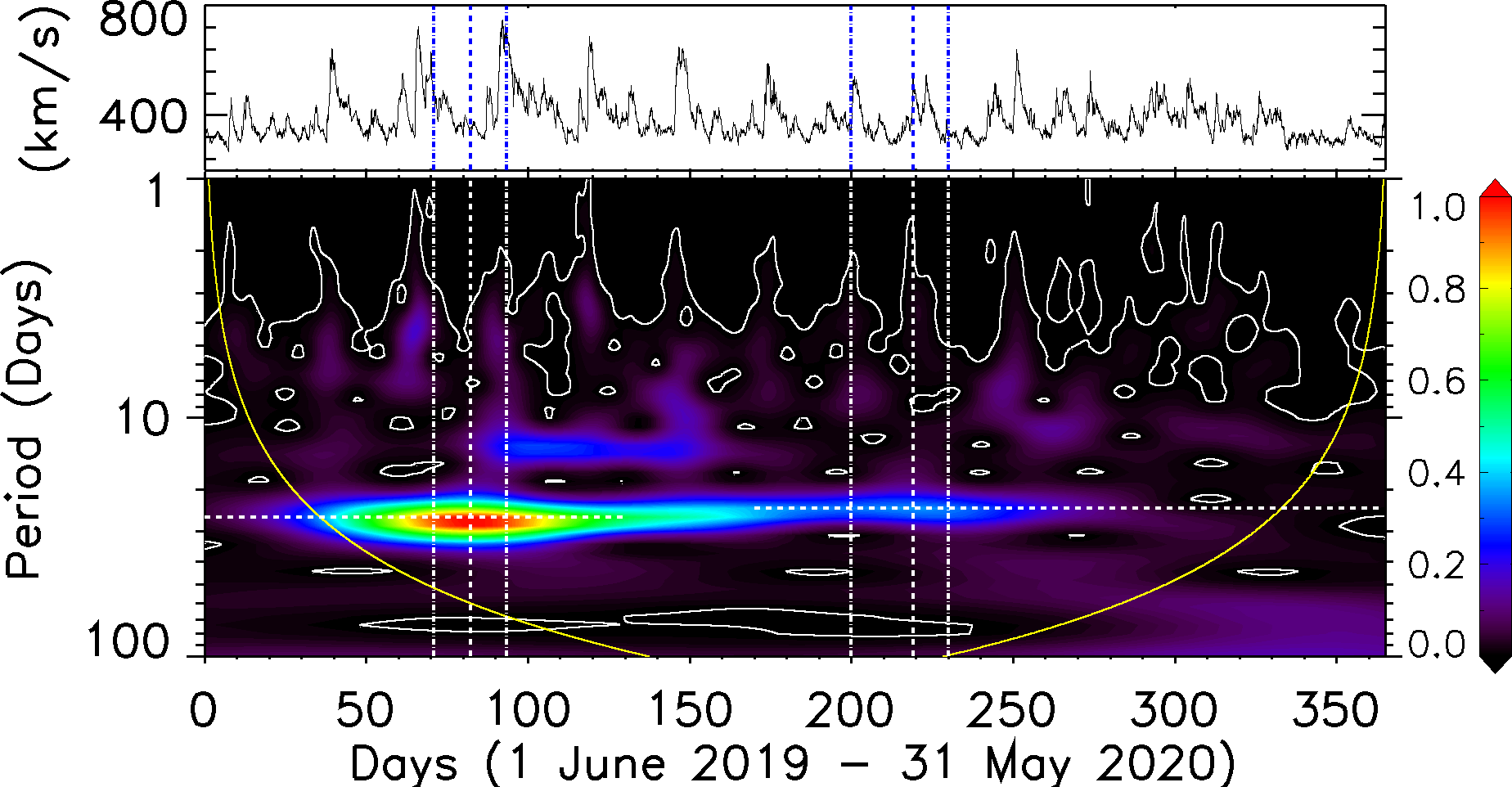}\label{fig:RSW2020}}  
  \subfigure[Normal solar wind velocity $V_n$]{\includegraphics[width=0.49\textwidth]{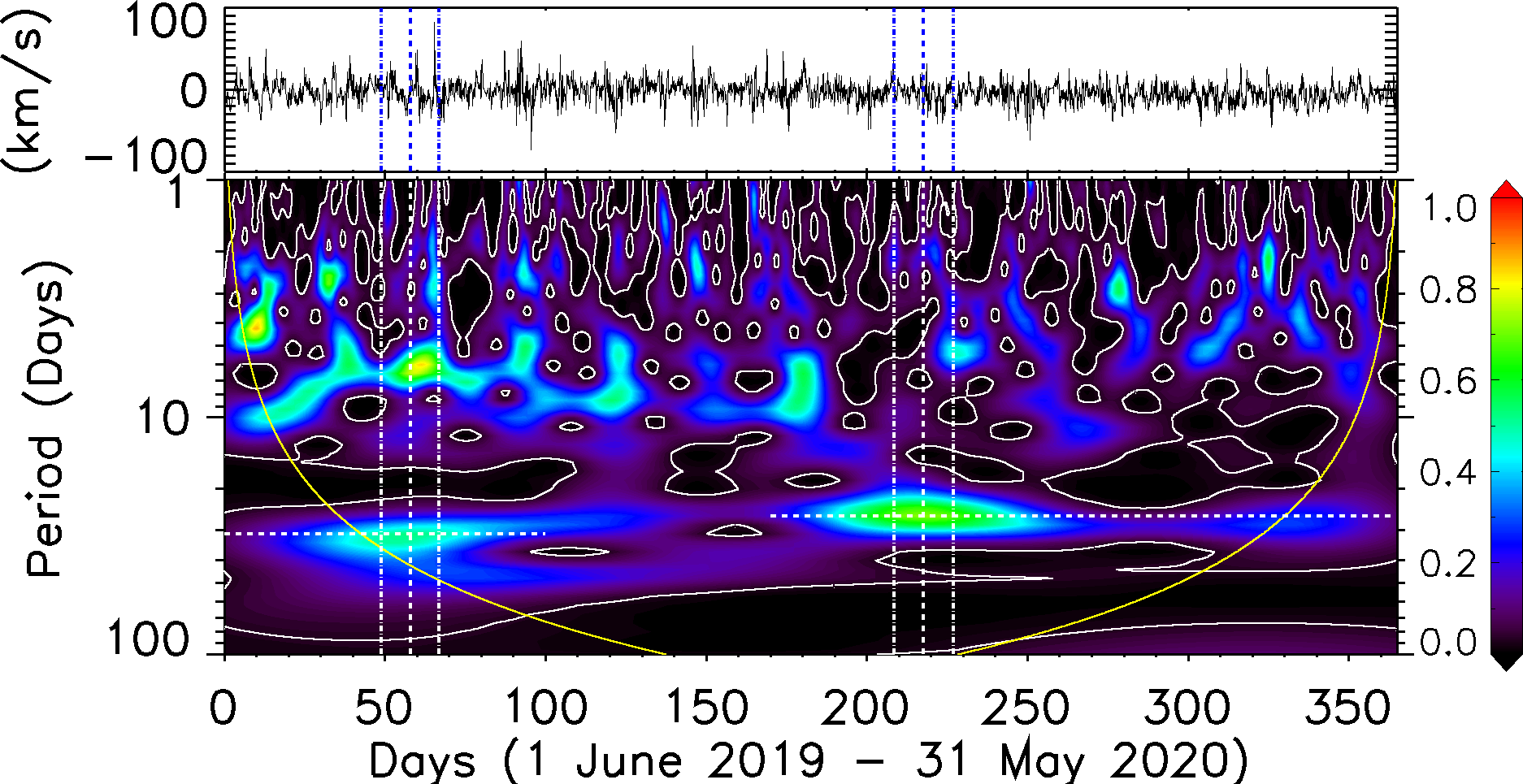}\label{fig:NSW2020}} 
  \subfigure[Proton number density $n_i$]{\includegraphics[width=0.49\textwidth]{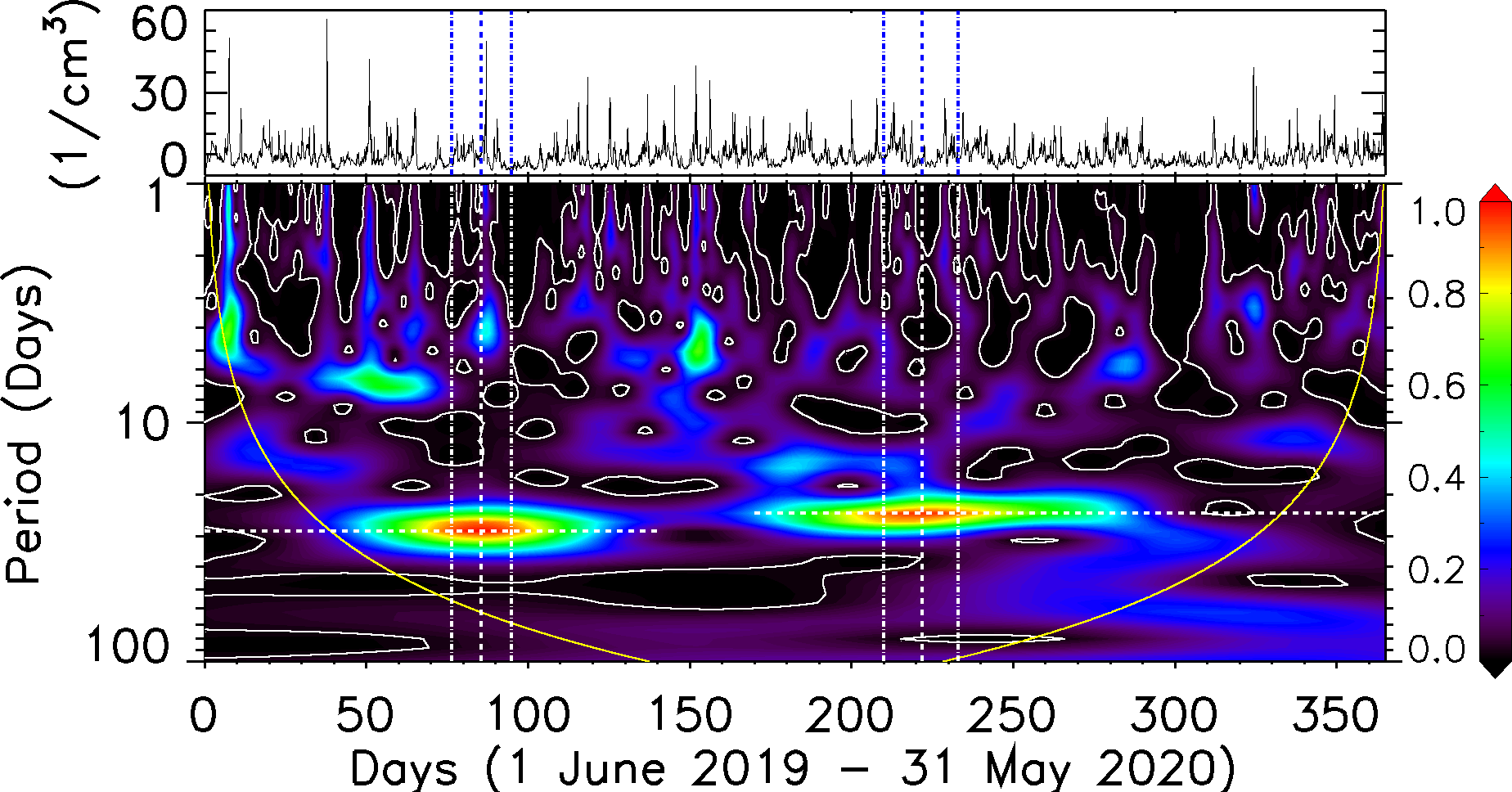}\label{fig:proton2020}} 
   \subfigure[Radial magnetic field $B_r$]{\includegraphics[width=0.49\textwidth]{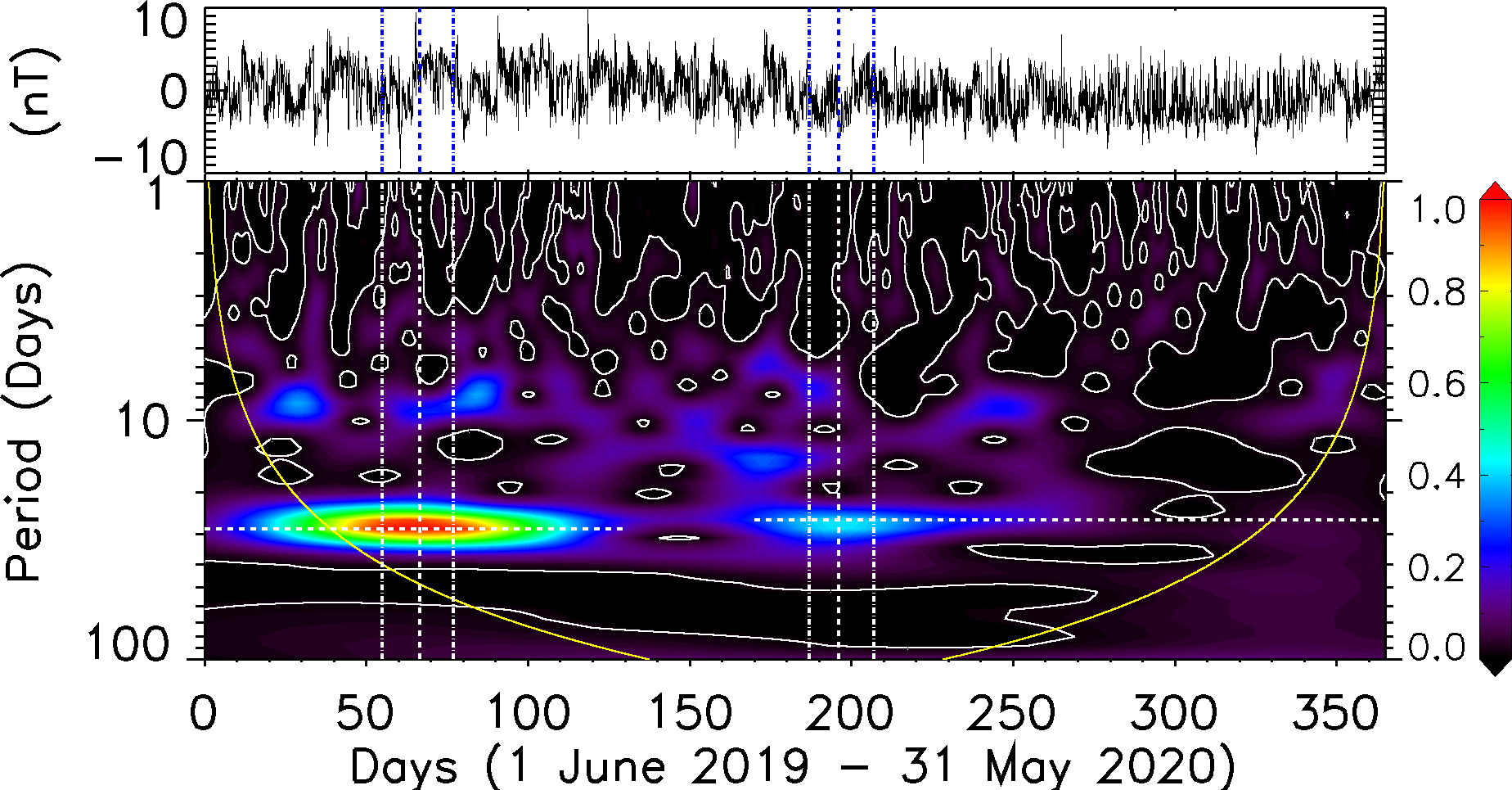}\label{fig:BR2020}}  
  \subfigure[Tangential magnetic field $B_t$]{\includegraphics[width=0.49\textwidth]{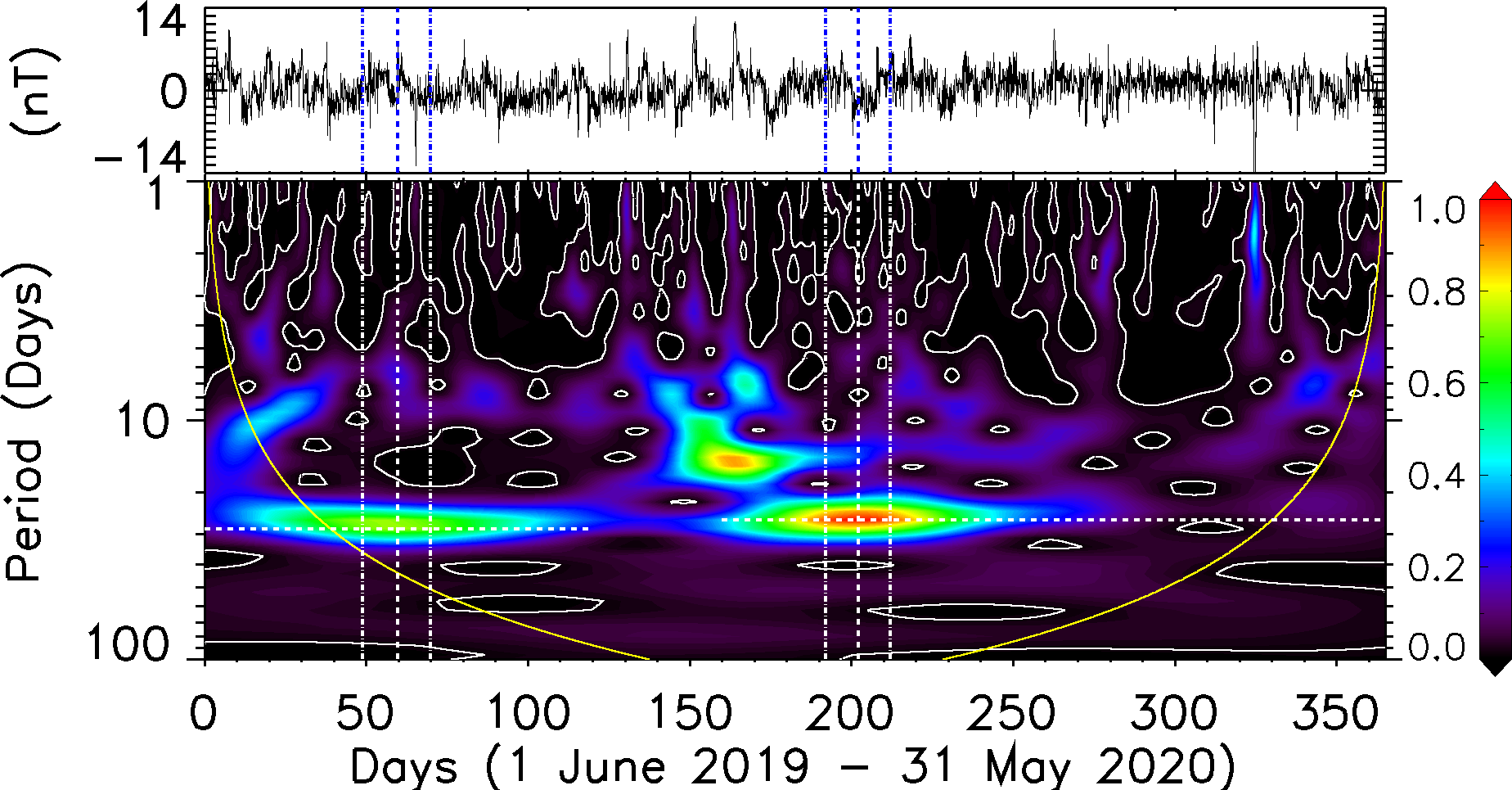}\label{fig:BT2020}} 
  \subfigure[Normal magnetic field $B_n$]{\includegraphics[width=0.49\textwidth]{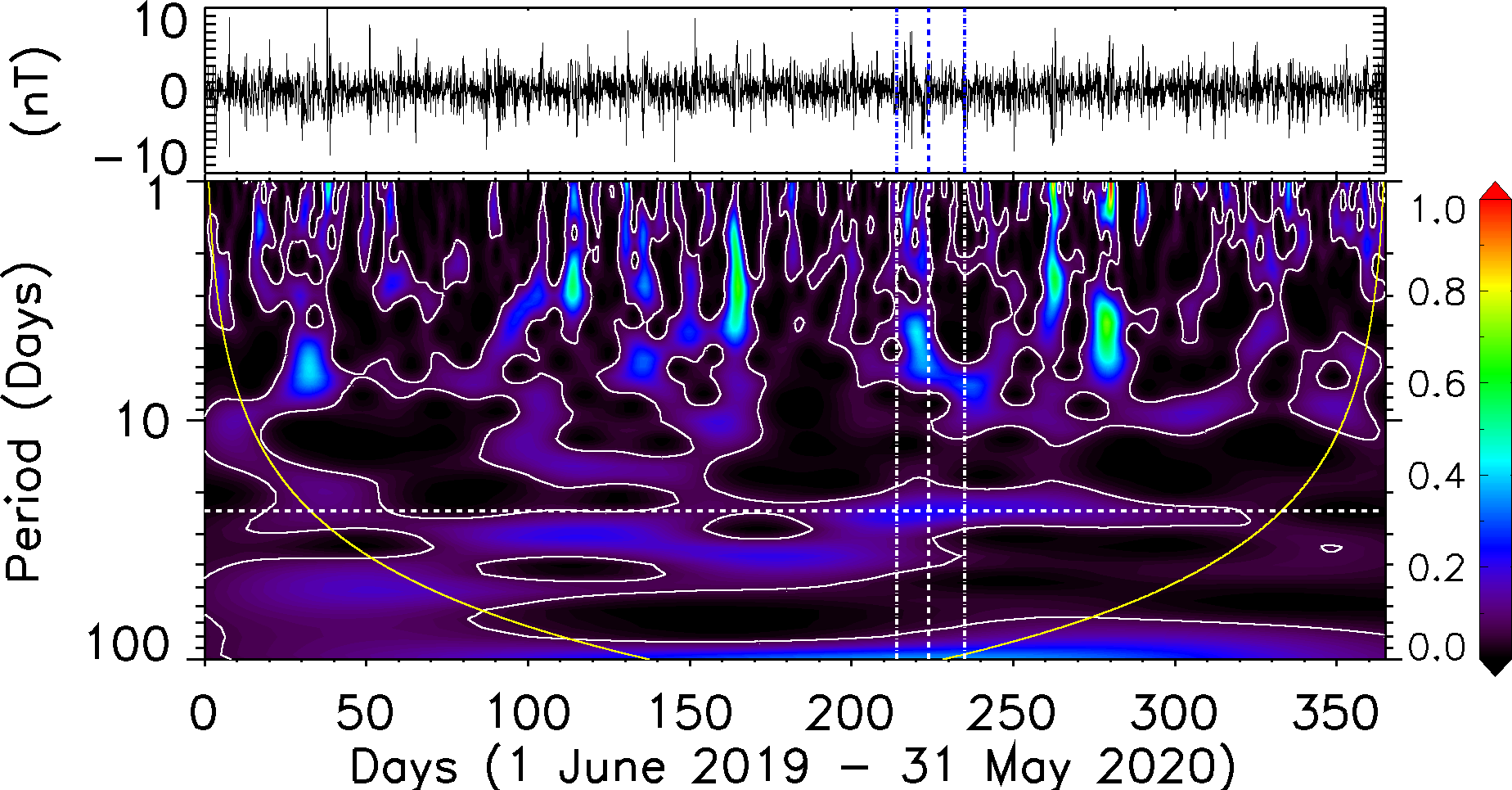}\label{fig:BN2020}} 
  \caption{Wavelets indicating the signals and periodicities of the (a) radial and (b) normal solar wind velocities, (c) proton number density, the (d) radial, (e) tangential and (f) normal magnetic field components between $1$ June $2019$ and $31$ May $2020$ during the $2019$/$2020$ solar minimum epoch. See Figure \ref{fig:1996} for a description of the dashed lines and contours.}
\label{fig:2020}
\end{figure*}

The same process as in Sections \ref{subsection: 1996/1997 Solar Min} and \ref{subsection: 2008/2009 Solar Min} is now followed for the $2019$/$2020$ solar minimum defined between $1$ June $2019$ and $31$ May $2020$ ($365$ days). Figure \ref{fig:2020} shows the wavelet results for $V_r$ (Figure \ref{fig:RSW2020}), $V_n$ (Figure \ref{fig:NSW2020}), $n_i$ (Figure \ref{fig:proton2020}), $B_r$ (Figure \ref{fig:BR2020}), $B_t$ (Figure \ref{fig:BT2020}) and $B_n$ (Figure \ref{fig:BN2020}). Table \ref{table:2019} summarises the details of this solar minimum epoch. The first pair of wavelets with maxima at approximately the same date during this solar minimum is $V_r$ on $22$ August $2019$ (day $82$) and $n_i$ on $25$ Augusts $2019$ (day $85$). The next trio of measurements grouped together are $V_n$ on $29$ July $2019$ (day $58$), $B_r$ on $6$ August $2019$ (day $66$) and $B_t$ on $31$ July $2019$ (day $60$). The third pair is $B_r$ on $14$ December $2019$ (day $196$) and $B_t$ on $20$ December $2019$ (day $202$). Lastly, the following four measurements show a local maximum 
 at approximately the same date: $V_r$ on $6$ January $2020$ (day $219$), $V_n$ on $4$ January $2020$ (day $217$), $n_i$ on $9$ January $2020$ (day $222$) and $B_t$ on $20$ December $2020$ (day $202$).

\begin{deluxetable}{ccccc}[t]
\tabletypesize{\footnotesize}
\tablecolumns{5}
\tablewidth{0pt}
\tablecaption{$2019$/$2020$ solar minimum with details of maxima in periodicity power at L$1$ as shown in Figure \ref{fig:2020}. \label{table:2019}}
\tablehead{
\colhead{} & \colhead{Date of max} &  \colhead{Period (Days)} &\colhead{CR} &\colhead{Figures} } 
\startdata
$V_r$       & $22$ Aug. $2019$    & $26.2$  & $2220$-$2221$ & \ref{fig:RSW2020} ; \ref{fig:CR2220_21} \\
       & $6$ Jan. $2020$    & $24.0$  & $2225$-$2226$ & \ref{fig:RSW2020} ; $--$ \\
$V_n$       & $29$ Jul. $2019$     & $31.17$  & $2220$        & \ref{fig:NSW2020} ; $--$\\
       & $4$ Jan. $2020$     & $26.2$  & $2225$        & \ref{fig:NSW2020} ; $--$\\
$n_i$  & $25$ Aug. $2019$    & $28.5$  & $2220$-$2221$ & \ref{fig:proton2020} ; $--$\\
  & $9$ Jan. $2020$    & $24.0$  & $2225$-$2226$ & \ref{fig:proton2020} ; $--$\\
$B_r$       & $6$ Aug. $2019$     & $28.5$  & $2219$-$2220$ & \ref{fig:BR2020} ; \ref{fig:CR2219},\ref{fig:CR2220_21}\\
       & $14$ Dec. $2019$     & $26.2$  & $2224$-$2225$ & \ref{fig:BR2020} ; $--$\\
$B_t$       & $30$ Jul. $2019$    & $28.5$  & $2219$-$2220$ & \ref{fig:BT2020} ; $--$\\
       & $20$ Dec. $2019$    & $26.2$  & $2224$-$2225$ & \ref{fig:BT2020} ; \ref{fig:CR2224_25}\\
$B_n$       & $10$ Jan. $2020$    & $24.0$  & $2225$-$2226$ & \ref{fig:BN2020} ; \ref{fig:CR2224_25} \ref{fig:CR2226}\\
\enddata
\end{deluxetable}

\section{Magnetic connection from L1 to the photosphere}
\label{section: Carrington Maps}

\begin{figure}[t]
\centering
\includegraphics[width=0.49\textwidth]{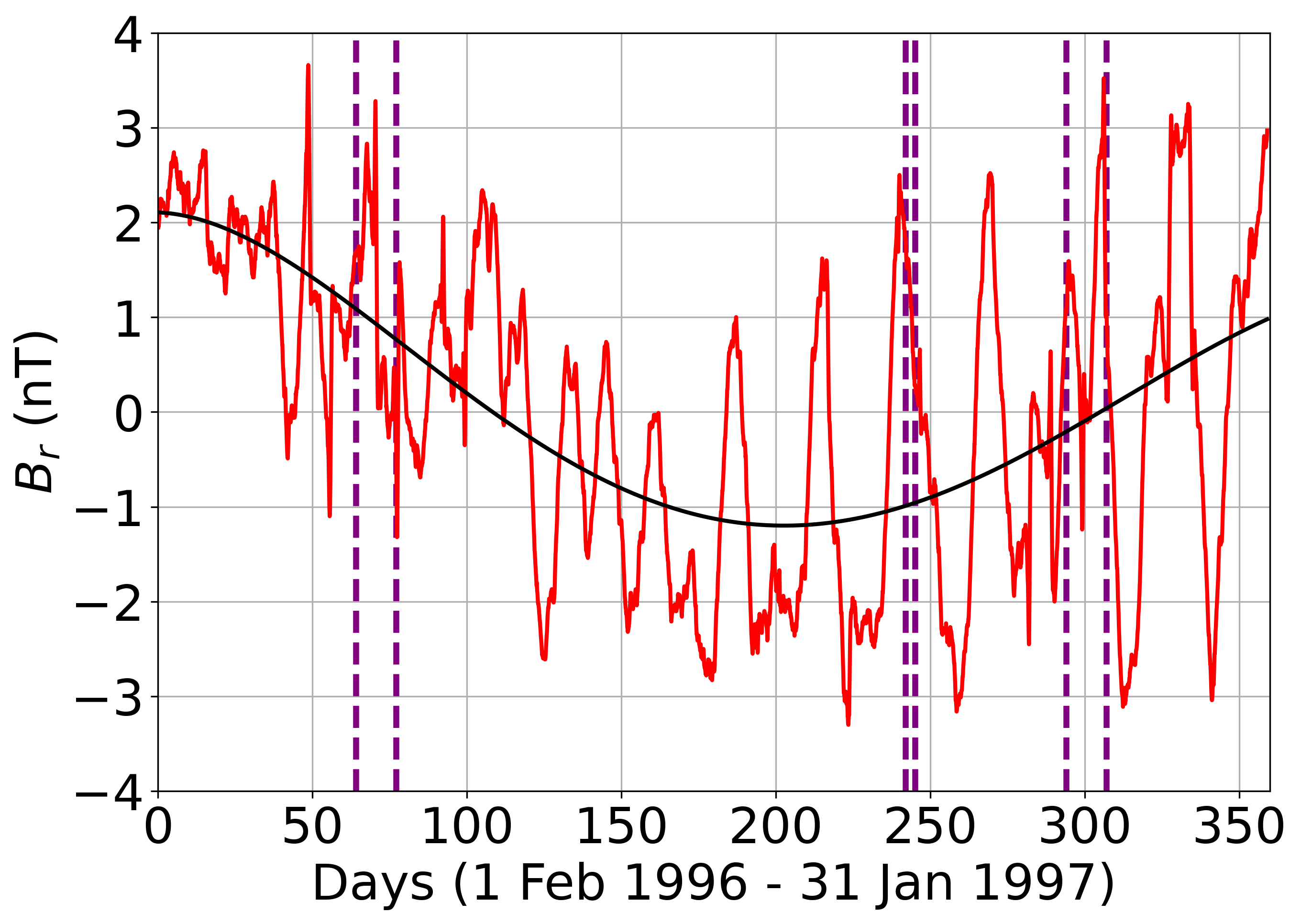}
\includegraphics[width=0.49\textwidth]{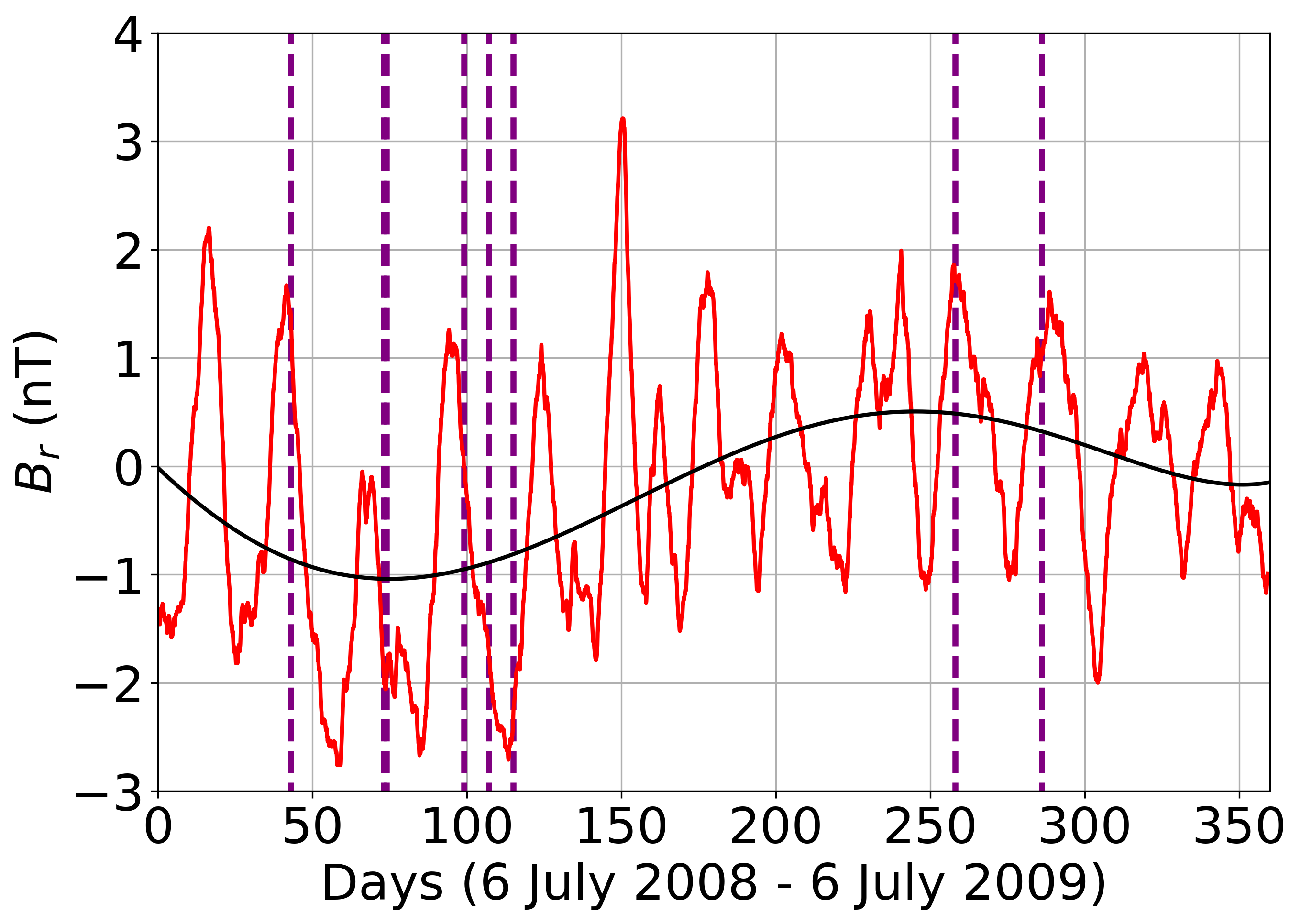}
\includegraphics[width=0.49\textwidth]{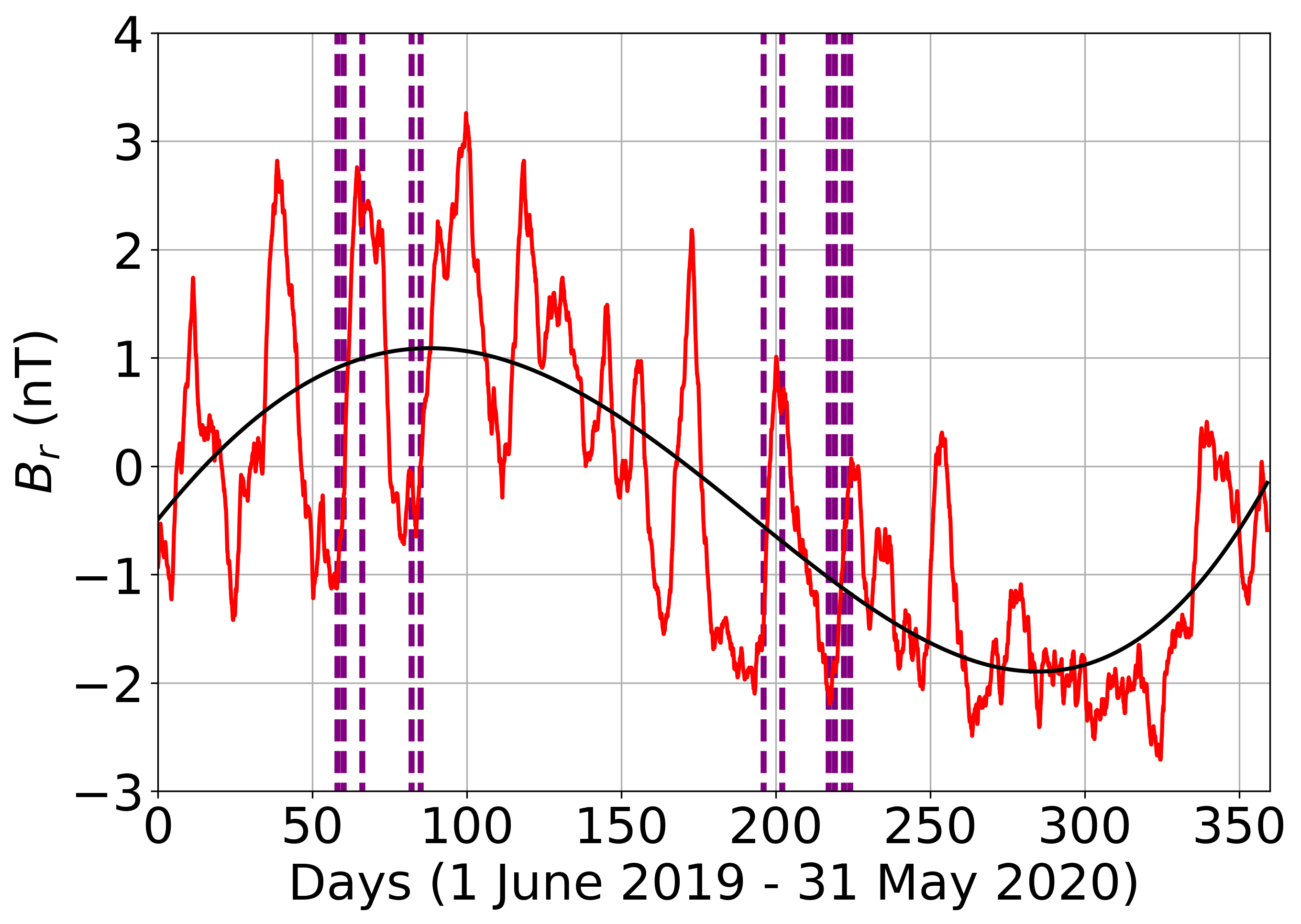}
\caption{One-week moving averages of $B_r$ (red) for the three investigated solar minima. A sixth-order polynomial approximation is fitted to $B_r$ in each panel (black). The vertical dashed lines indicate the dates of maximum signals according to Figures \ref{fig:1996}, \ref{fig:2009}, and \ref{fig:2020}.} 
\label{fig:Moving_Averages}
\end{figure}

The maximum periodicities observed at L$1$ in Figures \ref{fig:1996}, \ref{fig:2009}, and \ref{fig:2020} are now traced to their solar origins and shown on Carrington maps. The tracing is done along Fisk and Parker magnetic field lines connecting L$1$ with the SWSS (Figure \ref{fig:ArcLengths}). Next, the Fisk field lines are mapped to the photosphere using the transformations from Section \ref{subsection:Fisk HMF Model} while the Parker field lines are mapped to the photosphere using the PFSS model from Section \ref{subsection:PFSS model}. Figure \ref{fig:Moving_Averages} shows one-week moving averages (red) of $B_r$ for the $1996$/$1997$ (top), $2008$/$2009$ (middle), and $2019$/$2020$ (bottom) solar minima. A sixth-order polynomial approximation curve (black) is fitted to the moving average. The vertical dashed lines (purple) indicate the dates of the periodicities from Tables \ref{table:1996}, \ref{table:2008}, and \ref{table:2019}. The intersection between the polynomial approximation and the dates of peak periodicities indicate whether $B_r$ is positive or negative for that date. This informs whether the magnetic field line tracing is above or below the heliospheric current sheet (HCS).

During the $1996$/$1997$ solar minimum, the polarity of the sun was positive at the northern solar pole and negative at the southern solar pole ($A>0$ solar cycle). Therefore, a positive $B_r$ value in the top panel of Figure \ref{fig:Moving_Averages} is indicative of a magnetic field line with its origin at the northern solar pole while a negative $B_r$ value is indicative of a field line with its origin at the southern solar pole. The polarities of the PCHs are confirmed by the NOAA synoptic maps mentioned in Section \ref{subsection:Coronal Hole data}. During the $2008$/$2009$ solar minimum, the polarity of the sun was negative at the northern solar pole and positive at the southern solar pole ($A<0$ solar cycle). Therefore, a positive $B_r$ value in the middle panel of Figure \ref{fig:Moving_Averages} is indicative of a magnetic field line with its origin at the southern solar pole while a negative $B_r$ value is indicative of a field line with its origin at the northern solar pole. The $2019$/$2020$ solar minimum was an $A>0$ solar cycle again and therefore the polarities are the same as explained during the $1996$/$1997$ solar minimum.  

\subsection{1996/1997 solar minimum}
\label{subsection:1996/1997 solar minimum magnetic connection}

\begin{deluxetable}{ccccc}[t]
\tabletypesize{\footnotesize}
\tablecolumns{5}
\tablewidth{0pt}
\tablecaption{$1996$/$1997$ solar minimum temporal field line tracing details between L$1$ and the SWSS for the Fisk and Parker HMFs. \label{table:1996_Tracing}}
\tablehead{
\colhead{} & \colhead{L1} &  \colhead{$V_r$ (km/s)} &\colhead{Model} &\colhead{SWSS} } 
\startdata
$V_r$ & $3$ Oct. $1996$  & $614$ & Parker & $30$ Sept. $1996$  \\
      & $12$:$00$:$00$   &       &        & $05$:$00$:$00$    \\
      &                  &       & Fisk   & $30$ Sept. $1996$  \\
      &                  &       &        & $14$:$00$:$00$     \\
\hline
$n_i$& $5$ Dec. $1996$    & $288$  & Parker & $28$ Nov. $1996$ \\
     & $01$:$00$:$00$      &        &        & $10$:$00$:$00$    \\
     &                     &        & Fisk   & $29$ Nov. $1996$  \\
     &                     &        &        & $05$:$00$:$00$     \\
\hline
$B_t$& $21$ Nov. $1996$   & $410$   & Parker & $16$ Nov. $1996$ \\
     & $16$:$00$:$00$      &        &        & $17$:$00$:$00$    \\
     &                     &        & Fisk   & $17$ Nov. $1996$  \\
     &                     &        &        & $07$:$00$:$00$     \\   
\enddata
\end{deluxetable}

\begin{figure*}[t]
  \centering 
  \subfigure[]{\includegraphics[width=0.49\textwidth]{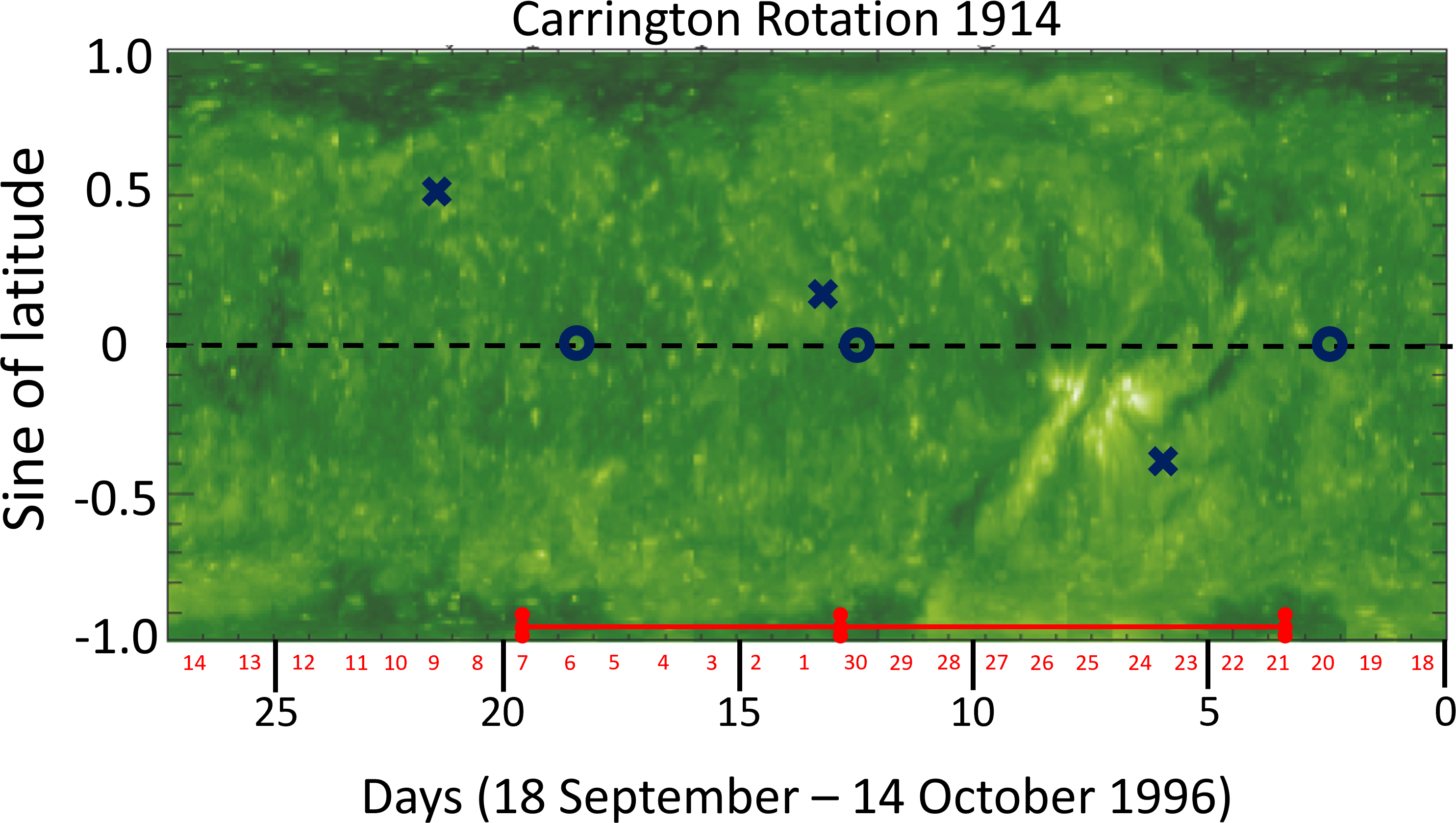}\label{fig:CR1914}} \\
  \subfigure[]{\includegraphics[width=0.49\textwidth]{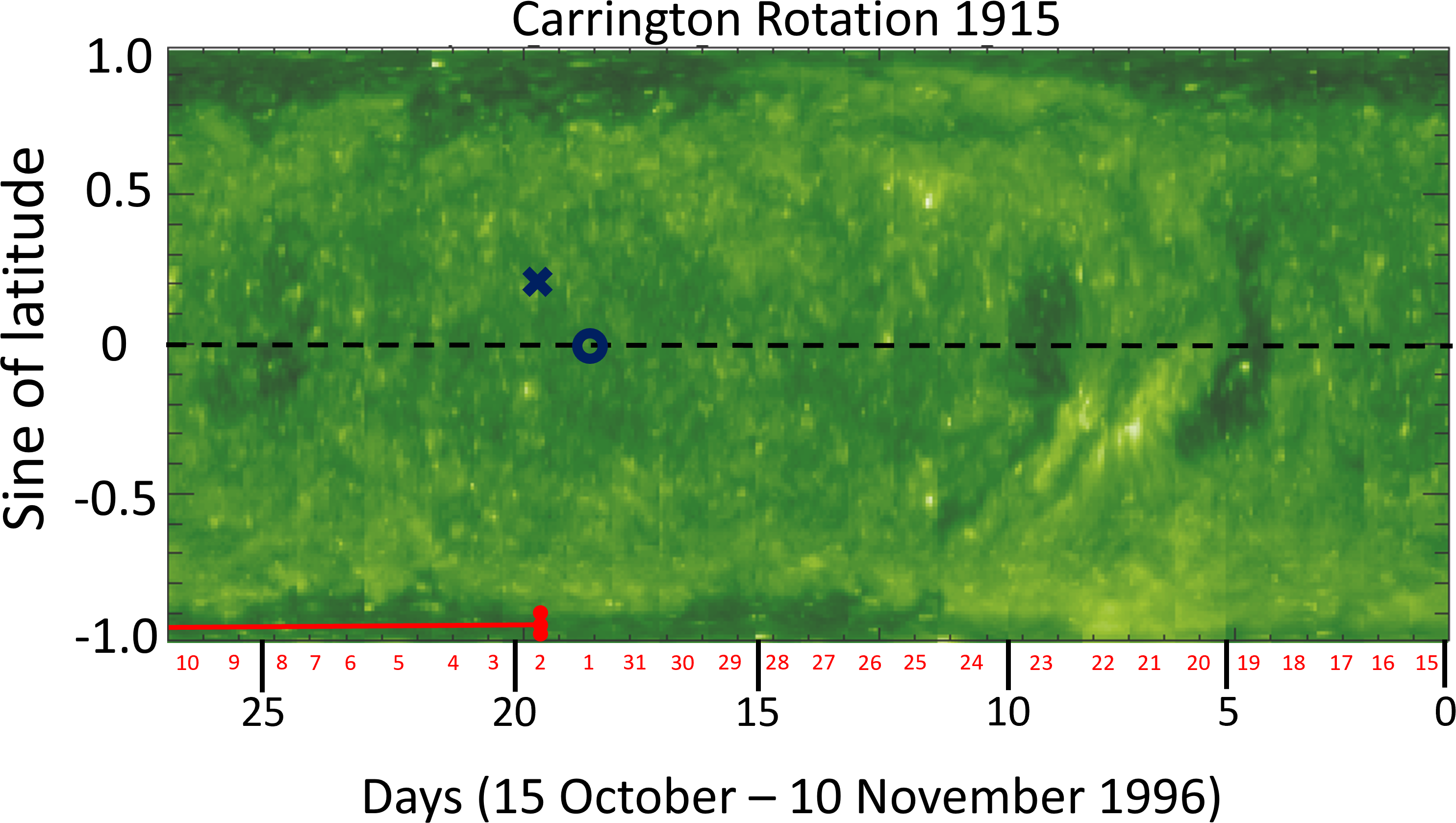}\label{fig:CR1915}} \\
  \subfigure[]{\includegraphics[width=1.0\textwidth]{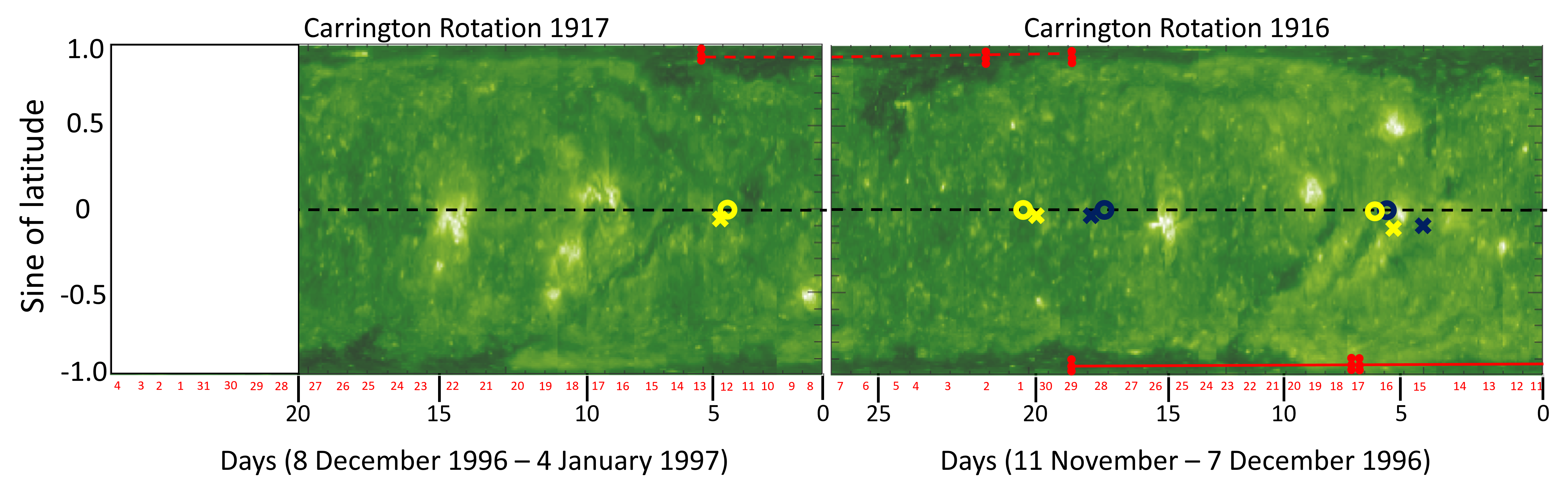}\label{fig:CR1916_1917}} 
  \caption{Carrington maps for Carrington rotations (a) $1914$, (b) $1915$, (c) $1916$ and $1917$. The horizontal axis runs chronologically from right to left (from the first day to the $27$th day of each Carrington rotation). The calendar dates are shown in red for ease of reference. The vertical axis is the sine of the solar latitude from $-1$ to $1$ with $0$ at the solar equator (indicated with a black vertical dashed line). Blue and yellow symbols refer to the mapping of the PFSS model. The circles are the entry points at the source surface and the crosses are the landing points at the photosphere. The red symbols refer to the mapping of the Fisk HMF. The vertical lines show the uncertainty in heliographic latitude, while the horizontal solid and dashed lines group together the traces from peak power and their $5\%$ reduction from Figure \ref{fig:1996}. The blank areas represent missing data, for example day $20$ - $27$ of CR$1917$.} 
\label{fig:CR1996}
\end{figure*}

The results of the mapping of magnetic field lines from the SWSS to the photosphere for the $1996$/$1997$ solar minimum of both the Fisk and PFSS models are shown in Figure \ref{fig:CR1996} and summarised in Table \ref{table:1996_Tracing}. Three ($V_r$, $n_i$, and $B_t$ in Table \ref{table:1996_Tracing}) of the six maxima from Table \ref{table:1996} are traced to the photosphere in Figure \ref{fig:CR1996}. The other maxima are not traced since their maxima occurs close to the already-traced maxima. The dates where the maximum power of the dominant periodicity occurred in Figure \ref{fig:1996}, together with the two dates corresponding to the $5\%$ decrease from the maxima are traced back from L$1$ to the photosphere using the Fisk model (Section \ref{subsection:Fisk HMF Model}), as well as a Parker spiral and PFSS model (Section \ref{subsection:PFSS model}). It is assumed that all signals observed at L$1$ pass the SWSS equator. The blue circles of Figure \ref{fig:CR1996} represent the entry points of the traced field lines at the SWSS equator ($2.5R_{\odot}$ above the photosphere) for both the Fisk and Parker HMFs.  Figure \ref{fig:ArcLengths} shows that the field lines lengths of the two HMF models are different and therefore, assuming a constant solar wind velocity (third column of Table \ref{table:1996_Tracing}) for both models, each HMF model will pass the SWSS at different dates (last column of Table \ref{table:1996_Tracing}). The differences between the times and dates are too small to differentiate on the Carrington maps and therefore one blue circle represents the entry point of both the Fisk and Parker field lines at the SWSS in Figure \ref{fig:CR1996}. The blue crosses show the heliographic latitude ($\theta_{hg}^{ph}$) to which each magnetic field line maps down from the SWSS equator to the photosphere using the PFSS model. 

The red vertical lines indicate the heliographic latitudes ($\theta_{hg}^{ph}$) to which the Fisk field lines map to on the photosphere. The length of the red vertical lines shows the uncertainty in the heliographic latitude during the mapping process brought on by the uncertainty of $\phi_{ss}^{hg}$ in equation (\ref{eq:FiskTrans1}). Cycling through $\phi_{ss}^{hg}=0^{\circ},90^{\circ},180^{\circ}$, and $270^{\circ}$ provides different $\theta_{hg}^{ph}$ coordinates to which the field line maps to the photosphere. This spread of values is represented by the red vertical lines connecting the different values of $\theta^{ph}_{hg}$. The red solid and dashed horizontal lines indicate the time duration of the signals represented by $\pm5\%$ from the maximum power of periodicity obtained from Figure \ref{fig:1996}. 

Note that the location of each blue cross symbol should not be interpreted in terms of time, since the time duration of the mapping between the SWSS and the photosphere is negligible. Rather, the location of each blue cross only represents the state of the corona when the mapping-down took place. The crosses and circles are not necessarily located on the same day on the Carrington map since there is a delay between when the mapping-down takes place and when that location is earth-facing in order to assemble the Carrington map. The same argument is true for the difference in longitude between the entry point of the Fisk field lines at the SWSS equator and the red vertical lines (indicative of $\phi_{hg}^{ph}$ in Equation (\ref{eq:FiskTrans6s})) for the results of the Fisk HMF. 

The already-assembled Carrington maps shown in Figure \ref{fig:CR1996} are from the SOHO/EIT Synoptic Map Database (Section \ref{subsection: carrington Maps Data}). These maps are not as accurate as those presented for the next two solar minima since they are constructed with wider longitudinal strips (approximately $16^{\circ}$), and there are instances where even wider strips are used from the previous day to cover missing data values \citep{Benevolenskaya2001}. Therefore, the calendar dates in red on the horizontal axis of Figure \ref{fig:CR1996} are only approximate dates. The darker green regions of each Carrington map show the locations of CHs. 

The tracing for $V_r$ (from Figure \ref{fig:RSW1996}) is shown in Figure \ref{fig:CR1914}. On $20$ September $1996$ (day $3$ of CR$1914$), the PFSS model maps to mid-latitudes close to the vicinity of an active region (AR) (which was earth-facing on $23$ September $1996$). On $30$ September $1996$ (day $13$ of CR$1914$) the field lines map close to the solar equator at a location with no CH activity. On $6$ October $1996$ (day $19$ of CR$1914$) the field lines map to the northern mid-latitudes which is in the vicinity of a well-developed northern PCH (which was earth-facing on $9$ October $1996$). The results of the Fisk model mapping are shown in red. On $21$ September $1996$ (day $4$ of CR$1914$) the Fisk model maps to a southern PCH. Furthermore, the Fisk field maps to two more southern PCHs on $30$ September $1996$ (day $13$) and $7$ October $1996$ (day $20$) of the same CR. Note that large northern PCHs are visible during CR$1914$, but the Fisk field do not map to these due to the orientation of the HCS shown in the top panel of Figure \ref{fig:Moving_Averages}. 

The blue circle in Figure \ref{fig:CR1915} represents the equatorial SWSS entry point of both Fisk and Parker field lines traced from the maximum of $B_t$ in Figure \ref{fig:BT1996} on day $19$ during CR$1915$ ($1$ November $1996$). The corresponding blue cross in Figure \ref{fig:CR1915} is not close to the location of visible CHs, while the red vertical line maps into a southern PCH (day $20$ during CR$1915$ on $2$ November $1996$) which extends into CR$1916$ (Figure \ref{fig:CR1916_1917}). The remaining blue circles and crosses in Figure \ref{fig:CR1916_1917} during CR$1916$ are from $B_t$ (Figure \ref{fig:BT1996}) crossing the SWSS equator on $16$ November $1996$ (day $6$) and $30$ November $1996$ (day $20$) where both field lines map near the photospheric equator. The yellow symbols in Figure \ref{fig:CR1916_1917} during CRs $1916$ and $1917$ refer to the maximum observed in $n_i$ (Figure \ref{fig:Proton1996}). The first SWSS entry of $n_i$ is made $17$ November $1996$ (day $7$ of CR$1916$), the maximum of $n_i$ is mapped on $28$ November $1996$ (day $18$ of CR$1916$), and the last field line from $n_i$ is mapped on $12$ December $1996$ (day $5$ of CR$1917$). All the PFSS results during CRs $1916$ and $1917$ map close to the solar equator and are not located near CH activity.    

The remaining Fisk field results are shown on $17$ November $1996$ (day $7$ of CR$1916$) and do not map to a PCH. On $29$ November $1996$ (day $19$ of CR$1916$), the Fisk model either maps to a southern PCH or a northern PCH which extends into CR$1917$. The top panel of Figure \ref{fig:Moving_Averages} shows the change of $B_r$ from negative to positive between the last two maxima and explains the change in mapping from the southern to the northern PCHs. The last two mapping results from the Fisk field are on $2$ December $1996$ (day $22$ during CR$1916$) on $13$ December $1996$ (day $6$ during CR$1917$), both of which are inside well-developed northern PCHs. 

\subsection{2008/2009 solar minimum}
\label{subsection:2008/2009 solar minimum magnetic connection}

\begin{deluxetable}{ccccc}[t]
\tabletypesize{\footnotesize}
\tablecolumns{5}
\tablewidth{0pt}
\tablecaption{$2008$/$2009$ solar minimum temporal field line tracing details between L$1$ and the SWSS for the Fisk and Parker HMFs. \label{table:2008_Tracing}}
\tablehead{
\colhead{} & \colhead{L1} &  \colhead{$V_r$ (km/s)} &\colhead{Model} &\colhead{SWSS} } 
\startdata
$V_r$ & $17$ Sept. $2008$  & $423$ & Parker & $12$ Sept. $2008$  \\
      & $15$:$00$:$00$     &       &        & $20$:$00$:$00$    \\
      &                    &       & Fisk   & $13$ Sept. $2008$  \\
      &                    &       &        & $09$:$00$:$00$     \\
\hline
$B_t$& $28$ Oct. $2008$    & $288$  & Parker & $21$ Oct. $2008$ \\
     & $20$:$00$:$00$      &        &        & $19$:$00$:$00$    \\
     &                     &        & Fisk   & $22$ Oct. $2008$  \\
     &                     &        &        & $15$:$00$:$00$     \\
\hline
$B_r$& $18$ Apr. $2009$   & $446$   & Parker & $13$ Apr. $2009$ \\
     & $05$:$00$:$00$      &        &        & $16$:$00$:$00$    \\
     &                     &        & Fisk   & $14$ Apr. $2009$  \\
     &                     &        &        & $02$:$00$:$00$     \\   
\hline
$V_t$& $13$ Oct. $2008$   & $498$   & Parker & $9$ Oct. $2008$ \\
     & $05$:$00$:$00$     &        &        & $03$:$00$:$00$    \\
     &                    &        & Fisk   & $9$ Oct. $2008$  \\
     &                    &        &        & $15$:$00$:$00$     \\ 
\hline
$V_n$& $21$ Mar. $2009$  & $432$   & Parker & $16$ Mar. $2009$ \\
     & $07$:$00$:$00$     &        &        & $14$:$00$:$00$    \\
     &                    &        & Fisk   & $17$ Mar. $2009$  \\
     &                    &        &        & $04$:$00$:$00$     \\ 
\enddata
\end{deluxetable}

\begin{figure*}[h]
  \centering
  \subfigure[]{\includegraphics[width=0.49\textwidth]{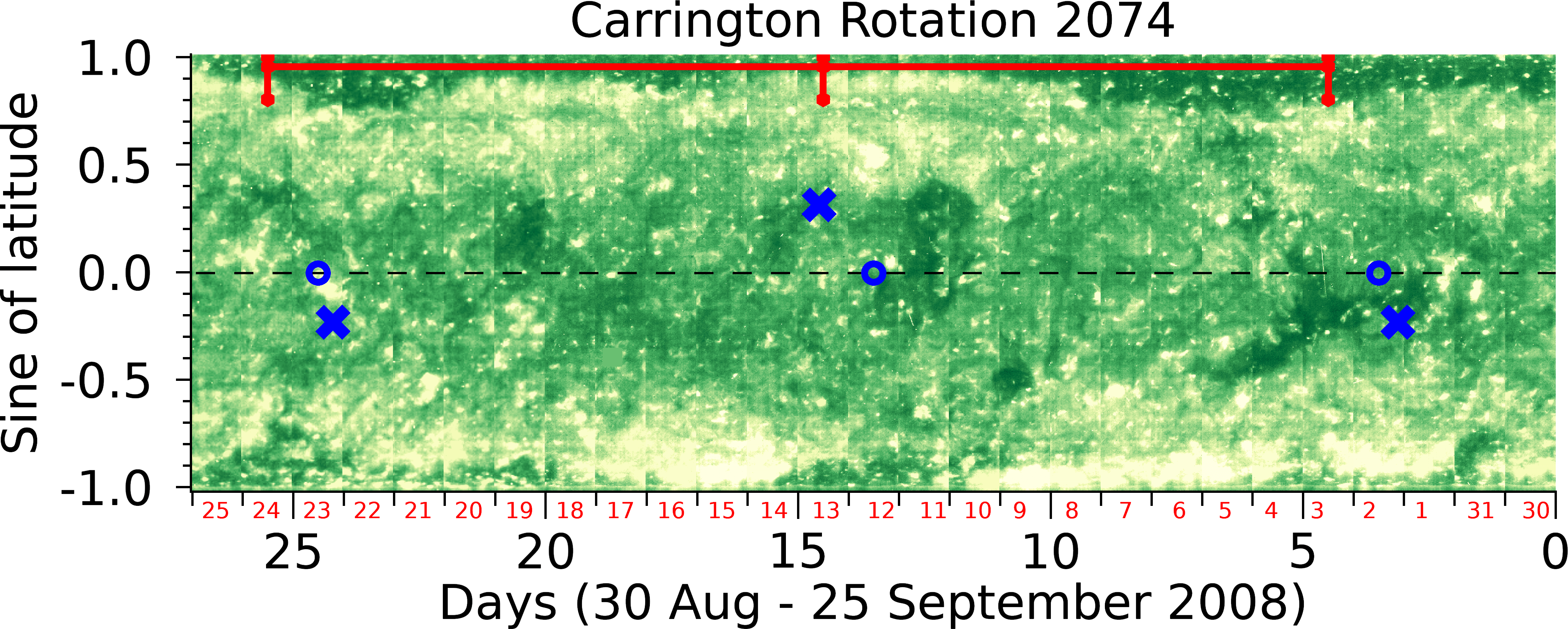}\label{fig:CR2074}} \\  
  \subfigure[]{\includegraphics[width=1\textwidth]{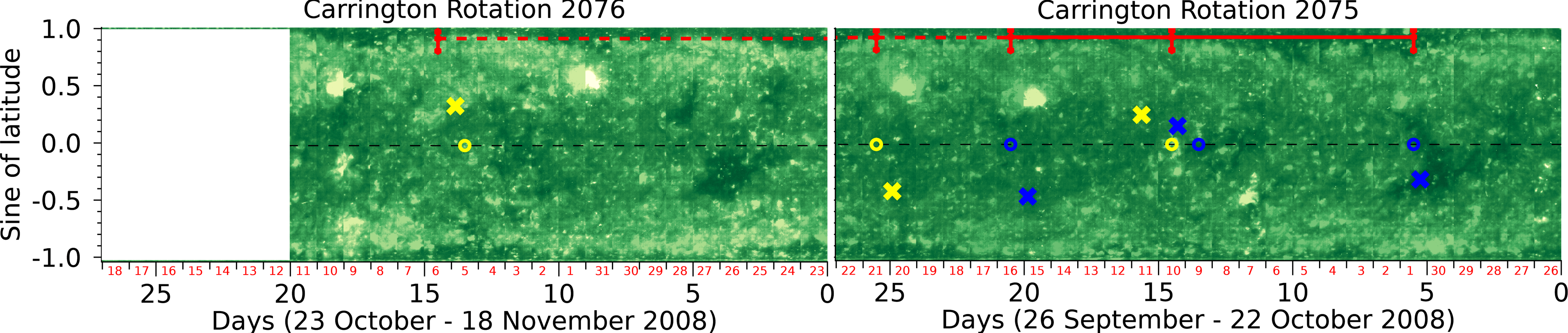}\label{fig:CR2075_76}} 
 \subfigure[]{\includegraphics[width=0.49\textwidth]{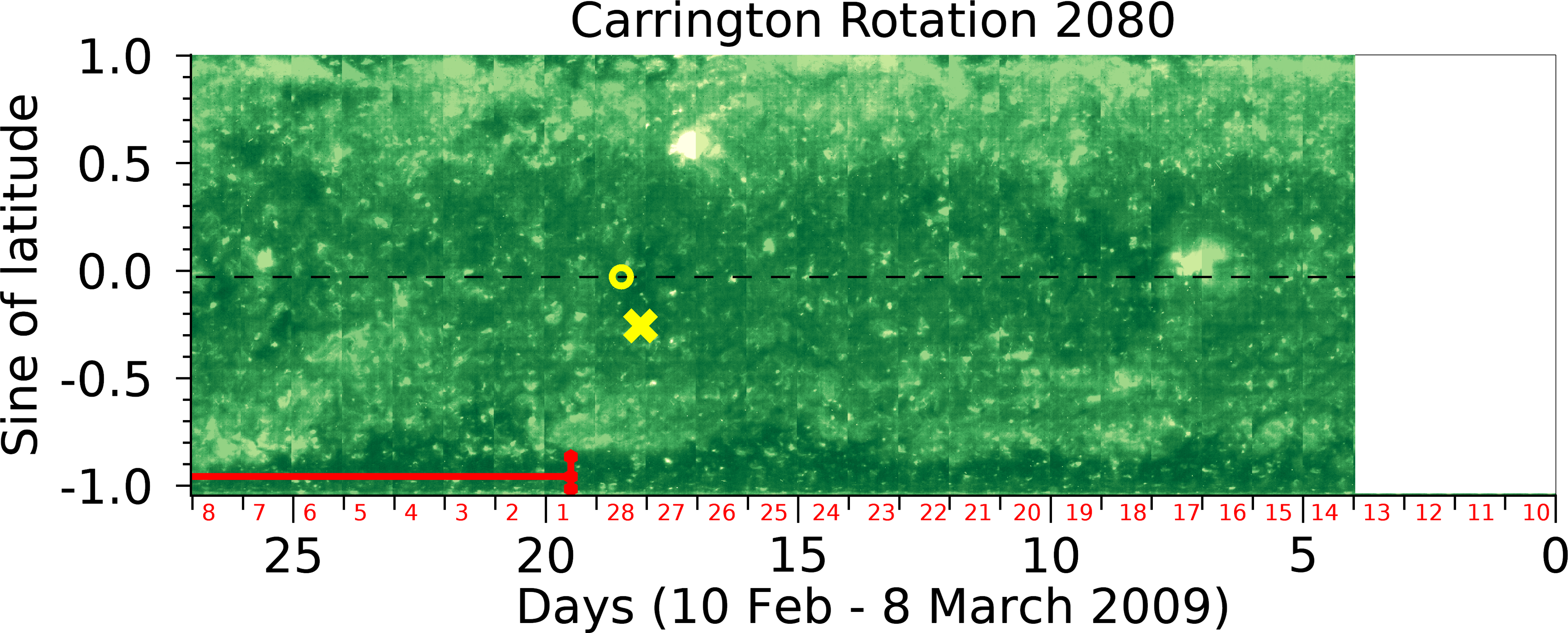}\label{fig:CR2080}} 
  \subfigure[]{\includegraphics[width=1\textwidth]{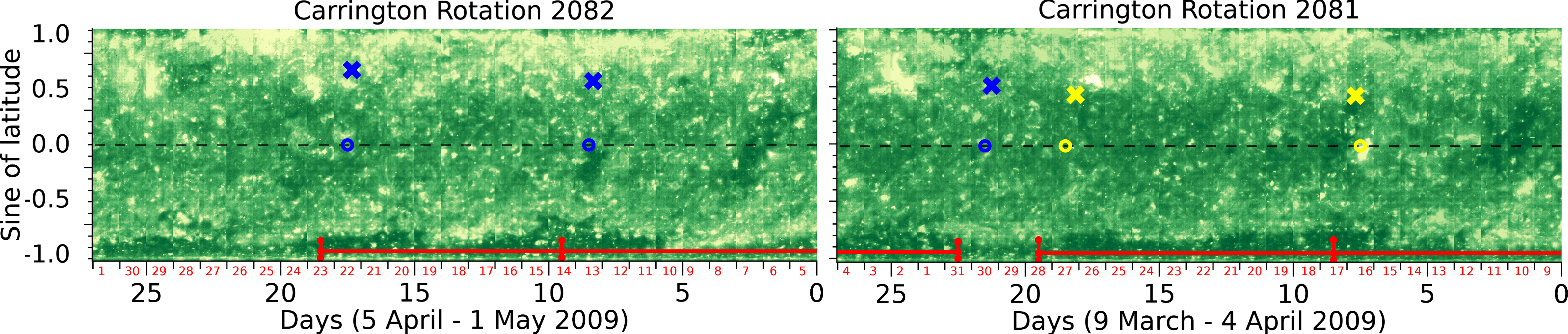}\label{fig:CR2081_82}} \\
 \caption{Carrington maps for Carrington rotations (a) $2074$, (b) $2075$ - $2076$, (c) $2080$, and (c) $2081$ - $2082$. See Figure \ref{fig:CR1996} for a description of the axes and the meaning of the colour labels.} 
\label{fig:CR2009}
\end{figure*}

Figure \ref{fig:CR2009} shows the Carrington maps for CRs $2074$ - $2076$ and $2080$ - $2082$ during the $2008$/$2009$ solar minimum of which the details are summarised in Table \ref{table:2008_Tracing}. Five ($V_r$, $B_t$, $B_r$, $V_t$ and $V_n$ in Table \ref{table:2008_Tracing}) of the eight maxima from Table \ref{table:2008} are traced to the photosphere in Figure \ref{fig:CR2009}. The other maxima are not traced since their maxima occurs close to the already-traced maxima. Figure \ref{fig:CR2074} shows three blue circles located on the SWSS equator which corresponds to the date of maximum power of the periodicity on $17$ September $2008$ (day $73$ of Figure \ref{fig:RSW2009}), and the two $5\%$ decreases from the maximum on $7$ September $2008$ and $29$ September $2008$ (days $63$ and $85$ in Figure \ref{fig:RSW2009}). Figure \ref{fig:CR2075_76} shows two CRs with the locations of the field line tracing from the maxima and $\pm5\%$ lines from Figures \ref{fig:TSW2009} and \ref{fig:BT2009} representing the blue and yellow circles and crosses respectively. The solid red horizontal lines show the mapping range using the Fisk model for the maximum shown in Figure \ref{fig:TSW2009}, while the dashed red horizontal lines (across the two Carrington maps) show the mapping for the maximum related to Figure \ref{fig:BR2009}. Figures \ref{fig:CR2080} and \ref{fig:CR2081_82} show the same as the previous figure, but now the blue symbols are related to Figure \ref{fig:BR2009} (maximum located on $17$ April $2009$ (day $285$)) and the yellow symbols are related to Figure \ref{fig:NSW2009} (maximum located on $21$ March $2009$ (day $258$)).

Figure \ref{fig:CR2074} shows that the PFSS model maps down to locations close to the solar equator, where the Fisk field maps exclusively to the north solar pole since $B_r$ is negative at the first maximum shown on the middle panel of Figure \ref{fig:Moving_Averages}. The Fisk model traces down into the northern PCHs between $3$ September $2008$ (day $5$ during CR$2074$) and $24$ August $2008$ (day $26$ during CR$2074$). There are two equatorial CHs visible between $2$ and $6$ September $2008$ (days $4$ to $8$ during CR$2074$) located a few degrees below the equator extending to mid-latitudes, and $10$ to $12$ September $2008$ (days $12$ to $14$ during the same CR) located a few degrees above the equator extending to mid-latitudes. The PFSS model maps down close to CH activity during these two instances, while no clear CH activity is observed at the location of the third and final mapping on $23$ September $2008$ (day $25$ during CR$2074$). 

Figure \ref{fig:CR2075_76} also shows the PFSS model mapping down close to the solar equator. A well-established equatorial CH is observed between $24$ and $28$ October $2008$ (days $2$ and $6$ during CR$2076$) which is a repetition of the CH observed in the previous CR between $29$ September and $2$ October $2008$ (days $4$ and $7$ during CR$2075$), which was captured by the tracing process. The Fisk model maps to well-developed northern PCHs during CR$2075$ and $2076$ due to the negative polarity of $B_r$ (middle panel of Figure \ref{fig:Moving_Averages}). The field line tracing results of the PFSS model in Figures \ref{fig:CR2080} and \ref{fig:CR2081_82} are further away from the solar equator than during previous CRs. The northern mid-latitude CHs visible on $16$ March $2009$ (day $8$ during CR$2081$) as well as the small northern mid-latitude CH visible on $26$ March $2009$ (day $18$ during the same CR) are well captured by the PFSS model. Mapping locations that are not close to CH activity include $30$ March $2009$ (day $22$ during CR$2081$), $13$ April $2009$ (day $9$ during CR$2082$), and $22$ April $2009$ (day $18$ during CR$2082$). Furthermore, the mapping of the Fisk model captures the southern PCHs (due to the positive $B_r$ polarity in the middle panel of Figure \ref{fig:Moving_Averages}) during the last three CRs of Figure \ref{fig:CR2009}.

\begin{figure*}[t]
  \centering
  \subfigure[]{\includegraphics[width=0.49\textwidth]{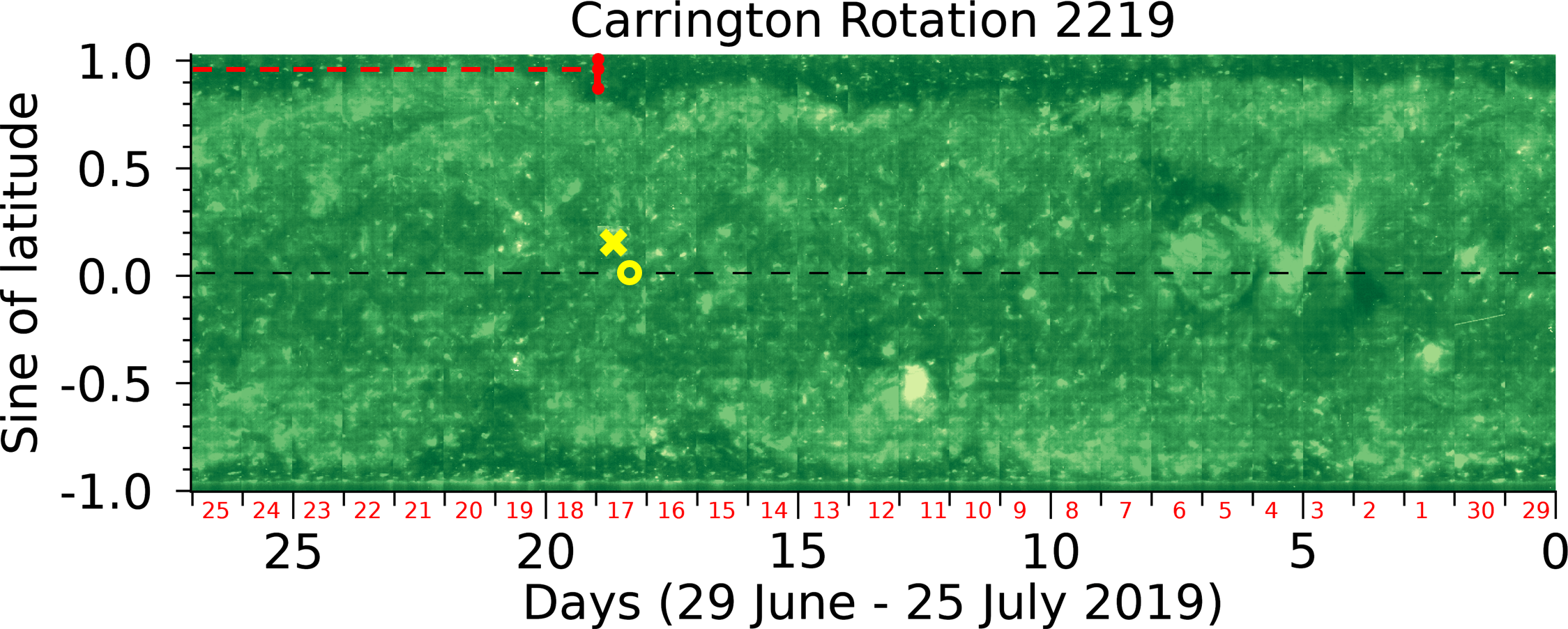}\label{fig:CR2219}} \\ 
  \subfigure[]{\includegraphics[width=1\textwidth]{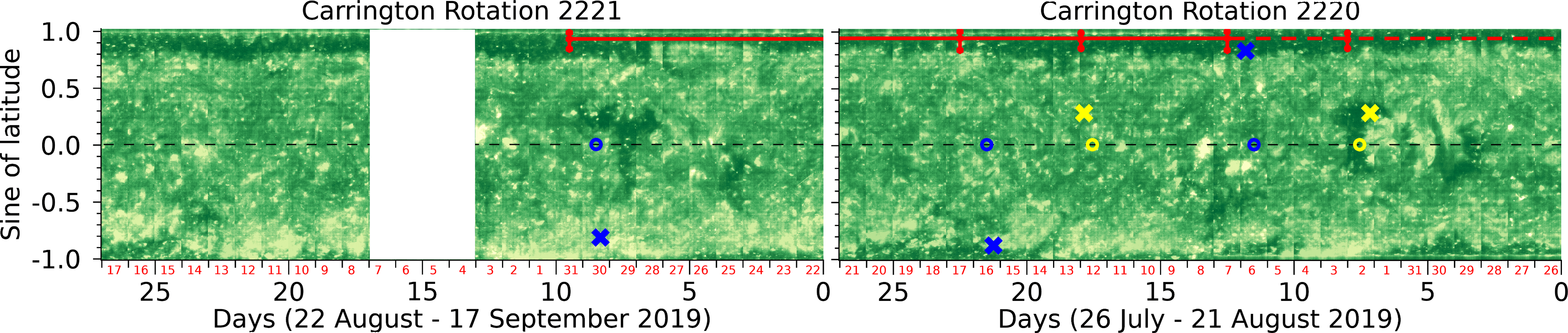}\label{fig:CR2220_21}} \\  
  \subfigure[]{\includegraphics[width=1\textwidth]{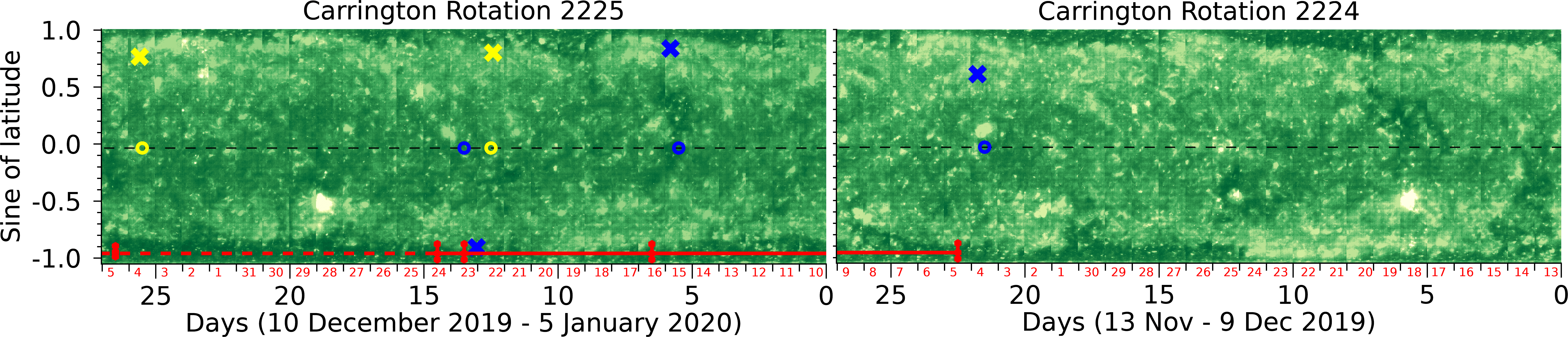}\label{fig:CR2224_25}} 
  \subfigure[]{\includegraphics[width=0.49\textwidth]{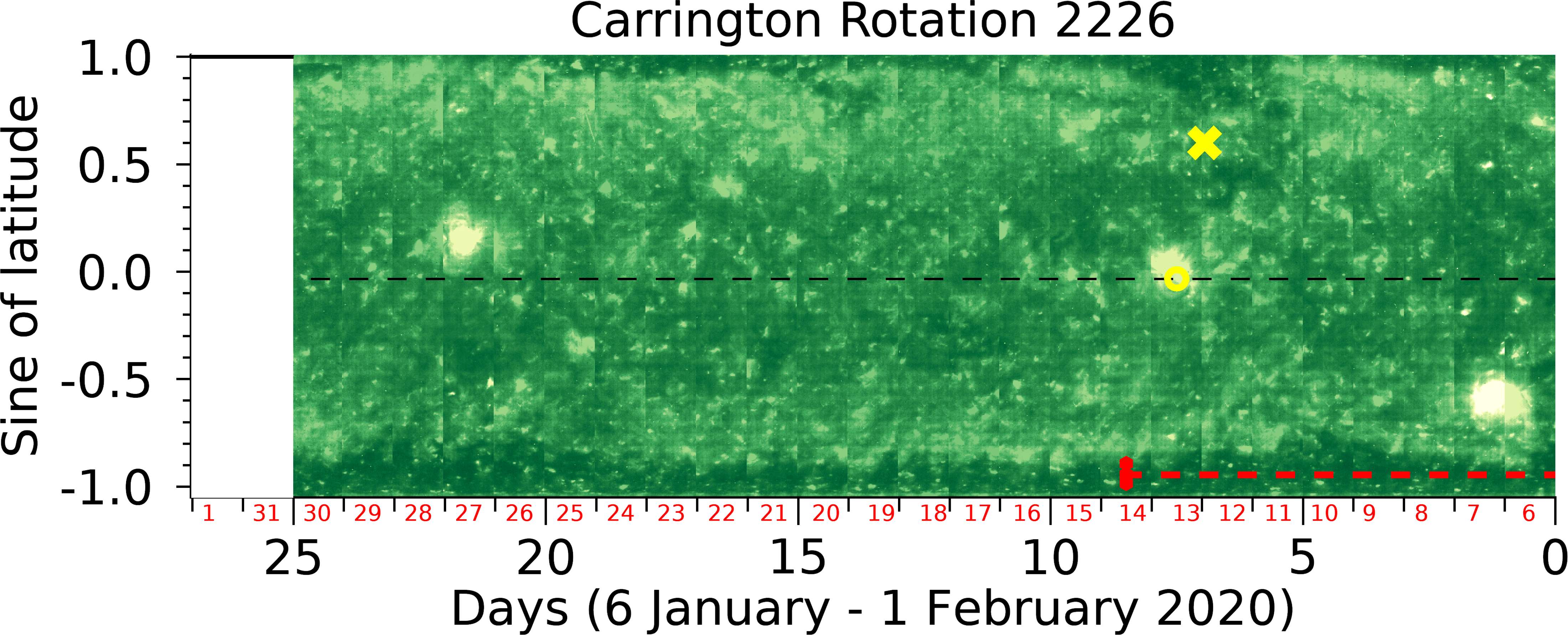}\label{fig:CR2226}} 
  \caption{Carrington maps for Carrington rotations (a) $2219$, (b) $2220$ - $2221$, (c) $2224$ - $2225$, and (d) $2226$. See Figure \ref{fig:CR1996} for a description of the axes and the meaning of the colour labels.} 
\label{fig:CR2020}
\end{figure*}

\subsection{2019/2020 solar minimum}
\label{subsection:2019/2020 solar minimum magnetic connection}

\begin{deluxetable}{ccccc}[t]
\tabletypesize{\footnotesize}
\tablecolumns{5}
\tablewidth{0pt}
\tablecaption{$2019$/$2020$ solar minimum temporal field line tracing details between L$1$ and the SWSS for the Fisk and Parker HMFs. \label{table:2019_Tracing}}
\tablehead{
\colhead{} & \colhead{L1} &  \colhead{$V_r$ (km/s)} &\colhead{Model} &\colhead{SWSS} } 
\startdata
$V_r$ & $22$ Aug. $2019$  & $345$ & Parker & $16$ Aug. $2019$  \\
      & $05$:$00$:$00$     &       &        & $10$:$00$:$00$    \\
      &                    &       & Fisk   & $17$ Aug. $2019$  \\
      &                    &       &        & $02$:$00$:$00$     \\
\hline
$B_r$& $6$ Aug. $2019$    & $502$  & Parker & $2$ Aug. $2019$ \\
     & $14$:$00$:$00$      &        &        & $13$:$00$:$00$    \\
     &                     &        & Fisk   & $3$ Aug. $2019$  \\
     &                     &        &        & $00$:$00$:$00$     \\
\hline
$B_t$& $20$ Dec. $2019$   & $430$   & Parker & $15$ Dec. $2019$ \\
     & $04$:$00$:$00$      &        &        & $11$:$00$:$00$    \\
     &                     &        & Fisk   & $16$ Dec. $2019$  \\
     &                     &        &        & $00$:$00$:$00$     \\   
\hline
$B_n$& $10$ Jan. $2020$   & $339$   & Parker & $4$ Jan. $2020$ \\
     & $20$:$00$:$00$     &        &        & $21$:$00$:$00$    \\
     &                    &        & Fisk   & $5$ Jan. $2020$  \\
     &                    &        &        & $14$:$00$:$00$     \\ 
\hline
\enddata
\end{deluxetable}

Figure \ref{fig:CR2020} shows the Carrington maps for CRs $2219$ - $2221$ and CRs $2224$ - $2226$ during the $2019$/$2020$ solar minimum of which the details are summarised in Table \ref{table:2019_Tracing}. Four ($V_r$, $B_r$, $B_t$ and $B_n$ in Table \ref{table:2019_Tracing}) of the eleven maxima from Table \ref{table:2019} are traced to the photosphere in Figure \ref{fig:CR2020}. The other maxima are not traced since their maxima occurs close to the already-traced maxima. Figures \ref{fig:CR2219} and \ref{fig:CR2220_21} show the locations of the mapping related to the first maximum observed for $B_r$ (day $66$ of Figure \ref{fig:BR2020}) and $V_r$ (day $82$ of Figure \ref{fig:RSW2020}). The second maximum power of periodicity for $B_t$ (day $202$ of Figure \ref{fig:BT2020}) and $B_n$ (day $223$ of \ref{fig:BN2020}) maps down to the photosphere during CRs $2224$ - $2226$ in Figures \ref{fig:CR2224_25} and \ref{fig:CR2226}. Figure \ref{fig:CR2220_21} (CR$2220$) shows two instances where the PFSS model maps near the solar equator and two instances where it maps near the solar poles. On $2$ August $2019$ (day $8$ during CR$2220$) the PFSS model maps into a well-developed CH extending from mid-latitudes to the solar equator. Northern PCHs are visible during CRs $2220$ and $2221$ (Figure \ref{fig:CR2220_21}) and are mapped by the Fisk HMF due to the positive polarity of the maxima shown in the first half of the bottom panel of Figure \ref{fig:Moving_Averages}. Both the PFSS and Fisk models map to a PCH on $6$ August $2019$ (day $12$ during CR$2220$). The remaining northern PCH activity, extending into CR$2221$ (up until $31$ August $2019$ (day $10$ during CR$2221$)) is captured by the Fisk field. On $16$ August $2019$ (day $22$ during CR$2220$) the PFSS maps down to a southern PCH while on $30$ August $2019$ (day $9$ during CR$2221$) it maps down to a location with little CH activity. It appears that the PFSS model does not map to the well-established equatorial-to-mid-latitude CH visible on $29$ August $2019$ (day $8$ during CR$2221$), although the PFSS model did map to this CH during the previous CR on $2$ August $2019$. During CRs $2224$ to $2226$ (Figures \ref{fig:CR2224_25} and \ref{fig:CR2226}) the PFSS model tends to map towards the polar regions as is the case for the Fisk model. Well-established southern PCHs starting on $5$ December $2019$ (day $23$ during CR$2224$) and extending to $14$ January $2020$ (day $9$ during CR$2226$) are captured by the Fisk field results. The Fisk field maps to the southern PCHs observed during CRs $2224$ and $2226$ due to the negative polarity of $B_r$ shown in the second half of the bottom panel of Figure \ref{fig:Moving_Averages}. During this solar minimum epoch, it seems the PFSS model is more likely to map to the solar poles than to the solar equator, in contrast to what was observed during the previous solar minimum epoch. 

\section{Discussion and Conclusion}
\label{Sec: Discussion and Conclusion}

\begin{deluxetable}{ccccc}[t]
\tabletypesize{\footnotesize}
\tablecolumns{5}
\tablewidth{0pt}
\tablecaption{Details of magnetic field line tracing to CHs for the two HMF models. Y indicates a field line mapped into a CH, while N indicates a field lines mapped outside a CH.   \label{table:Summary}}
\tablehead{
\colhead{Model} & \colhead{SWSS} &  \colhead{$-5\%$ (L)} &\colhead{Max} &\colhead{$-5\%$ (R)} } 
\startdata
Fisk       & $30$ Sept. $1996$  & Y & Y & Y  \\
P \& PFSS  & $30$ Sept. $1996$  & Y & N & N    \\
\hline
Fisk          & $17$ Nov. $1996$    & Y  & N & Y \\
P \& PFSS     & $16$ Nov. $1996$    & N  & N & N    \\
\hline
Fisk          & $29$ Nov. $1996$    & N  & Y & Y \\
P \& PFSS     & $28$ Nov. $1996$    & Y  & N & N    \\
\hline
\hline
Fisk          & $13$ Sept. $2008$    & Y  & Y & Y \\
P \& PFSS     & $12$ Sept. $2008$    & Y  & N & N    \\
\hline
Fisk          & $9$ Oct. $2008$    & Y  & Y & Y \\
P \& PFSS     & $9$ Oct. $2008$    & Y  & N & N    \\
\hline
Fisk          & $22$ Oct. $2008$    & Y  & N & Y \\
P \& PFSS     & $21$ Oct. $2008$    & N  & Y & N    \\
\hline
Fisk          & $17$ Mar. $2009$    & Y  & Y & N \\
P \& PFSS     & $16$ Mar. $2009$    & Y  & Y & N    \\
\hline
Fisk          & $14$ Apr. $2009$    & Y  & Y & Y \\
P \& PFSS     & $13$ Apr. $2009$    & N  & N & N    \\
\hline
\hline
Fisk          & $3$ Aug. $2019$    & Y  & Y & Y \\
P \& PFSS     & $2$ Aug. $2019$    & N  & Y & N    \\
\hline
Fisk          & $17$ Aug. $2019$    & Y  & Y & Y \\
P \& PFSS     & $16$ Aug. $2019$    & Y  & Y & N    \\
\hline
Fisk          & $16$ Dec. $2019$    & Y  & Y & Y \\
P \& PFSS     & $15$ Dec. $2019$    & N  & Y & Y    \\
\hline
Fisk          & $4$ Jan. $2020$    & Y  & Y & Y \\
P \& PFSS     & $5$ Jan. $2020$    & Y  & N & Y    \\
\hline
\enddata
\end{deluxetable}

This study focused on the identification of periodic signals at the L$1$ position using the solar wind velocity and proton number density data from the WIND spacecraft, and the IMF data from the ACE spacecraft during the $1996$/$1997$, $2008$/$2009$, and $2019$/$2020$ solar minima. We focused on periodicities between $26$ and $28$ days which is close to the solar synodic equatorial rotation rate. This is also the rotation rate of rigidly rotating CHs. To establish whether the periodic signals observed at L$1$ are related to the locations of CHs, the magnetic field lines are traced in the ecliptic plane from L$1$ to the SWSS assuming a Fisk field and for comparison also a Parker field. The tracing of field lines between the SWSS and the photosphere is done using the Fisk HMF model for the Fisk field lines and compared with the results of a PFSS model which arrives at the SWSS trough a Parker spiral. Table \ref{table:Summary} summarises the details of whether a field line, either Fisk or Parker (PFSS), maps to a CH or not. The HMF models (column $1$), the date at which each field line passed the SWSS (column $2$), the field line trace from the $-5\%$ reduction to the left of the maximum (column $3$), the field line trace at the maximum (column $4$), and the field line trace from the  $-5\%$ reduction to the right of the maximum (column $5$) are shown in the header of Table \ref{table:Summary}. Y indicates the field line traced to a CH and N indicates the field line did not trace to a CH. The normalised O$^{7+}$/O$^{6+}$ ratio is also used in the identification of the solar wind originating from CHs. \cite{Zhao2017} explain that a low heavy ion charge state ratio (O$^{7+}$/O$^{6+}$) is expected from CHs. The Appendix shows the analysis of these oxygen ratios for the $2008$/$2009$ and $2019$/$2020$ solar minima. Our results confirm the magnetic field lines traced in this study originate from PCHs and PCH boundaries as shown in the Carrington maps of Figures \ref{fig:CR2009} and \ref{fig:CR2020}.

The solar minimum of $1996$/$1997$ had a total of $9$ mappings for both the PFSS and Fisk models respectively. Note the change in magnetic field polarity (from negative to positive) between the last two maxima shown in the top panel of Figure \ref{fig:Moving_Averages} (day $19$ of CR$1916$ in Figure \ref{fig:CR1916_1917}. The PFSS model did not map to CHs in any of the cases during this solar minimum, but did map to locations close to (within $10^{\circ}$ in both latitude and longitude) CH activity in $2$/$9$ ($\sim 22\%$) cases (day $6$ of CR$1914$ and day $6$ of CR$1916$). The Fisk model mapped down to locations of CH activity in $7$/$9$ ($\sim 78\%$) cases. The two instances where the Fisk field did not map into a PCH are found on day $7$ of CR$1916$, in contrast to the PFSS model which mapped close to a CH on this day. During this time a combination of equatorial, mid-latitude, and polar mappings were obtained from the PFSS model, although during CRs $1916$ and $1917$ (Figure \ref{fig:CR1916_1917}) the PFSS model maps exclusively to the equatorial region. This epoch had CHs missed by the PFSS model. These include the northern mid-latitude CH visible on $22$ September $1996$ (day $5$ during CR$1914$ shown in Figure \ref{fig:CR1996}), a CH crossing the equator and visible on $19$ October $1996$ (day $5$ during CR$1915$ shown in Figure \ref{fig:CR1915}), and another CH crossing the equator during the same CR visible on $23$ October $1996$ (day $10$ during CR$1915$).

During the $2008$/$2009$ solar minimum, the PFSS model maps to equatorial to mid-latitude locations on the photosphere, while the Fisk model maps exclusively to the polar regions. During this solar minima, a total of $15$ instances of each model were traced from their respective maxima at L$1$ to the photosphere. Comparing the photospheric mapping coordinates with the locations of CHs using both Carrington maps and solar synoptic maps, the PFSS model maps to the locations of CHs in $5$/$15$ ($\sim33\%$) cases and the Fisk traces to $13$/$15$ ($\sim87\%$) cases. The two instances where the Fisk model mapped outside the PCH are found on day $26$ of CR$2075$ and day $20$ of CR$2081$. Note that the former instance is an example where the Fisk field does not map to a PCH, but the PFSS model maps to a mid-latitude CH. Therefore, the periodicity observed on that day was dominated by a mid-latitude CH identified by the PFSS model. There are examples where CHs are visible, but neither of the models mapped to these locations. These examples include a northern PCH visible between $30$ August to $1$ September $2008$ (days $1$ to $3$ during CR$2074$ shown in Figure \ref{fig:CR2074}), a CH bordering the solar equator from the south between $25$ to $27$ October $2008$ (days $3$ to $5$ during CR$2076$ shown in the left panel of Figure \ref{fig:CR2075_76}), a southern PCH visible between $14$ and $28$ February $2009$ (days $5$ to $19$ during CR$2080$ shown in Figure \ref{fig:CR2080}), and a CH crossing the equator between $10$ and $11$ March $2009$ (days $2$ to $3$ during CR$2081$ shown in the right panel of Figure \ref{fig:CR2081_82}). 

During the $2019$/$2020$ solar minimum epoch, a total of $12$ mapping results were obtained from both the PFSS and Fisk models respectively. The PFSS model mapped to CH locations in $5$/$12$ cases ($\sim 42\%$) and the Fisk model mapped down to $12$/$12$ ($100\%$) cases. Note that the results from the PFSS model deviates from the previous solar minimum in the observation that it maps much closer to the polar regions. This could be attributed to using more accurate magnetograms in the PFSS model during the $2019$/$2020$ solar minimum as mentioned in Section \ref{subsection:Photospheric BField}. A long lasting northern PCH between $1$ and $17$ September $2019$ (days $11$ to $27$ during CR$2221$ shown in the left panel of Figure \ref{fig:CR2220_21}) was not mapped by either model. 

Considering the results from the three solar minima, it seems that a PFSS model is successful in connecting magnetic field lines to periodicities observed at L$1$ to CH locations in less than half of the investigated cases. 

The Fisk model successfully connected magnetic field lines to the periodicities observed at L$1$ to CH locations in more than $78\%$ of the cases during all three solar minima investigated. It is important to note the main difference between the models. Magnetograms are used as input for the PFSS model which ensures its results are based on the magnetic configuration of the photosphere. From the results, it seems that when a magnetic field is traced from the SWSS equator, the PFSS model maps close to the solar equator and mid-latitudes in most instances. In contrast, the dominant input parameter of the Fisk model is the tilt angle $\alpha$. Field lines starting at the equator on the SWSS map exclusively to the solar poles. 

The most important qualitative point in this study is the relationship shown between the theory of Fisk-type fields and spacecraft observations at L$1$. In most cases investigated, the Fisk model traces field lines to inside PCHs. This emphasises the influence of PCHs on plasma measurements in the ecliptic. Furthermore, there are examples where very little to no PCH activity is observed, and the Fisk field does not trace to these polar regions due to the orientation of the HCS (Figure \ref{fig:Moving_Averages}). These examples include: CR$1917$ between $14$ and $19$ December $1996$, the northern pole during CR$2080$ between $14$ and $27$ February $2009$, CR$2081$ between $29$ and $30$ March $2009$, CR$2081$ between $24$ and $28$ April $2009$, the southern pole during CR$2221$ between $1$ and $13$ September $2019$ and lastly, CR$2224$ between $21$ and $24$ November $2019$. 

There are examples where the Fisk model traces field lines to PCHs on a specific date, but in contrast, the PFSS tracing on that same date does not map to existing CHs. One example include $13$ September $2008$ during CR$2074$ (Figure \ref{fig:CR2074}) where the northern PCH is mapped by the Fisk model, but the PFSS model maps to $\sim 30^{\circ}$ east (left) of a well-established CH. Another example include $13$ and $22$ April $2009$ during CR$2082$ (left panel of Figure \ref{fig:CR2081_82}) where the Fisk model maps to the southern PCH, and the PFSS model maps to the northern mid-latitudes with no CH activity. Similar examples are found during both the $1996$/$1997$ and $2019$/$2020$ solar minima. There are also two instances where the Fisk model did not map to a PCH, but the PFSS model mapped to a CH on that date. This asserts our notion that the Fisk and PFSS models are complimentary and should be viewed together when investigating solar wind disturbances at L$1$.  

Since the Fisk model cannot, by design, map to the equatorial and mid-latitude regions, it will not map to CHs in these regions. The Fisk model maps exclusively to PCHs, while the PFSS model is more successful in mapping to CHs in the equatorial region. This study emphasises the importance of combining results from the PFSS and the Fisk models to provide a more complete description of the connectivity between solar wind and magnetic field disturbances observed at L$1$ with their CH origin. Lastly, this study illustrates the large influence of PCHs on solar wind and magnetic field components within $1$ AU of the sun.  

\FloatBarrier
\begin{acknowledgments}
We thank the anonymous reviewer for their insightful and helpful comments that have improved this manuscript. We thank Micah J. Weberg for his advice in obtaining WIND and ACE data. All authors acknowledge UK Science and Technology Facilities Council (STFC) for IDL support. PJS thanks the National Research Foundation of South Africa (Ref Numbers TTK2204183656) for research support. Opinions expressed and conclusions arrived at are those of the authors and are not necessarily to be attributed to the NRF. The responsibility of the contents of this work is with the authors. GJJB and SR acknowledge the UK Science and Technology Facilities Council (STFC) for support from grant no. ST/X001008/1.
The data that support the findings of this study are available from the corresponding author upon reasonable request.
\end{acknowledgments}

\facilities{ACE, WIND, SOHO, SDO, NOAA}

\software{Python \citep{Hunter2007, Matplotlib2023},  
          Additional data analyses were done using IDL version 9.0 (Exelis Visual Information Solutions, Boulder, Colorado).
          }

\bibliography{sample631}{}

\begin{thebibliography}{}
\expandafter\ifx\csname natexlab\endcsname\relax\def\natexlab#1{#1}\fi
\providecommand{\url}[1]{\href{#1}{#1}}
\providecommand{\dodoi}[1]{doi:~\href{http://doi.org/#1}{\nolinkurl{#1}}}
\providecommand{\doeprint}[1]{\href{http://ascl.net/#1}{\nolinkurl{http://ascl.net/#1}}}
\providecommand{\doarXiv}[1]{\href{https://arxiv.org/abs/#1}{\nolinkurl{https://arxiv.org/abs/#1}}}

\bibitem[{{Altschuler} \& {Newkirk}(1969)}]{Altschuler1969}
{Altschuler}, M.~D., \& {Newkirk}, G. 1969, \solphys, 9, 131,
  \dodoi{10.1007/BF00145734}

\bibitem[{{Balogh} \& {Erd{\~o}s}(2013)}]{Balogh2013}
{Balogh}, A., \& {Erd{\~o}s}, G. 2013, \ssr, 176, 177,
  \dodoi{10.1007/s11214-011-9835-3}

\bibitem[{{Benevolenskaya}(2001)}]{CarringtonMaps1996}
{Benevolenskaya}, E. 2001, EIT Synoptic Maps.
\newblock \url{http://sun.stanford.edu/synop/EIT/}

\bibitem[{{Benevolenskaya} {et~al.}(2001){Benevolenskaya}, {Kosovichev}, \&
  {Scherrer}}]{Benevolenskaya2001}
{Benevolenskaya}, E.~E., {Kosovichev}, A.~G., \& {Scherrer}, P.~H. 2001, \apjl,
  554, L107, \dodoi{10.1086/320925}

\bibitem[{{Carrasco} \& {Vaquero}(2021)}]{Carrasco2021}
{Carrasco}, V.~M.~S., \& {Vaquero}, J.~M. 2021, Research Notes of the American
  Astronomical Society, 5, 181, \dodoi{10.3847/2515-5172/ac19a2}

\bibitem[{{Caswell} {et~al.}(2023){Caswell}, {Sales De Andrade}, {Lee},
  {Droettboom}, {Hoffmann}, {Klymak}, {Hunter}, {Firing}, {Stansby},
  {Varoquaux}, {Hedegaard Nielsen}, {Gustafsson}, {Root}, {May}, {Elson},
  {Sepp{\"a}nen}, {Lee}, {Dale}, {Sunden}, {Hannah}, {McDougall}, {Straw},
  {Hobson}, {Lucas}, {Gohlke}, {Vincent}, {Yu}, {Ma}, {Silvester}, \&
  {Moad}}]{Matplotlib2023}
{Caswell}, T.~A., {Sales De Andrade}, E., {Lee}, A., {et~al.} 2023,
  {matplotlib/matplotlib: REL: v3.7.2}, v3.7.2, Zenodo,  Zenodo,
  \dodoi{10.5281/zenodo.8118151}

\bibitem[{{Contopoulos}(2013)}]{Contopoulos2013}
{Contopoulos}, I. 2013, \solphys, 282, 419, \dodoi{10.1007/s11207-012-0154-y}

\bibitem[{{Domingo} {et~al.}(1995){Domingo}, {Fleck}, \& {Poland}}]{SOHO1995}
{Domingo}, V., {Fleck}, B., \& {Poland}, A.~I. 1995, \solphys, 162, 1,
  \dodoi{10.1007/BF00733425}

\bibitem[{ESA(2023)}]{CoronalHoles}
ESA. 2023, Soho Science Archive.
\newblock \url{https://ssa.esac.esa.int/ssa/#/pages/search}

\bibitem[{{Fisk}(1996)}]{Fisk1996}
{Fisk}, L.~A. 1996, \jgr, 101, 15547, \dodoi{10.1029/96JA01005}

\bibitem[{{Fisk} {et~al.}(1999){Fisk}, {Zurbuchen}, \& {Schwadron}}]{Fisk1999}
{Fisk}, L.~A., {Zurbuchen}, T.~H., \& {Schwadron}, N.~A. 1999, \apj, 521, 868,
  \dodoi{10.1086/307556}

\bibitem[{{Freeland} \& {Handy}(1998)}]{SSW1998}
{Freeland}, S.~L., \& {Handy}, B.~N. 1998, \solphys, 182, 497,
  \dodoi{10.1023/A:1005038224881}

\bibitem[{{Garraffo} {et~al.}(2013){Garraffo}, {Cohen}, {Drake}, \&
  {Downs}}]{Garraffo2013}
{Garraffo}, C., {Cohen}, O., {Drake}, J.~J., \& {Downs}, C. 2013, \apj, 764,
  32, \dodoi{10.1088/0004-637X/764/1/32}

\bibitem[{{Gilbert} {et~al.}(2007){Gilbert}, {Zurbuchen}, \&
  {Fisk}}]{Gilbert2007}
{Gilbert}, J.~A., {Zurbuchen}, T.~H., \& {Fisk}, L.~A. 2007, \apj, 663, 583,
  \dodoi{10.1086/518099}

\bibitem[{{Gloeckler} {et~al.}(1998){Gloeckler}, {Cain}, {Ipavich}, {Tums},
  {Bedini}, {Fisk}, {Zurbuchen}, {Bochsler}, {Fischer}, {Wimmer-Schweingruber},
  {Geiss}, \& {Kallenbach}}]{SWICS1998}
{Gloeckler}, G., {Cain}, J., {Ipavich}, F.~M., {et~al.} 1998, \ssr, 86, 497,
  \dodoi{10.1023/A:1005036131689}

\bibitem[{{Hamada} {et~al.}(2018){Hamada}, {Asikainen}, {Virtanen}, \&
  {Mursula}}]{Hamada2018}
{Hamada}, A., {Asikainen}, T., {Virtanen}, I., \& {Mursula}, K. 2018, \solphys,
  293, 71, \dodoi{10.1007/s11207-018-1289-2}

\bibitem[{{Hoeksema}(1984)}]{Hoeksema1984}
{Hoeksema}, J.~T. 1984, PhD thesis, Stanford University, California

\bibitem[{{Hoeksema}(1995)}]{Hoeksema1995_Tiltangles}
---. 1995, \ssr, 72, 137, \dodoi{10.1007/BF00768770}

\bibitem[{{Hofmeister} {et~al.}(2019){Hofmeister}, {Utz}, {Heinemann},
  {Veronig}, \& {Temmer}}]{Hofmeister2019}
{Hofmeister}, S.~J., {Utz}, D., {Heinemann}, S.~G., {Veronig}, A., \& {Temmer},
  M. 2019, \aap, 629, A22, \dodoi{10.1051/0004-6361/201935918}

\bibitem[{{Hunter}(2007)}]{Hunter2007}
{Hunter}, J.~D. 2007, Computing in Science and Engineering, 9, 90,
  \dodoi{10.1109/MCSE.2007.55}

\bibitem[{{Jain} {et~al.}(2024){Jain}, {Choudhary}, \& {Imamura}}]{Jain2024}
{Jain}, R.~N., {Choudhary}, R.~K., \& {Imamura}, T. 2024, \mnras, 529, L123,
  \dodoi{10.1093/mnrasl/slae008}

\bibitem[{{Jian} {et~al.}(2011){Jian}, {Russell}, \& {Luhmann}}]{Jian2011}
{Jian}, L.~K., {Russell}, C.~T., \& {Luhmann}, J.~G. 2011, \solphys, 274, 321,
  \dodoi{10.1007/s11207-011-9737-2}

\bibitem[{{Lepping} {et~al.}(1995){Lepping}, {Ac{\~{u}}na}, {Burlaga},
  {Farrell}, {Slavin}, {Schatten}, {Mariani}, {Ness}, {Neubauer}, {Whang},
  {Byrnes}, {Kennon}, {Panetta}, {Scheifele}, \& {Worley}}]{Lepping1995}
{Lepping}, R.~P., {Ac{\~{u}}na}, M.~H., {Burlaga}, L.~F., {et~al.} 1995, \ssr,
  71, 207, \dodoi{10.1007/BF00751330}

\bibitem[{{Li} {et~al.}(2016){Li}, {Cairns}, {Gosling}, {Steward}, {Francis},
  {Neudegg}, {Schulte in den B{\"a}umen}, {Player}, \& {Milne}}]{Li_2016}
{Li}, B., {Cairns}, I.~H., {Gosling}, J.~T., {et~al.} 2016, Journal of
  Geophysical Research (Space Physics), 121, 925, \dodoi{10.1002/2015JA021853}

\bibitem[{{Luhmann} {et~al.}(2022){Luhmann}, {Li}, {Lee}, {Jian}, {Arge}, \&
  {Riley}}]{SpaceWeather2022}
{Luhmann}, J.~G., {Li}, Y., {Lee}, C.~O., {et~al.} 2022, Space Weather, 20,
  e2022SW003110, \dodoi{10.1029/2022SW003110}

\bibitem[{{Mursula} \& {Zieger}(1996)}]{Murula1996}
{Mursula}, K., \& {Zieger}, B. 1996, \jgr, 101, 27077,
  \dodoi{10.1029/96JA02470}

\bibitem[{{Neugebauer} \& {Snyder}(1962)}]{Neugebauer1962}
{Neugebauer}, M., \& {Snyder}, C.~W. 1962, Science, 138, 1095,
  \dodoi{10.1126/science.138.3545.1095.a}

\bibitem[{NOAA(1996-2020)}]{SynopticCharts}
NOAA. 1996-2020, Solar Synoptic Analysis.
\newblock
  \url{https://www.ngdc.noaa.gov/stp/space-weather/solar-data/solar-imagery/composites/full-sun-drawings/boulder/}

\bibitem[{{Ogilvie} {et~al.}(1995){Ogilvie}, {Chornay}, {Fritzenreiter},
  {Hunsaker}, {Keller}, {Lobell}, {Miller}, {Scudder}, {Sittler}, {Torbert},
  {Bodet}, {Needell}, {Lazarus}, {Steinberg}, {Tappan}, {Mavretic}, \&
  {Gergin}}]{Ogilvie1995}
{Ogilvie}, K.~W., {Chornay}, D.~J., {Fritzenreiter}, R.~J., {et~al.} 1995,
  \ssr, 71, 55, \dodoi{10.1007/BF00751326}

\bibitem[{{Owens} \& {Forsyth}(2013)}]{Owens2013}
{Owens}, M.~J., \& {Forsyth}, R.~J. 2013, Living Reviews in Solar Physics, 10,
  5, \dodoi{10.12942/lrsp-2013-5}

\bibitem[{{Parenti} {et~al.}(2021){Parenti}, {Chifu}, {Del Zanna}, {Edmondson},
  {Giunta}, {Hansteen}, {Higginson}, {Laming}, {Lepri}, {Lynch}, {Rivera}, {von
  Steiger}, {Wiegelmann}, {Wimmer-Schweingruber}, {Zambrana Prado}, \&
  {Pelouze}}]{Parenti2021}
{Parenti}, S., {Chifu}, I., {Del Zanna}, G., {et~al.} 2021, \ssr, 217, 78,
  \dodoi{10.1007/s11214-021-00856-1}

\bibitem[{{Parker}(1958)}]{Parker1958}
{Parker}, E.~N. 1958, \apj, 128, 664, \dodoi{10.1086/146579}

\bibitem[{{Pesnell} {et~al.}(2012){Pesnell}, {Thompson}, \& {Chamberlin}}]{SDO}
{Pesnell}, W.~D., {Thompson}, B.~J., \& {Chamberlin}, P.~C. 2012, \solphys,
  275, 3, \dodoi{10.1007/s11207-011-9841-3}

\bibitem[{{Posner} {et~al.}(2001){Posner}, {Zurbuchen}, {Schwadron}, {Fisk},
  {Gloeckler}, {Linker}, {Miki{\'c}}, \& {Riley}}]{Posner2001}
{Posner}, A., {Zurbuchen}, T.~H., {Schwadron}, N.~A., {et~al.} 2001, \jgr, 106,
  15869, \dodoi{10.1029/2000JA000112}

\bibitem[{{Prabhakaran Nayar} {et~al.}(2002){Prabhakaran Nayar}, {Radhika},
  {Revathy}, \& {Ramadas}}]{Prabhakaran2002}
{Prabhakaran Nayar}, S.~R., {Radhika}, V.~N., {Revathy}, K., \& {Ramadas}, V.
  2002, \solphys, 208, 359, \dodoi{10.1023/A:1020565831926}

\bibitem[{{Richardson} \& {Cane}(2010)}]{Richardson2010}
{Richardson}, I.~G., \& {Cane}, H.~V. 2010, \solphys, 264, 189,
  \dodoi{10.1007/s11207-010-9568-6}

\bibitem[{{Riley} {et~al.}(2022){Riley}, {Caplan}, {Downs}, {Linker}, \&
  {Lionello}}]{Riley2022}
{Riley}, P., {Caplan}, R.~M., {Downs}, C., {Linker}, J.~A., \& {Lionello}, R.
  2022, Journal of Geophysical Research (Space Physics), 127, e30261,
  \dodoi{10.1029/2022JA030261}

\bibitem[{{Schatten} {et~al.}(1969){Schatten}, {Wilcox}, \&
  {Ness}}]{Schatten1969}
{Schatten}, K.~H., {Wilcox}, J.~M., \& {Ness}, N.~F. 1969, \solphys, 6, 442,
  \dodoi{10.1007/BF00146478}

\bibitem[{{Scherrer} {et~al.}(1995){Scherrer}, {Bogart}, {Bush}, {Hoeksema},
  {Kosovichev}, {Schou}, {Rosenberg}, {Springer}, {Tarbell}, {Title},
  {Wolfson}, {Zayer}, \& {MDI Engineering Team}}]{SOHO_MDI}
{Scherrer}, P.~H., {Bogart}, R.~S., {Bush}, R.~I., {et~al.} 1995, \solphys,
  162, 129, \dodoi{10.1007/BF00733429}

\bibitem[{{Scherrer} {et~al.}(2012){Scherrer}, {Schou}, {Bush}, {Kosovichev},
  {Bogart}, {Hoeksema}, {Liu}, {Duvall}, {Zhao}, {Title}, {Schrijver},
  {Tarbell}, \& {Tomczyk}}]{SDO_HMI}
{Scherrer}, P.~H., {Schou}, J., {Bush}, R.~I., {et~al.} 2012, \solphys, 275,
  207, \dodoi{10.1007/s11207-011-9834-2}

\bibitem[{{Schrijver} \& {De Rosa}(2003)}]{DeRosa2003}
{Schrijver}, C.~J., \& {De Rosa}, M.~L. 2003, \solphys, 212, 165,
  \dodoi{10.1023/A:1022908504100}

\bibitem[{{SILSO World Data Center}(1996-2020)}]{SILSO}
{SILSO World Data Center}. 1996-2020, International Sunspot Number Monthly
  Bulletin and online catalogue

\bibitem[{{Simpson} {et~al.}(1995){Simpson}, {Anglin}, {Bothmer}, {Connell},
  {Ferrando}, {Heber}, {Kunow}, {Lopate}, {Marsden}, {McKibben},
  {Muller-Mellin}, {Paizis}, {Rastoin}, {Raviart}, {Sanderson}, {Sierks},
  {Trattner}, {Wenzel}, {Wibberenz}, \& {Zhang}}]{Simpson1995}
{Simpson}, J.~A., {Anglin}, J.~D., {Bothmer}, V., {et~al.} 1995, Science, 268,
  1019, \dodoi{10.1126/science.268.5213.1019}

\bibitem[{{Singh} \& {Badruddin}(2019)}]{Singh2019}
{Singh}, Y.~P., \& {Badruddin}. 2019, \solphys, 294, 27,
  \dodoi{10.1007/s11207-019-1413-y}

\bibitem[{{Smith} {et~al.}(1998){Smith}, {L'Heureux}, {Ness}, {Acu{\~n}a},
  {Burlaga}, \& {Scheifele}}]{Smith1998}
{Smith}, C.~W., {L'Heureux}, J., {Ness}, N.~F., {et~al.} 1998, \ssr, 86, 613,
  \dodoi{10.1023/A:1005092216668}

\bibitem[{{Stakhiv} {et~al.}(2015){Stakhiv}, {Landi}, {Lepri}, {Oran}, \&
  {Zurbuchen}}]{Stakhiv2015}
{Stakhiv}, M., {Landi}, E., {Lepri}, S.~T., {Oran}, R., \& {Zurbuchen}, T.~H.
  2015, \apj, 801, 100, \dodoi{10.1088/0004-637X/801/2/100}

\bibitem[{{Stansby} {et~al.}(2019){Stansby}, {Horbury}, \&
  {Matteini}}]{Stansby2019}
{Stansby}, D., {Horbury}, T.~S., \& {Matteini}, L. 2019, \mnras, 482, 1706,
  \dodoi{10.1093/mnras/sty2814}

\bibitem[{{Steyn} \& {Burger}(2020)}]{Steyn2020}
{Steyn}, P.~J., \& {Burger}, R.~A. 2020, \apj, 902, 33,
  \dodoi{10.3847/1538-4357/abb2a5}

\bibitem[{{Strauss} {et~al.}(2017){Strauss}, {Dresing}, \&
  {Engelbrecht}}]{Strauss_2017}
{Strauss}, R. D.~T., {Dresing}, N., \& {Engelbrecht}, N.~E. 2017, \apj, 837,
  43, \dodoi{10.3847/1538-4357/aa5df5}

\bibitem[{{Thompson} {et~al.}(1997){Thompson}, {Newmark}, {Gurman},
  {Delaboudiniere}, {Clette}, \& {Gibson}}]{Thompson1997}
{Thompson}, B.~J., {Newmark}, J.~S., {Gurman}, J.~B., {et~al.} 1997, in ESA
  Special Publication, Vol. 404, Fifth SOHO Workshop: The Corona and Solar Wind
  Near Minimum Activity, ed. A.~{Wilson}, 779

\bibitem[{{Thompson}(2006)}]{Thompson2006}
{Thompson}, W.~T. 2006, \aap, 449, 791, \dodoi{10.1051/0004-6361:20054262}

\bibitem[{{Tsichla} {et~al.}(2019){Tsichla}, {Gerontidou}, \&
  {Mavromichalaki}}]{Tsichla2019}
{Tsichla}, M., {Gerontidou}, M., \& {Mavromichalaki}, H. 2019, \solphys, 294,
  15, \dodoi{10.1007/s11207-019-1403-0}

\bibitem[{{Wang} {et~al.}(1996){Wang}, {Hawley}, \& {Sheeley}}]{Wang1996}
{Wang}, Y.-M., {Hawley}, S.~H., \& {Sheeley}, Neil~R., J. 1996, Science, 271,
  464, \dodoi{10.1126/science.271.5248.464}

\bibitem[{{Wang} {et~al.}(2009){Wang}, {Ko}, \& {Grappin}}]{Wang2009}
{Wang}, Y.~M., {Ko}, Y.~K., \& {Grappin}, R. 2009, \apj, 691, 760,
  \dodoi{10.1088/0004-637X/691/1/760}

\bibitem[{{Wang} \& {Sheeley}(1992)}]{Wang1992}
{Wang}, Y.~M., \& {Sheeley}, N.~R., J. 1992, \apj, 392, 310,
  \dodoi{10.1086/171430}

\bibitem[{{Wilson} {et~al.}(2021){Wilson}, {Brosius}, {Gopalswamy},
  {Nieves-Chinchilla}, {Szabo}, {Hurley}, {Phan}, {Kasper}, {Lugaz},
  {Richardson}, {Chen}, {Verscharen}, {Wicks}, \& {TenBarge}}]{WIND2021}
{Wilson}, Lynn~B., I., {Brosius}, A.~L., {Gopalswamy}, N., {et~al.} 2021,
  Reviews of Geophysics, 59, e2020RG000714, \dodoi{10.1029/2020RG000714}

\bibitem[{{Yang} {et~al.}(2012){Yang}, {Feng}, {Xiang}, {Liu}, {Zhao}, \&
  {Wu}}]{Feng2012}
{Yang}, L.~P., {Feng}, X.~S., {Xiang}, C.~Q., {et~al.} 2012, Journal of
  Geophysical Research (Space Physics), 117, A08110,
  \dodoi{10.1029/2011JA017494}

\bibitem[{{Yeates} {et~al.}(2018){Yeates}, {Amari}, {Contopoulos}, {Feng},
  {Mackay}, {Miki{\'c}}, {Wiegelmann}, {Hutton}, {Lowder}, {Morgan}, {Petrie},
  {Rachmeler}, {Upton}, {Canou}, {Chopin}, {Downs}, {Druckm{\"u}ller},
  {Linker}, {Seaton}, \& {T{\"o}r{\"o}k}}]{Yeates2018}
{Yeates}, A.~R., {Amari}, T., {Contopoulos}, I., {et~al.} 2018, \ssr, 214, 99,
  \dodoi{10.1007/s11214-018-0534-1}

\bibitem[{{Zhao} {et~al.}(2017){Zhao}, {Landi}, {Lepri}, {Gilbert},
  {Zurbuchen}, {Fisk}, \& {Raines}}]{Zhao2017}
{Zhao}, L., {Landi}, E., {Lepri}, S.~T., {et~al.} 2017, \apj, 846, 135,
  \dodoi{10.3847/1538-4357/aa850c}

\bibitem[{{Zurbuchen} {et~al.}(1997){Zurbuchen}, {Schwadron}, \&
  {Fisk}}]{Zurbuchen1997}
{Zurbuchen}, T.~H., {Schwadron}, N.~A., \& {Fisk}, L.~A. 1997, \jgr, 102,
  24175, \dodoi{10.1029/97JA02194}

\end{thebibliography}
\bibliographystyle{aasjournal}

\appendix
\label{appendix}

\begin{figure*}[h]
  \centering
  \subfigure[]{\includegraphics[width=0.49\textwidth]{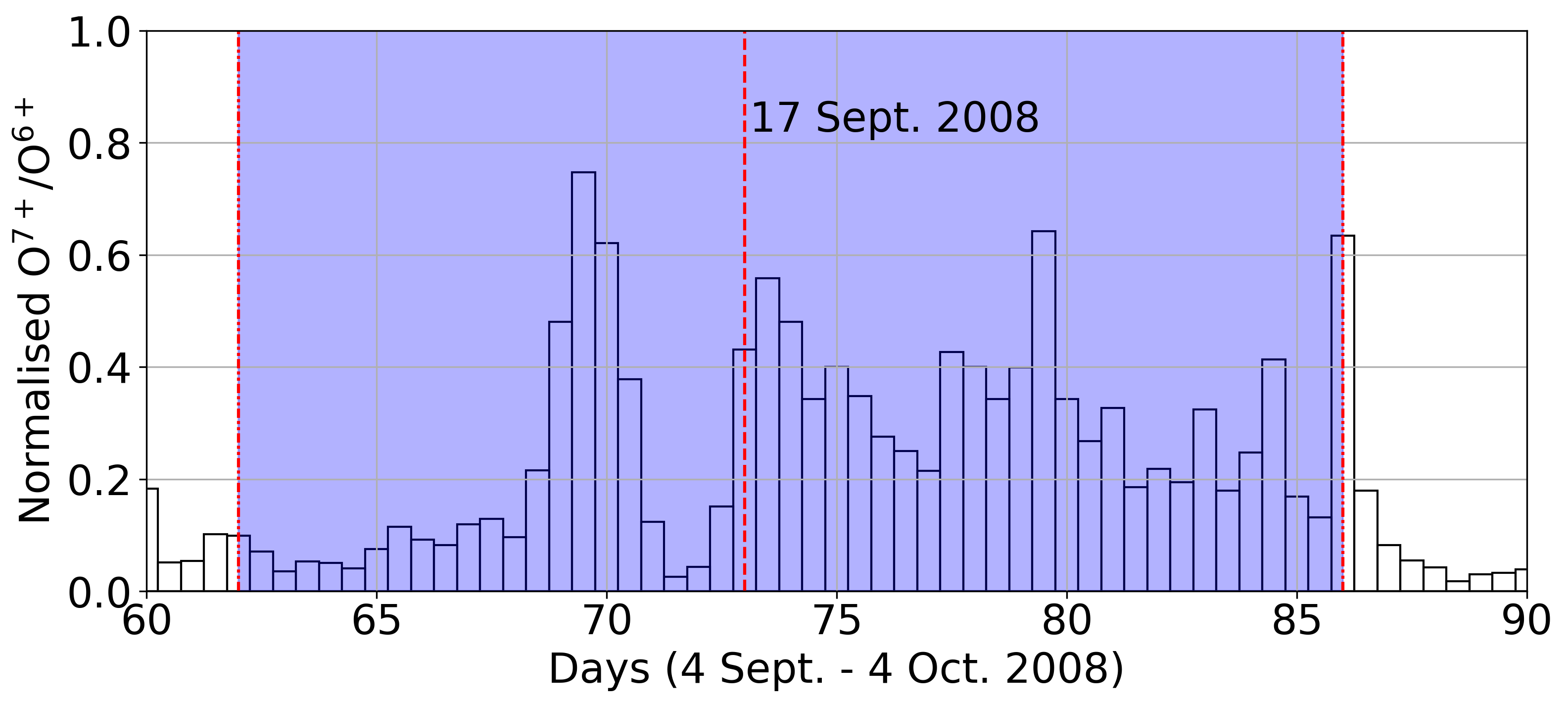}\label{fig:17Sept2008}} 
  \subfigure[]{\includegraphics[width=0.49\textwidth]{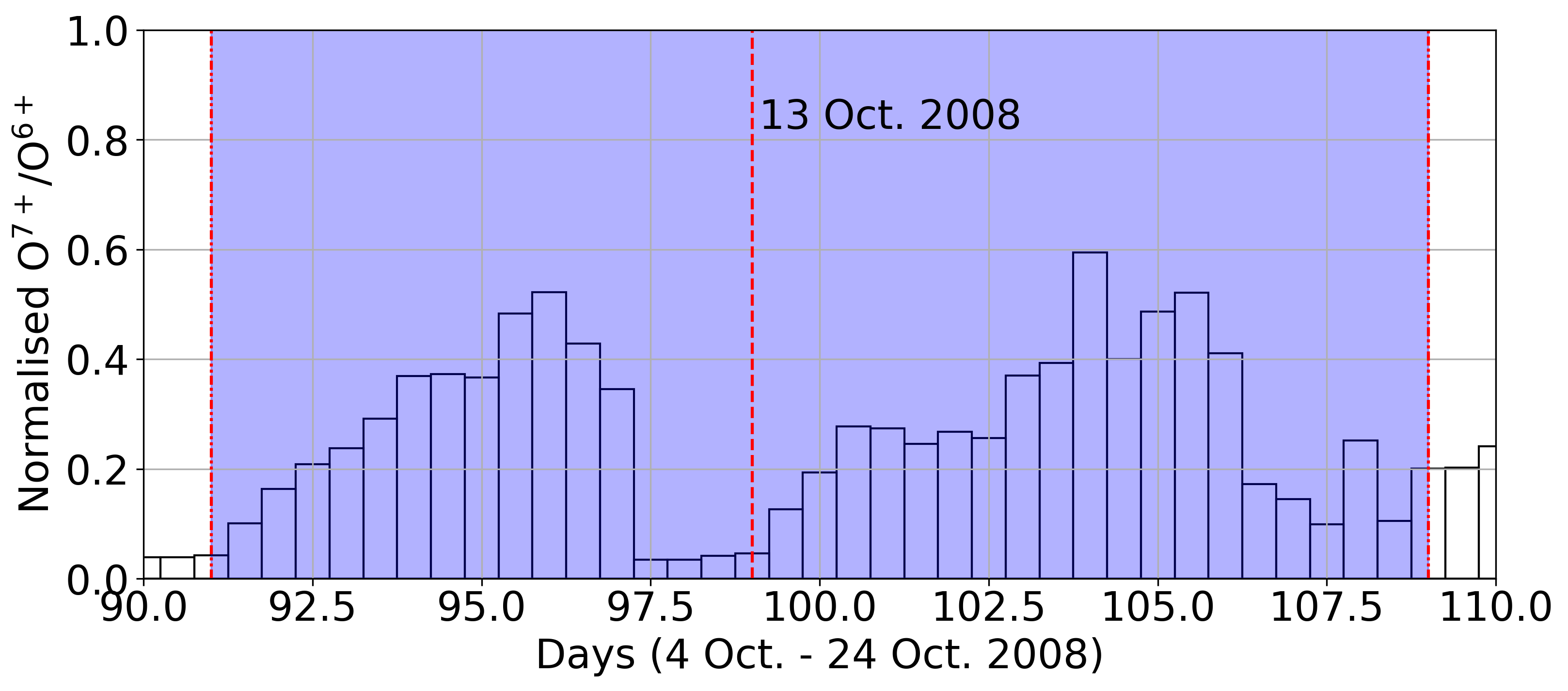}\label{fig:13Oct2008}}
  \subfigure[]{\includegraphics[width=0.49\textwidth]{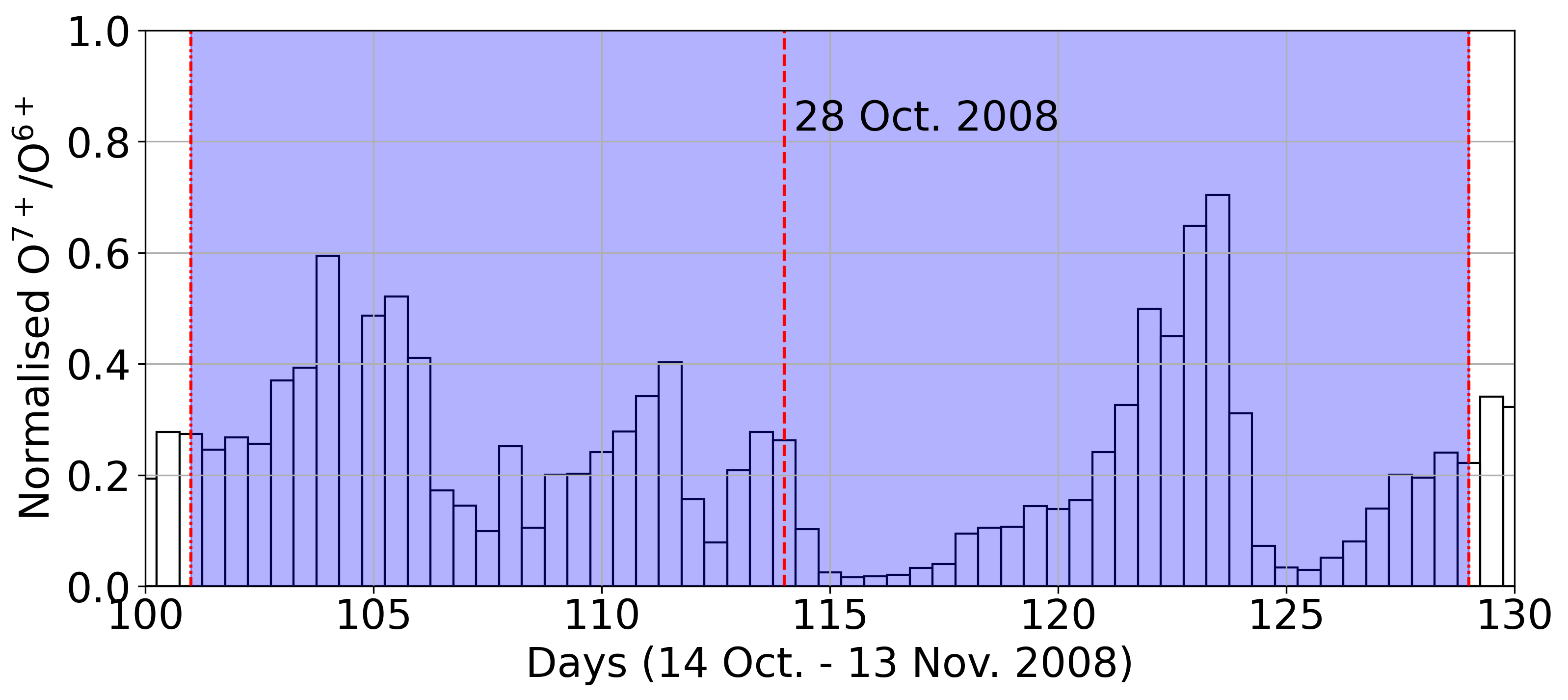}\label{fig:28Oct2008}} 
  \subfigure[]{\includegraphics[width=0.49\textwidth]{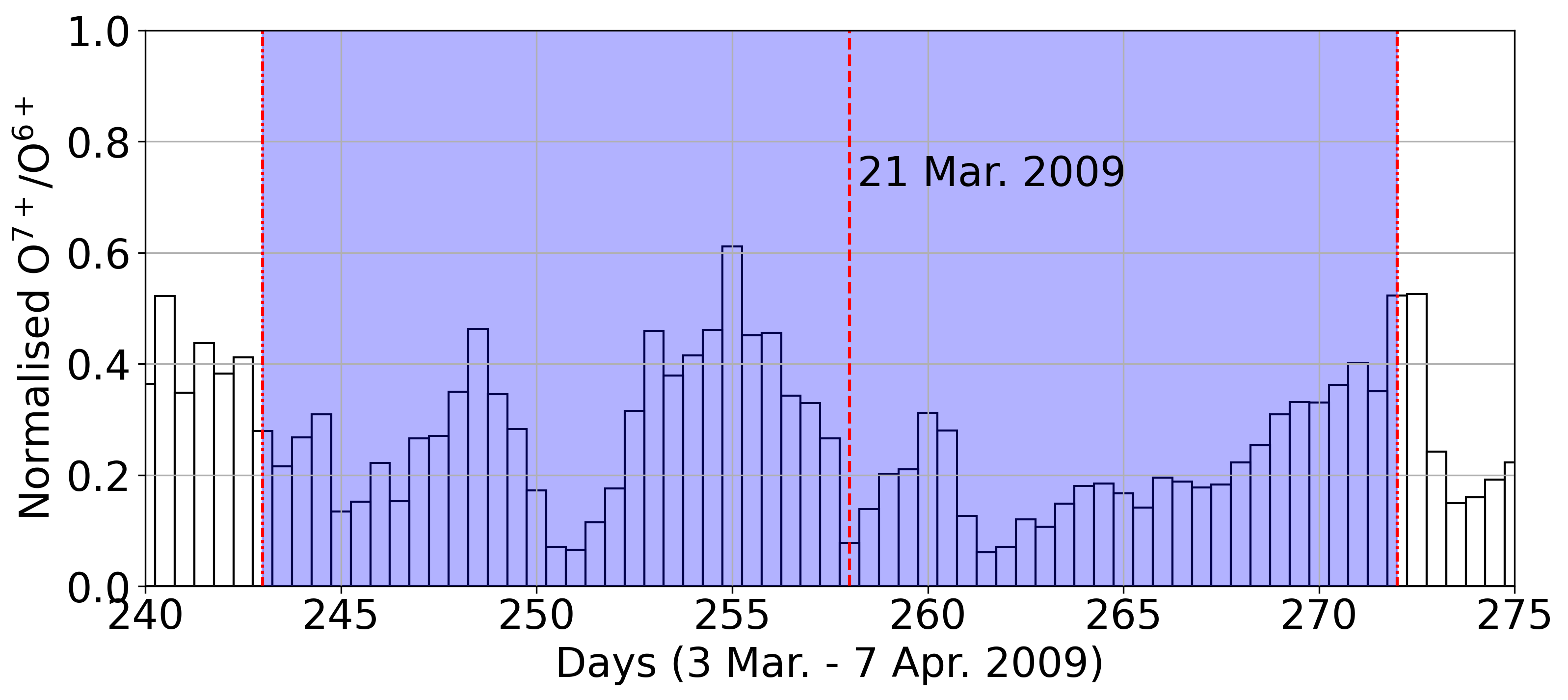}\label{fig:21March2009}}
  \subfigure[]{\includegraphics[width=0.49\textwidth]{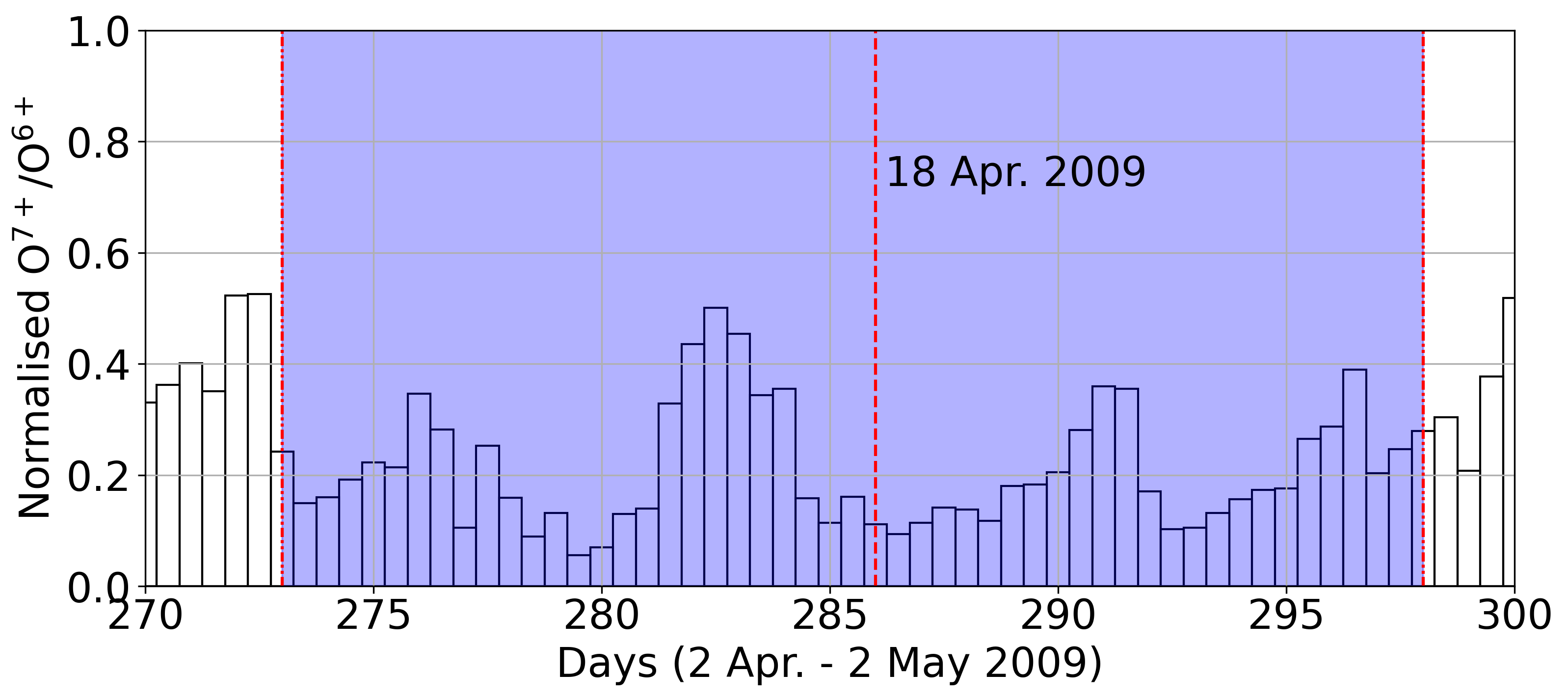}\label{fig:18Apr2009}}
  \caption{Normalised O$^{7+}$/O$^{6+}$ ratios for the $2008$-$2009$ solar minimum. The bottom axis of each panel indicates the number of days since $6$ July $2008$. The red dashed vertical line (in the middle of the blue shaded area) shows the date of the maximum signal according to the wavelet analysis of Section \ref{subsection: 2008/2009 Solar Min} while the left and right edges of the blue area indicate the $5$\% decrease from the maximum signal (see Figure \ref{fig:2009}). The corresponding Carrington Rotation of each panel, according to Section \ref{subsection:2008/2009 solar minimum magnetic connection} and Figure \ref{fig:CR2009}, is (a) CR$2074$, (b) CR$2075$, (c) CR$2075$ - $2076$, (d) CR$2080$ - $2081$ and (e) CR$2081$ - $2082$} 
\label{fig:BarChart2008_2009}
\end{figure*}

\cite{Zhao2017} investigated the proton speed and heavy ion charge state ratios (O$^{7+}$/O$^{6+}$) of different coronal source solar wind types between $1998$ to $2011$, including $2000$ to $2002$ (solar maximum) and $2007$ to $2009$ (solar minimum.) Their results show that proton speed alone is not sufficient to identify PCHs since solar wind speeds from PCHs range between $300$ kms$^{-1}$ and $800$ kms$^{-1}$. Furthermore, \cite{Zhao2017} show by correlating EIT Carrington maps with the oxygen ratio using PFSS that the O$^{7+}$/O$^{6+}$ ratio of the solar wind types is ordered such that the helmet-streamer and active region winds have the highest ratio (normalised over the dataset) while CH-boundary and CH winds have the lowest ratios.

\begin{figure*}[t!]
  \centering
  \subfigure[]{\includegraphics[width=0.49\textwidth]{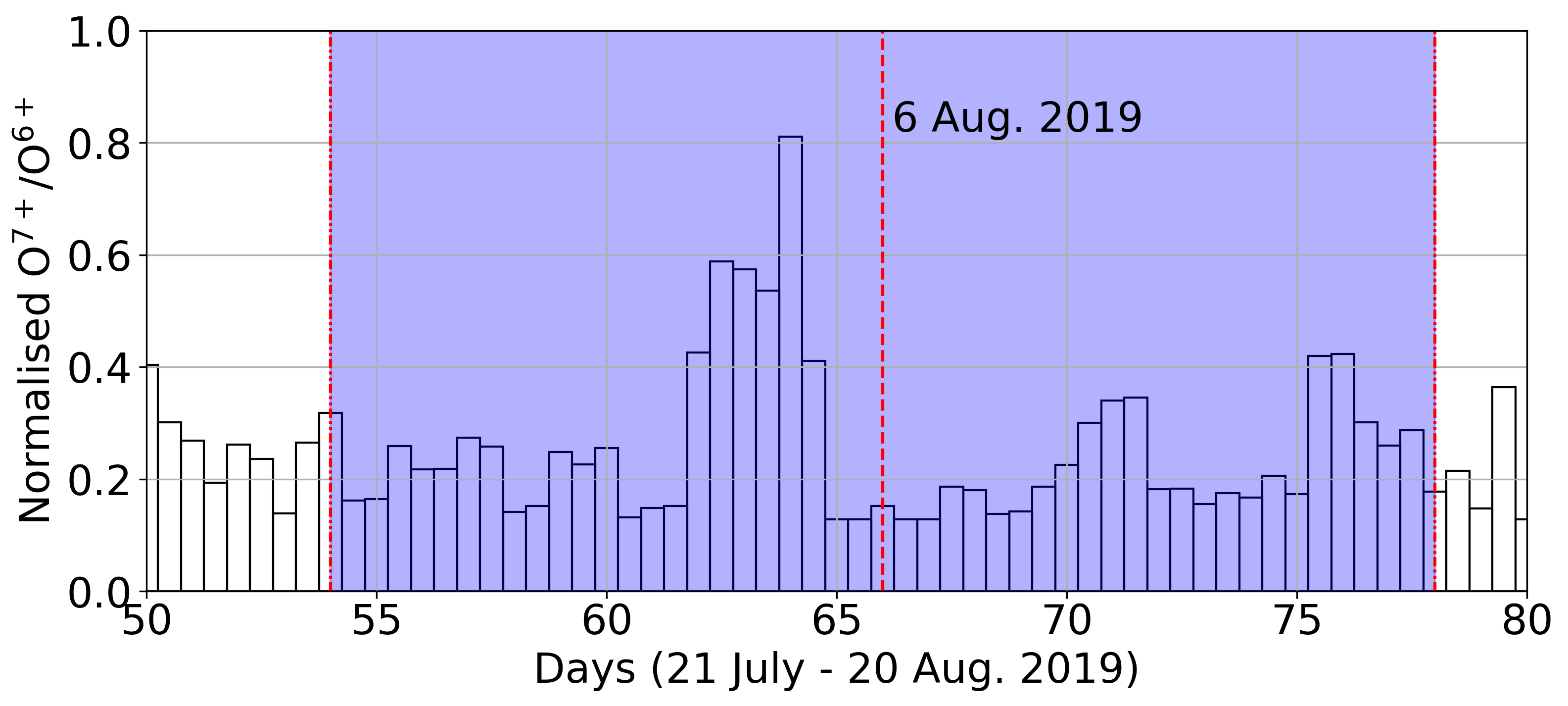}\label{fig:6Aug2019}} 
  \subfigure[]{\includegraphics[width=0.49\textwidth]{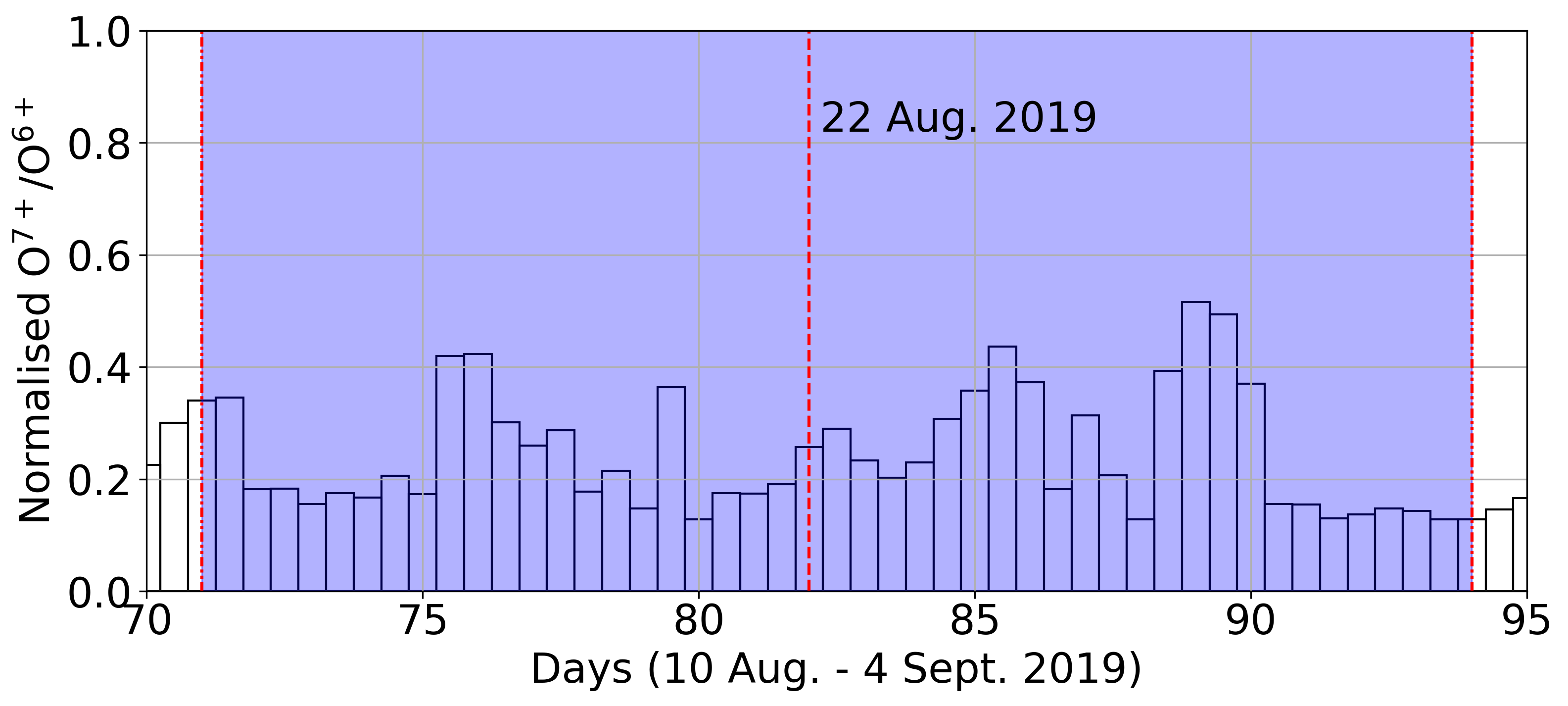}\label{fig:22Aug2019}}
  \subfigure[]{\includegraphics[width=0.49\textwidth]{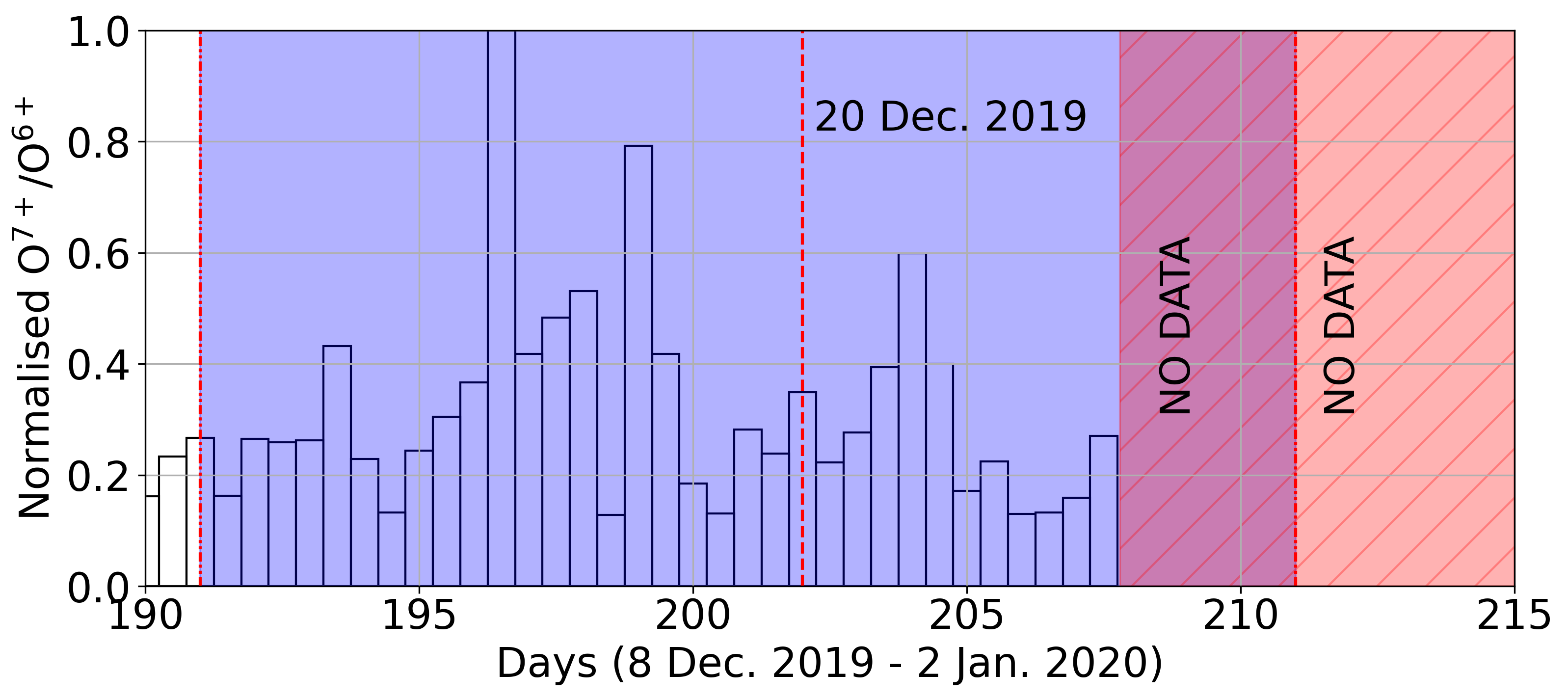}\label{fig:20Dec2019}} 
  \subfigure[]{\includegraphics[width=0.49\textwidth]{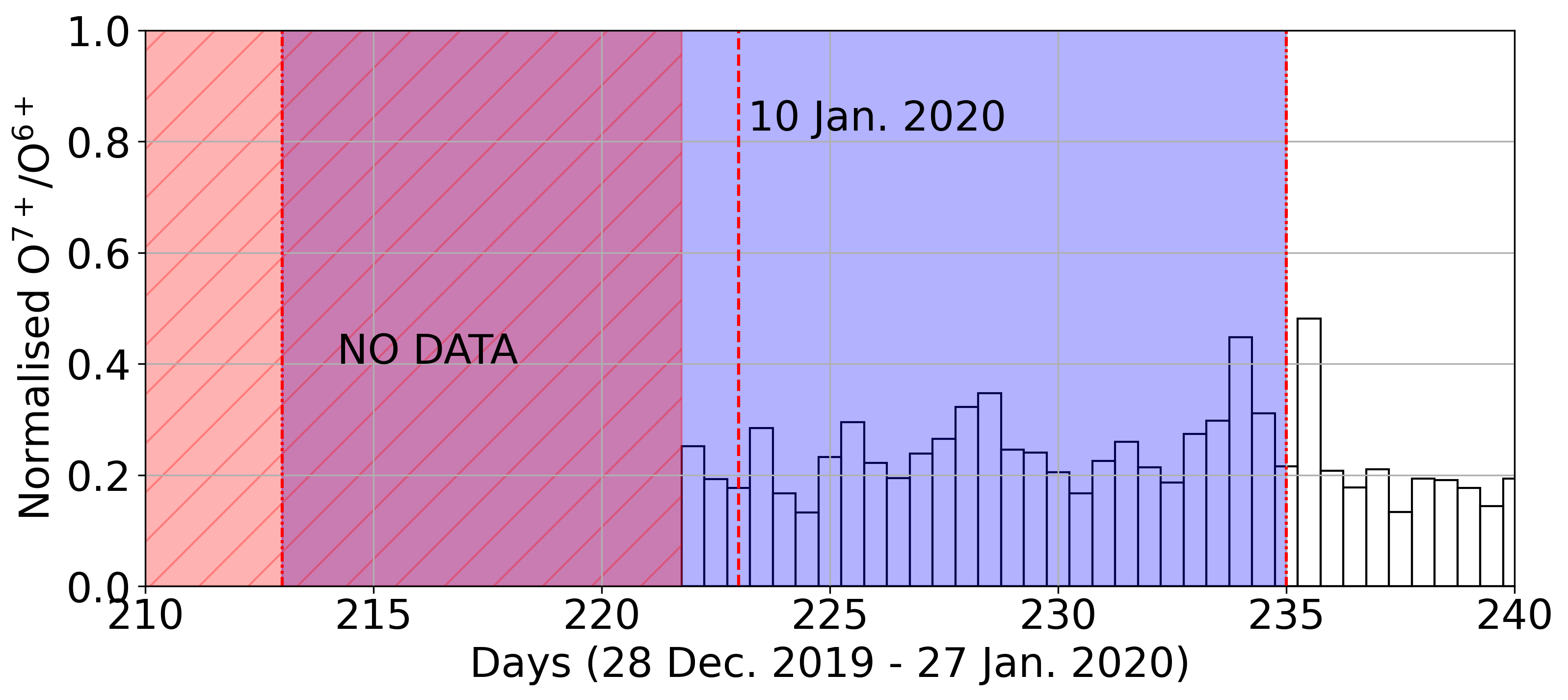}\label{fig:10Jan202}}
  \caption{Normalised O$^{7+}$/O$^{6+}$ ratios for the $2019$-$2020$ solar minimum. The bottom axis of each panel indicates the number of days since $1$ June $2019$. The red dashed vertical line (in the middle of the blue shaded area) shows the date of the maximum signal according to the wavelet analysis of Section \ref{subsection: 2019/2020 Solar Min} while the left and right edges of the blue area indicate the $5$\% decrease from the maximum signal (see Figure \ref{fig:2020}). The corresponding Carrington Rotation of each panel, according to Section \ref{subsection:2019/2020 solar minimum magnetic connection} and Figure \ref{fig:CR2020}, is (a) CR$2219$ - $2220$, (b) CR$2220$ - $2221$, (c) CR$2224$ - $2225$, and (d) CR$2225$ - $2226$.} 
\label{fig:BarChart2019_2020}
\end{figure*}

The two-hour averaged O$^{7+}$/O$^{6+}$ ratios (normalised in terms of the maximum value observed during one year) measured by ACE/SWICS \citep{SWICS1998} during the $2008$-$2009$ and $2019$-$2020$ solar minima are presented in Figures \ref{fig:BarChart2008_2009} and \ref{fig:BarChart2019_2020}. The maximum O$^{7+}$/O$^{6+}$ ratio during the $2008$-$2009$ and $2019$-$2020$ solar minima is $0.27$ and $0.41$, respectively. Since the two maxima are not the same, the results from the two solar minima are not directly compared with each other. Each panel in Figures \ref{fig:BarChart2008_2009} and \ref{fig:BarChart2019_2020} is linked to its corresponding Carrington Rotation from Sections \ref{subsection:2008/2009 solar minimum magnetic connection} and \ref{subsection:2019/2020 solar minimum magnetic connection}.

A low O$^{7+}$/O$^{6+}$ ratio is observed at the start of the blue shaded region in Figure \ref{fig:17Sept2008} which corresponds to Days $5$ to $10$ in Figure \ref{fig:CR2074} where a PCH is observed. Next, the PCH boundary is reached and the O$^{7+}$/O$^{6+}$ ratio increases (Day $69$ in Figure \ref{fig:17Sept2008}) after which it decreases (Day $72$), increases again (Day $74$), and continues on a relative high level up until the $5\%$ decrease from the maximum value is reached. The low O$^{7+}$/O$^{6+}$ ratios approximately correspond to visible PCHs in Figure \ref{fig:CR2074}. Although Figure \ref{fig:13Oct2008} shows a drop in O$^{7+}$/O$^{6+}$ ratio on Day $98$, only a comparatively small PCH (dark green pixels) is observed during CR$2075$ in Figure \ref{fig:CR2075_76}. There is a larger PCH earlier during this CR on Day $10$ of CR$2075$ which remains within the window of the $5\%$ decrease from the maximum signal observed. Figure \ref{fig:28Oct2008} shows an increasing and decreasing trend which corresponds to the scattered PCHs and PCH boundaries shown in Figure \ref{fig:CR2075_76} during CR$2076$. A similar increase and decrease of O$^{7+}$/O$^{6+}$ ratios are observed in Figures \ref{fig:21March2009} and \ref{fig:18Apr2009} corresponding to the PCH boundaries and PCHs from Figures \ref{fig:CR2080} and \ref{fig:CR2081_82}. The assumption of a constant solar wind speed between L$1$ and the photosphere (see Section \ref{section:Magnetic field models}) is idealistic (also see comments made by \cite{Zhao2017}) and therefore the exact dates at which the maximum signal is observed in the wavelet analyses will not necessarily correspond to the exact dates where the  O$^{7+}$/O$^{6+}$ ratio is a minimum. However, the results show that the minima in the O$^{7+}$/O$^{6+}$ ratios occur in the vicinity of the date of the maximum observed signal.

CR$2219$ and $2220$ show long lasting, continuous PCHs in Figures \ref{fig:CR2219} and \ref{fig:CR2220_21} which corresponds to the relative low O$^{7+}$/O$^{6+}$ ratios in Figure \ref{fig:6Aug2019}. This trend continues into CR$2221$ which corresponds to the Figure \ref{fig:22Aug2019}. The PCH in the southern hemisphere during CR$2224$ and $2225$ vary in size and intensity (Figure \ref{fig:CR2224_25}) which corresponds to the increasing and decreasing O$^{7+}$/O$^{6+}$ ratios shown in Figure \ref{fig:20Dec2019}. Note there is a data gap in the latter part of Figure \ref{fig:20Dec2019} and at the start of Figure \ref{fig:10Jan202}. A large, continuous PCH during the latter part of CR$2225$ and at the start of CR$2226$ corresponds well with the low O$^{7+}$/O$^{6+}$ ratios observed in Figure \ref{fig:10Jan202}.

Combining the results from the wavelet analyses (Sections \ref{subsection: 2008/2009 Solar Min} and \ref{subsection: 2019/2020 Solar Min}), the Carrington maps (Sections \ref{subsection:2008/2009 solar minimum magnetic connection} and \ref{subsection:2019/2020 solar minimum magnetic connection}) and the results from the O$^{7+}$/O$^{6+}$ ratios, provide a comprehensive description about whether the investigated solar wind originates from a PCH or not.

\end{document}